\documentclass[a4paper,fleqn,usenatbib]{mnras}

\usepackage{newtxtext,newtxmath}

\usepackage[T1]{fontenc}
\usepackage{ae,aecompl}

\usepackage{graphicx}	
\usepackage{amsmath}	
\usepackage{amssymb}	

\newcommand{\mtrx}[1]{\mathsf{\mathbf{#1}}}
\newcommand{\vctr}[1]{\mathbf{#1}}

\title[WL Mass Calibration of eROSITA Cluster Counts]{Impact of Weak Lensing Mass Calibration on eROSITA 
Galaxy Cluster Cosmological Studies -- a Forecast}

\author[S. Grandis et al.]{S. Grandis,$^{1,2}$\thanks{E-mail: s.grandis@physik.lmu.de}
J. J. Mohr,$^{1,2,3}$
J. P. Dietrich,$^{1,2}$
S. Bocquet,$^{1, 2}$
A. Saro,$^{4}$ 
\newauthor
M. Klein,$^{1,3}$
M. Paulus,$^{1,3}$ and R. Capasso$^{1,2}$
\\
$^{1}$Faculty of Physics, Ludwig-Maximilians-Universit\"at, Scheinerstr. 1, 81679, Munich,  Germany\\
$^{2}$Excellence Cluster Universe, Boltzmannstr. 2, 85748, Garching, Germany\\
$^{3}$Max Planck Institute for Extraterrestrial Physics, Giessenbachstr. 85748, Garching, Germany \\
$^{4}$INAF-Osservatorio Astronomico di Trieste, Via G.B. Tiepolo 11, I-34143 Trieste, Italy}

\date{Accepted XXX. Received YYY; in original form ZZZ}

\pubyear{2018}

\hypersetup{draft}
\begin{document}
\label{firstpage}
\pagerange{\pageref{firstpage}--\pageref{lastpage}}
\maketitle

\begin{abstract}
We forecast the impact of weak lensing (WL) cluster mass calibration on 
the cosmological constraints from the 
X-ray selected galaxy cluster counts in the upcoming eROSITA survey.  We employ a prototype
cosmology pipeline to analyze mock cluster catalogs.
Each cluster is sampled from the mass function in a fiducial cosmology and given an 
eROSITA count rate and redshift, where count 
rates are modeled using the eROSITA effective area, a typical exposure time, Poisson noise and 
the scatter and form of the observed X-ray luminosity-- and temperature--mass--redshift 
relations.  
A subset of clusters have mock shear profiles to mimic either those from 
DES and HSC
or from the future Euclid and LSST surveys.  
Using a count rate selection, we generate a baseline cluster cosmology catalog that contains 
13k clusters over 14,892~deg$^2$ of extragalactic sky.  Low mass groups are excluded 
using raised count rate thresholds at low redshift.
Forecast parameter uncertainties for $\Omega_\mathrm{M}$, $\sigma_8$ and $w$ are 0.023 (0.016; 0.014),  0.017 
(0.012; 0.010), and 0.085 (0.074; 0.071), respectively, when adopting DES+HSC WL (Euclid; LSST), 
while marginalizing over the sum of the neutrino masses. 
A degeneracy between the distance--redshift relation and the parameters of the observable--mass 
scaling relation limits the impact of the WL calibration on the $w$ constraints, but with
BAO measurements from DESI an improved determination of $w$ to 
0.043 becomes possible. With Planck CMB priors, 
$\Omega_\text{M}$ ($\sigma_8$) can be determined to $0.005$ ($0.007$), 
and the summed neutrino mass limited to $\sum m_\nu < 0.241$ eV (at 95\%).
If systematics on the group mass scale can be controlled, the eROSITA group and cluster sample 
with 43k objects and LSST WL could constrain
$\Omega_\mathrm{M}$ and $\sigma_8$ to 0.007 and $w$ to 0.050.
\end{abstract}




\section{Introduction}

Over the last decade, measuring the number density of galaxy clusters as a function of observable and redshift 
has proven to be a potent way to determine not only the density and clustering of matter in the Universe, but 
also to shed light on the yet unknown source of the late time accelerated expansion of the Universe 
\citep{Koester07, vikhlinin08, mantz10, rozo10, benson13, Mantz15, bocquet15, planck16cluster_cosmo, 
dehaan16, bocquet18}. To this end, ever larger samples of galaxy clusters have been selected in X-rays 
\citep[][]{vikhlinin98,boehringer01,romer01,clerc14,klein19}, at millimeter wavelengths \citep{Hasselfield13, 
bleem15, planck16_sze}, and in the optical \citep{Koester07, rykoff16}. Extracting accurate cosmological 
constraints from these samples depends critically on the ability to determine the mapping between the 
observable in which the samples have been selected, and the halo mass over the relevant range of redshifts.  
This aspect is commonly referred to as \textit{mass calibration}.

Two main methods have been developed for this purpose. 
The first-- weak lensing (hereafter WL)-- the coherent distortion of the shapes of galaxies behind galaxy clusters 
by the cluster gravitational potential has 
proven to be the method of choice to calibrate masses \citep[e.g.,][]{bardeau07, okabe10, hoekstra12, 
applegateetal14, israel14, Melchior15, okabesmith16, melchior17,  schrabback18a, dietrich19}. Alternatively, 
the dynamics of the cluster galaxies themselves has been used within recent cluster surveys to calibrate the 
cluster halo masses \citep{sifon13,bocquet15,capasso19, zhang17}.
On individual clusters, these methods characteristically provide a low signal to noise mass constraint with low bias 
that, importantly, can be reliably characterized using numerical structure formation simulations.  For example, 
the scaling between the mass observed through WL (hereafter the WL mass) and the halo 
mass can be calibrated to robustly characterize the biases and  scatter \citep[e.g.,][]{becker11}.  With modern 
hydrodynamical simulations it is now possible to include baryon physics in this calibration \citep{lee18}.  
Similarly, the biases and scatter in dynamical mass estimators can be characterized using numerical 
simulations \citep[e.g.,][]{evrard08,mamon13} in a manner that includes the impact of the (red) galaxy sample 
selection \citep{saro13}.

A third method-- hydrostatic masses using X-ray observations-- has played an important role in the development of 
our understanding of galaxy clusters, but through simulation studies and comparison with WL 
masses, these hydrostatic masses have been shown to be biased at the $\sim20$\% level or more \citep[see, 
e.g.,][]{nagai07,rasia12,vonderlinden14,hoekstra15,shi15,planck16cluster_cosmo,planck_cosmo_legacy18}, although the 
scale of the 
bias remains a topic of ongoing research \citep{smith16,gupta17}.  This hydrostatic mass bias together with 
the availability of shear catalogs from deep, multiband surveys and the increasingly large wide field 
spectroscopic datasets, have created a situation where the X-ray hydrostatic masses no longer offer clear 
benefits within the context of large scale cluster cosmological studies.

The low signal to noise of individual cluster WL mass measurements is compensated to some degree by 
the larger number of galaxy clusters that can be studied. This stems from the fact that, in addition to the 
cluster observables of redshift and position, cluster weak lensing mass calibration requires the same data as 
cosmic shear experiments. The advent of deep, large area photometric imaging surveys with a well controlled 
point spread function correction for accurate shape measurements and high quality photometric redshifts now 
enables the WL study of large samples of galaxy clusters \citep{Melchior15, murata18, miyatake18,
mcclintock19,stern19}.

It is in this context that we investigate the impact of WL mass calibration on the cluster cosmology results from the 
X-ray selected sample that will be extracted from the all sky X-ray survey undertaken with the forthcoming 
eROSITA\footnote{\url{http://www.mpe.mpg.de/eROSITA}} telescope \citep{Predehl10,merloni12} on board 
the Russian "Spectrum-Roentgen-Gamma" satellite.  Previous analyses adopting a Fisher matrix approach have explored
the constraining power of the eROSITA cluster sample on non Gaussianities \citep{pillepich12} and the dark 
energy equation of state parameter \citep{pillepich18}, further underscoring the promise of 
cluster number counts as a cosmological probe \citep[e.g.,][]{haiman01}.  

In this work, we create a mock cluster catalog with 
characteristics of the expected eROSITA catalog, and we use a prototype of the eROSITA cluster cosmology 
analysis code to perform the number counts experiment. We consider the improvement in constraining 
power when the eROSITA X-ray cluster catalog is calibrated with realistic WL lensing shear profiles from the 
ongoing Dark Energy Survey\footnote{\url{https://www.darkenergysurvey.org}} \citep[DES,][]{DES16} 
and Hyper-Suprime-Cam Survey\footnote{\url{https://www.naoj.org/Projects/HSC/}}
\citep[HSC,][]{HSC},
and the forthcoming Euclid\footnote{\url{http://sci.esa.int/euclid/42266-summary/}} \citep{laureijs11} 
and Large Synoptic Survey Telescope\footnote{\url{https://www.lsst.org/}}  \citep[LSST,][]{Ivezic08} surveys.  
We explore parameter sensitivities and probe for limiting degeneracies in the analysis.
Finally, we explore the synergies of combining the eROSITA cluster counts cosmological constraints with 
those from existing CMB temperature anisotropy measurements \citep{planck16_cosmo} and with those from 
the future DESI BAO measurements \citep{levi13}.

The paper is organized as follows: in Section~\ref{sec:setup} we discuss how we create the mock data. In 
Section~\ref{sec:method} we discuss the modeling used to determine the cosmological parameters, and we 
present and validate a prototype of the eROSITA cosmological analysis pipeline. In Section~\ref{sec:results}, 
we present the results of the impact of WL mass calibration on our knowledge of the cosmological 
parameters and the observable mass relation. Various aspects of these results together with the parameter 
sensitivities and important degeneracies are then discussed in 
Section~\ref{sec:discussion}. We conclude this work by summarizing the main results in 
Section~\ref{sec:conclusions}.

\section{Experimental setup}\label{sec:setup}

To constrain the impact of direct mass calibration through WL tangential shear measurements on 
eROSITA cluster cosmology, we create an eROSITA mock cluster catalog. The actual eROSITA cluster candidate 
catalog will be extracted from the eROSITA X-ray sky survey using specially designed 
detection and characterization tools \citep{brunner18}.  

Each candidate source will be assigned a detection significance, an extent 
significance, an X-ray count rate and uncertainty, and other more physical parameters such as the flux within 
various observing bands \citep{merloni12}. For a subset of this sample, precise X-ray temperatures and 
rough X-ray redshifts will also be available \citep{borm14, hofmann17}.

This X-ray cluster candidate catalog will then be studied in the optical to identify one or more optical counterparts 
(assigning a probability to each) and to estimate a photometric redshift.  
A special purpose Multi-Component-Matched-Filter (MCMF) optical followup tool \citep{klein18}
 has been designed for eROSITA cluster analysis 
and has been tested on available X-ray and SZE catalogs.  It has been 
shown in RASS+DES analyses that one can reliably obtain both cluster and group redshifts over the 
relevant ranges of redshift \citep{klein19}, 
and thus for the analysis undertaken here 
we assume redshifts are available for all the eROSITA clusters.

The MCMF tool also allows one to quantify the 
probability of chance superposition between X-ray cluster candidates and optical counterparts, using the 
statistics of optical systems along random lines of sight together with estimates of the initial contamination in 
the X-ray cluster candidate catalog.  
Synthetic sky simulations by \citet{clerc18} have shown that the initial X-ray cluster candidate list selected on 
both detection and extent significance will be contaminated at the 10\% level, consistent with experience 
in X-ray selection from archival ROSAT PSPC data that have a similar angular resolution to eROSITA \citep{vikhlinin98}. 
 After processing with MCMF the resulting eROSITA X-ray cluster catalog is expected to have contamination at 
the sub-percent level. Therefore, we do not include contamination in the mock catalogs produced for 
this study. 

For the WL mass calibration we will be using shear and photometric redshift catalogs from wide field, deep 
extragalactic surveys, including DES and HSC in the near term and Euclid and LSST on the longer term.   
The label ``Euclid'' refers to the nominal requirements for Euclid \citep{laureijs11}, although
these requirements will realistically be met when combining Euclid with LSST, where the LSST data
would be used for the photometric redshifts.
We also explore the impact of LSST WL alone, where we 
adopt the requirements described in the following references\citep{lsst_desc12,lsst_desc18}. 
There is also the promise of CMB lensing 
as another method of mass calibration that is expected to be especially helpful for the highest redshift end of 
our cluster sample, but in our current analysis we do not model the impact of CMB lensing.  

\begin{figure*}
	\includegraphics[width=\columnwidth]{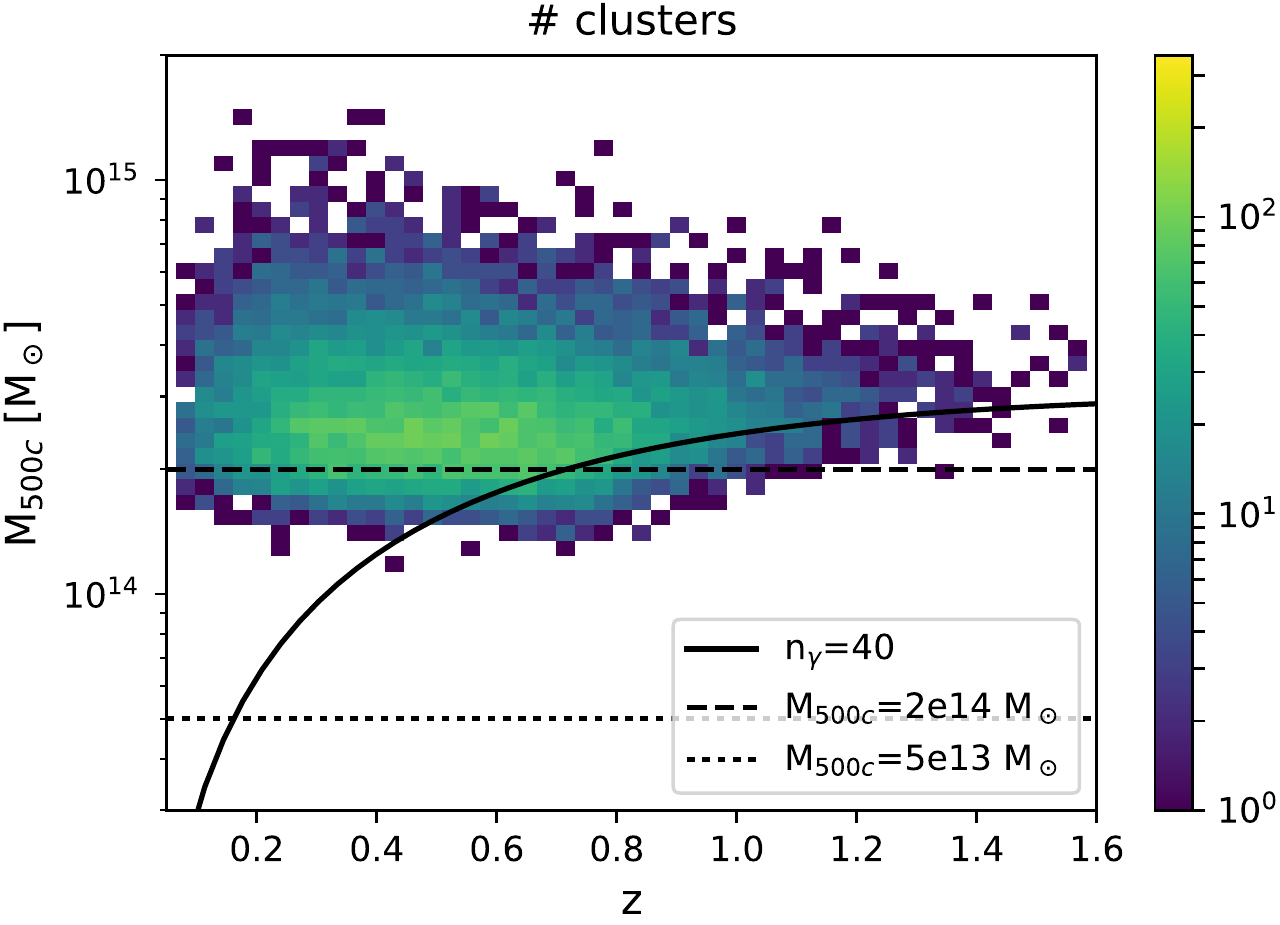}
	\includegraphics[width=\columnwidth]{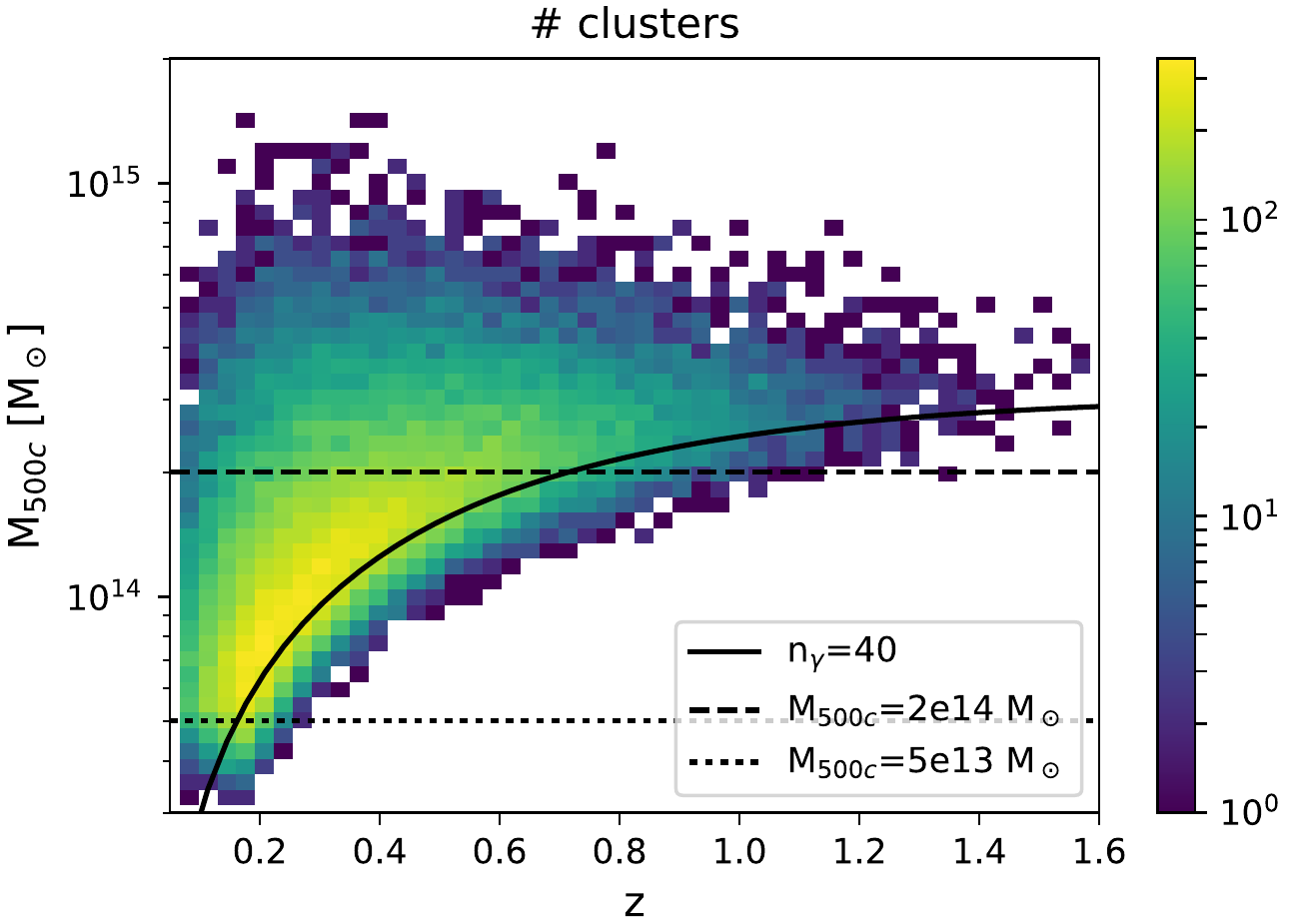}
	\vskip-0.15in
    \caption{Distribution in halo mass $M_\text{500c}$ and cluster redshift $z$ of the mock, X-ray selected 
cluster catalogs used in this analysis. \textit{Left:}  Above redshift $\sim 0.7$, the 13k cluster baseline 
sample is selected using the fiducial count 
rate cut $\eta=2.5\times10^{-2}$~cts~s$^{-1}$ that corresponds approximately to 
40 photons at the median exposure 
time and a signal to noise $\xi_\mathrm{det}>7$. 
Below that redshift the observable cut is pushed upward to mimic a mass exclusion at $M_\text{500c}\sim2\times 10^{14} 
\text{M}_{\sun}$.
Due to intrinsic and observational scatter between halo mass and 
the observable count rate, the cuts in observable used to create these samples appear 
smoothed in halo mass-redshift space.
 \textit{Right:} The 43k sample that includes groups is selected similarly but the count rate cut is adjusted to mimic a mass 
exclusion at  
$M_\text{500c}\sim5\times 10^{13} \text{M}_{\sun}$.}
\label{fig:hist_mz}
\end{figure*}

Our strategy in the analysis that follows is to adopt direct, cosmology independent cluster observables, 
including the cluster (1) X-ray 
detection significance or count rate, (2) photometric redshift, (3) WL tangential shear profile and (4) 
shear source redshift distributions for use in the cosmological analysis of the cluster sample.  A 
benefit of using the count rate rather than
the physical flux is that uncertainties in effective area and the temperature dependence of the conversion from
count rate to physical flux do not contribute to cosmological uncertainties.

Empirically mapping these 
observables to mass as a function of redshift and testing consistency of observed and theoretical cluster 
distributions as a function of cosmological parameters is described in Section~\ref{sec:method}.
Below, in Section~\ref{sec:x-ray_mock}, we describe how the mock cluster catalog is generated and how the X-ray and 
optical cluster properties are assigned. In Section~\ref{sec:wl_signal} we describe how we model the shear 
profiles that are produced for an appropriate subset of the mock eROSITA cluster sample.  We discuss briefly 
our choice of fiducial cosmology and input X-ray scaling relations in Section~\ref{sec:fiducial}.

\subsection{Creating the mock cluster catalog} \label{sec:x-ray_mock}

To create the X-ray catalog, we perform the following calculations.

\begin{enumerate}
    \item For our choice of input cosmology (see Table~\ref{tab:input_parmas} and Section~\ref{sec:fiducial}), we 
compute the number of expected clusters as a function of halo mass $M_{500 \text{c}}$ and redshift $z$ 
using the halo mass function  \citep{tinker08}. We then draw a Poisson realization of the number of 
expected clusters, obtaining a mass selected cluster sample with $ M_{500\text{c}} > 1.3\times 10^{13} 
\text{M}_{\sun} $ and $0.05<z<1.8$. For this calculation we assume a survey solid angle of $
\text{Area}_\text{DE} = 0.361 \times 4 \upi$, corresponding to regions of the western
galactic hemisphere with a galactic hydrogen column $N_{\text{H}}<10^{21}$ cm$^{-2}$  \citep{kalberla05}. 

This corresponds approximately to a galactic latitude cut of $|b|>20$ deg. 
We adopt the cluster true redshift as the photometric redshift, because the MCMF optical followup tool has 
been demonstrated to achieve photometric redshift uncertainties with the DES dataset with an accuracy of $
\sigma_z/(z+1)\loa 0.01$ \citep{klein18,klein19} out to redshifts $z\sim1.1$.  Photometric redshift uncertainties at 
this level are small enough to play no role in the cosmological analysis of the eROSITA cluster counts.
    \item We use the scaling between X-ray luminosity $L_{[0.5-2]\text{keV}}$ ($L_X$ hereafter) in the rest frame 
$0.5-2$ keV band and halo mass   
    \begin{equation} \label{eq:l_mz}
\frac{L_X}{L_0}= e^{\ln A_\text{L}}\left(\frac{M_{500\text{c}}}{M_0}\right)^{B_\text{L}} \left(\frac{E(z)}{E_0}\right)^2 
\left(\frac{1+z}{1+z_0}\right)^{\gamma_\text{L}} e^{\Delta_\text{L}},
\end{equation}
that was extracted from a large sample of SPT selected clusters 
with pointed XMM-{\it Newton} observations \citep{bulbul19}.  In this relation 
$E(z)=H(z)/H_0$ encodes the expansion history of the universe and is used to calculate the impact of changes in the 
critical density of the Universe ($\rho_\mathrm{crit}\propto E^2(z)$), $\ln A_\text{L}$, $B_\text{L}$ and $
\gamma_\text{L}$ are the amplitude, the mass trend and the non-self-similar redshift trend parameters
of the luminosity--mass scaling relation, and $\Delta_\text{L}\sim \mathcal{N}(0, 
\sigma^2_\text{L})$ is a random number drawn from a Gaussian with standard deviation $\sigma_\text{L}$, 
which models the log-normal intrinsic scatter of the relation.  

The \citet{bulbul19} X-ray scaling relations are 
derived from the Sunyaev-Zel'dovich effect (SZE) selected cluster sample from the SPT-SZ 2500 deg$^2$ 
survey \citep{carlstrom11,bleem15} that have available XMM-{\it Newton} observations.  This is a sample of 
59 clusters with $0.2\le z\le 1.5$ and masses $ M_{500\text{c}} > 3\times 10^{14} \text{M}_{\sun}$.  These halo 
masses have been 
calibrated separately in a cosmological analysis \citep{dehaan16} and exhibit a characteristic 
uncertainty of $\sim$20\% (statistical) and $\sim15$\% (systematic).  
The scaling relation parameter uncertainties from \citet{bulbul19} include both statistical and systematic 
uncertainties.

We also utilize the temperature mass relation  
    \begin{equation}\label{eq:t_mz}
\frac{T}{T_0}= e^{\ln A_\text{T}}\left(\frac{M_{500\text{c}}}{M_0}\right)^{B_\text{T}} \left(\frac{E(z)}
{E_0}\right)^{\frac{2}{3}} \left(\frac{1+z}{1+z_0}\right)^{\gamma_\text{T}} e^{\Delta_\text{T}},
\end{equation}
from the same analysis \citep{bulbul19}, where the parameters $(\ln A_\text{T}, B_\text{T}, \gamma_\text{T})$ 
have the same meaning as in the luminosity 
scaling relation, with $\Delta_\text{T}\sim \mathcal{N}(0, \sigma^2_\text{T})$ for the scatter $\sigma_\text{T}$. 
The only difference is the scaling with the critical density, derived from self similar collapse theory.  

Following these relations, we attribute to each cluster an X-ray luminosity $L_\text{X}$ and a temperature $T$, 
randomly applying the respective intrinsic log normal scatter and assuming that the two scatters are 
uncorrelated.

    \item Given the cluster rest frame 0.5-2 keV luminosity $L_\text{X}$ and its redshift $z$, we compute the rest 
frame 0.5-2 keV flux
\begin{equation} \label{eq:flux}
    f_\text{X} = \frac{L_X}{4 \upi d_\mathrm{L}^2(z)},
 \end{equation}
    where $d_\mathrm{L}(z)$ is the luminosity distance.
    \item For each cluster we calculate the X-ray spectrum assuming an APEC plasma emission model \citep{apec} 
with temperature $T$ and metallicity $Z = 0.3$ Z$_{\sun}$\footnote{For simplicity, we do not apply any scatter 
to the metallicity, and assume it is constant as a function of redshift, as recent measurements of the 
metallicity of SPT selected clusters suggest \citep{McDonald16}. We assume the solar abundances model of 
\citet{andersgrevesse89}}. This spectrum is normalized to the cluster rest frame 0.5-2 keV flux. 

    \item We compute the eROSITA count rate $\eta$ for each cluster by shifting the spectrum to the observed 
frame and by averaging it with the eROSITA Ancillary Response Function (hereafter ARF) in the observed 
frame 0.5-2 keV band\footnote{Of the seven eROSITA cameras, two have a 100 nm Al and 200 nm Pl filter, 
while the remaining five have a 200 nm Al and 200 nm Pl filter \citep{Predehl10,merloni12}. Consequently, 
the total ARF is the sum of two (100 nm Al + 200 nm Pl)-ARFs and five (200 nm Al + 200 nm Pl)-ARFs.}. For 
simplicity, we do not follow the variation in neutral hydrogen column across the eROSITA-DE field.  
In fact, we ignore the impact of Galactic absorption altogether in our count rate calculation, which for the median neutral  
hydrogen column density in our footprint, $N_\text{H}=3\times10^{20}$~cm$^{-2}$ would lead on average to 5\% 
lower rates.

    \item To model the measurement uncertainty on the rate, we draw a Poisson realization of the expected
rate $\hat \eta = \eta \pm \sqrt{\eta / t_\text{exp}}$, where $t_\text{exp}=1600$~s is the expected median 
exposure time of the 4 year eROSITA survey \citep{pillepich12}. With this we account for the Poisson noise in 
the rate measurement. The count rate uncertainty for each cluster will be included in the real eROSITA 
cluster catalogs.
    
\item Finally, we select our baseline cluster sample using the count rate $\eta>2.5\times10^{-2}$~ct~s$^{-1}$
(corresponding for our median exposure to $\hat n_\gamma > 40$).
For reference, given the background expectations, survey PSF and clusters modeled as $\beta$ models with 
core radii that are 20\% of the virial radius $r_{500}$, this selection threshold corresponds approximately 
to a cut in detection significance of $\xi_\text{det}>7$, 
irrespective of the cluster redshift.  Simple mock observations (see discussion in Appendix~\ref{app:selection}) indicate 
that at this threshold and above the extent likelihood for the eROSITA sample is
$\xi_\text{ext}>2.5$, enabling an initial eROSITA cluster 
candidate list after X-ray selection (but prior to optical followup) that is contaminated at the $\sim$10\% level.
At low redshift ($z<0.7$), we raise the detection threshold above the nominal level in such a way 
as to exclude most clusters with masses $M_\text{500c}\lessapprox 2\times 10^{14} \text{M}_{\sun}$ at each redshift.
We create a second sample to examine the impact of 
lower mass clusters and groups (see Section~\ref{sec:low_mass}) by adjusting the 
low redshift count rate cut so that systems with masses 
$M_\text{500c}\lessapprox 5\times 10^{13} \text{M}_{\sun}$ are excluded at each redshift.
We discuss the X-ray selection in more detail in 
Appendix~\ref{app:selection}. 
The reasons for excluding lower mass systems are discussed below (cf. 
Section~\ref{sec:low_mass}).
\end{enumerate}

The procedure described above provides us with a baseline cosmology catalog of $\sim 13$k clusters. 
Their distribution in halo 
mass\footnote{We use this binning in mass just to visualize our sample, the number counts analysis will be 
performed on a fixed grid of observed rate $\hat \eta$ and redshift, as specified in 
Section~\ref{sec:number_counts}. The corresponding mass grid depends on the cosmological and the scaling 
relation parameters, and is thus recomputed every time the likelihood function is called on a specific set of 
parameters.} and redshift is shown in the left panel of Fig.~\ref{fig:hist_mz}. They span a redshift range $z\in 
(0.05, 1.6)$. The total number of clusters and their redshift range are mainly impacted by the choice of the 
input cosmology, the observed luminosity mass relation, and the choice of cut in eROSITA count rate for selection. The 
sample 
has a median redshift $\bar z=0.51$ and median halo mass of $\bar M_{500\text{c}} =2.5\times 10^{14} 
\text{M}_{\sun}$. This sample extends to high redshift with 3\% of the sample, corresponding to 420 clusters, at $z>1$.

The sample of 43k objects with the count rate cut that only excludes lower mass systems 
with $M_\text{500c}\le5\times 10^{13} \text{M}_{\sun}$ is shown
in Fig.~\ref{fig:hist_mz} (right).  The bulk of the additional low mass systems in this sample appear at redshifts $z\le0.7$.
As with the overall number of clusters, the median mass and redshift depend on the observable cut used to 
exclude low mass objects, with these being $\bar z=0.30$, and $\bar M_{500\text{c}} =1.4\times 10^{14} 
\text{M}_{\sun}$. 
We discuss the implications of lowering the mass limit in Section~\ref{sec:low_mass}.  

The number of objects in this $\xi_\mathrm{det}>7$ group dominated sample is in good agreement with the
numbers presented in previous discussions of the eROSITA cluster sample \citep{merloni12,pillepich12,pillepich18}.  
Importantly, there are significantly more eROSITA clusters that can be detected if one reduces the detection 
threshold below $\sim7\sigma$.  But at that level there will be little extent information for 
each X-ray source, and so the candidate sample 
will be highly contaminated by AGN.  Interestingly, \citet{klein18} have demonstrated that for the
RASS faint source catalog where the survey PSF was so poor that little extent information is available, 
it is possible to filter out the non-cluster sources to produce low contamination cluster catalogs.  The price for this
filtering is that one introduces incompleteness for those systems that contain few galaxies 
\citep[i.e., low mass clusters and groups at each redshift; see][]{klein19}.

\subsection{Forecasting the WL signal}\label{sec:wl_signal}

We adopt the cosmology independent tangential reduced shear profile $\hat g_\text{t}(\theta_i)$ 
in radial bins $\theta_i$ around the cluster as the observable for cluster WL mass calibration.  
A crucial complementary observable is the redshift distribution of the source galaxies $N(z_\text{s}, 
z_\text{cl})$ behind the galaxy cluster, where $z_\text{s}$ is the source redshift, and $z_\text{cl}$ the cluster 
redshift. Assuming that the galaxy cluster mass profile is consistent with a Navarro-Frenk-White model \citep[]
[hereafter NFW]{NFW}, these two observables can be combined into a measurement of the halo mass. 

Although, in theory, WL mass calibration provides a direct mass measurement, in practice we refer to the mass 
resulting from an NFW fit to the shear profile as the WL mass $M_\text{WL}$. Following \citet{becker11}, the 
WL mass is related to the halo mass by 
\begin{equation}\label{eq:mwl}
M_\text{WL} = b_\text{WL} M_\text{200c} e^{\Delta_\text{WL}},
\end{equation}
with $\Delta_\text{WL}\sim\mathcal{N}(0, \sigma_\text{WL}^2)$, where $\sigma_\text{WL}$ is the intrinsic log-
normal scatter between WL mass and halo mass, induced by the morphological deviation of observed galaxy cluster 
mass profiles from the NFW profile, and $b_\text{WL}$ is the WL mass bias describing the characteristic bias
in the WL mass compared to the halo mass.  This bias encodes 
several theoretical and observational systematics, as discussed below in Section~\ref{sec:WL_syst}.

Given that DES, HSC, Euclid and LSST will not overlap completely with the German eROSITA sky, only a fraction 
$f_\text{WL}$ of the galaxy clusters of our X-ray mock catalog will have WL information available. Comparing 
the survey footprints, we estimate $f_\text{WL}=0.3$ for DES,  $f_\text{WL}=0.05$ for HSC, $f_\text{WL}=0.5$ for 
Euclid, 
and $f_\text{WL}=0.62$ for LSST.  For the LSST case we also assume that the northern celestial hemisphere 
portion of the German eROSITA sky with $0^\circ<\delta<30^\circ$ will be observed. 
For this northern extension of LSST, we adopt $f_\text{WL}=0.2$ 
and treat it as if it has the equivalent of DES depth.
Therefore, we assign a WL mass only to a corresponding fraction of the eROSITA clusters in our mock catalogs, 
by drawing from equation~(\ref{eq:mwl}).

Besides the WL mass and the cluster redshift, the background source distribution of the survey $N(z_\text{s})
$ in redshift and the background source density $n_\epsilon$ are necessary to predict the WL signal. For 
DES, we project $n_\epsilon=10 \, \text{arcmin}^{-2}$ and utilize the redshift distribution  presented in 
\citet{stern19}, whose median redshift is $z_\text{s,m}=0.74$. These parameters are derived from the Science Verification 
Data and their extrapolation to Y5 
data will depend on the details of the future calibration (Gruen, priv. comm.).  
For HSC we assume $n_\epsilon=21 \, \text{arcmin}^{-2}$, and for the redshift distribution of HSC sources 
we adapt the parametrization by \citet{smail94} with a median redshift $z_\text{s,m}=1.1$.
For Euclid, we use 
$n_\epsilon=30 \, \text{arcmin}^{-2}$ \citep{laureijs11}. For the source redshift distribution we assume the 
parametric form proposed by \citet{smail94} and utilized by \citet{Giannantonio14}, adopting a median redshift 
of $z_\text{s,m}=0.9$ \citep{laureijs11}. 
For LSST we assume $n_\epsilon=40 \, \text{arcmin}^{-2}$ and parametrise the source redshift distribution as 
$p(z_\text{s})=1./(2 z_0) (z_\text{s}/z_0)^2 \exp(-z_\text{s}/z_0)$ with median redshift $z_\text{m,s} = 2.67 z_0 = 0.8$\footnote{These specification are taken from \url{https://www.lsst.org/sites/default/files/docs/sciencebook/SB_3.pdf}, Section 3.7.2}. 

\begin{figure*}
	\includegraphics[width=0.49\textwidth]{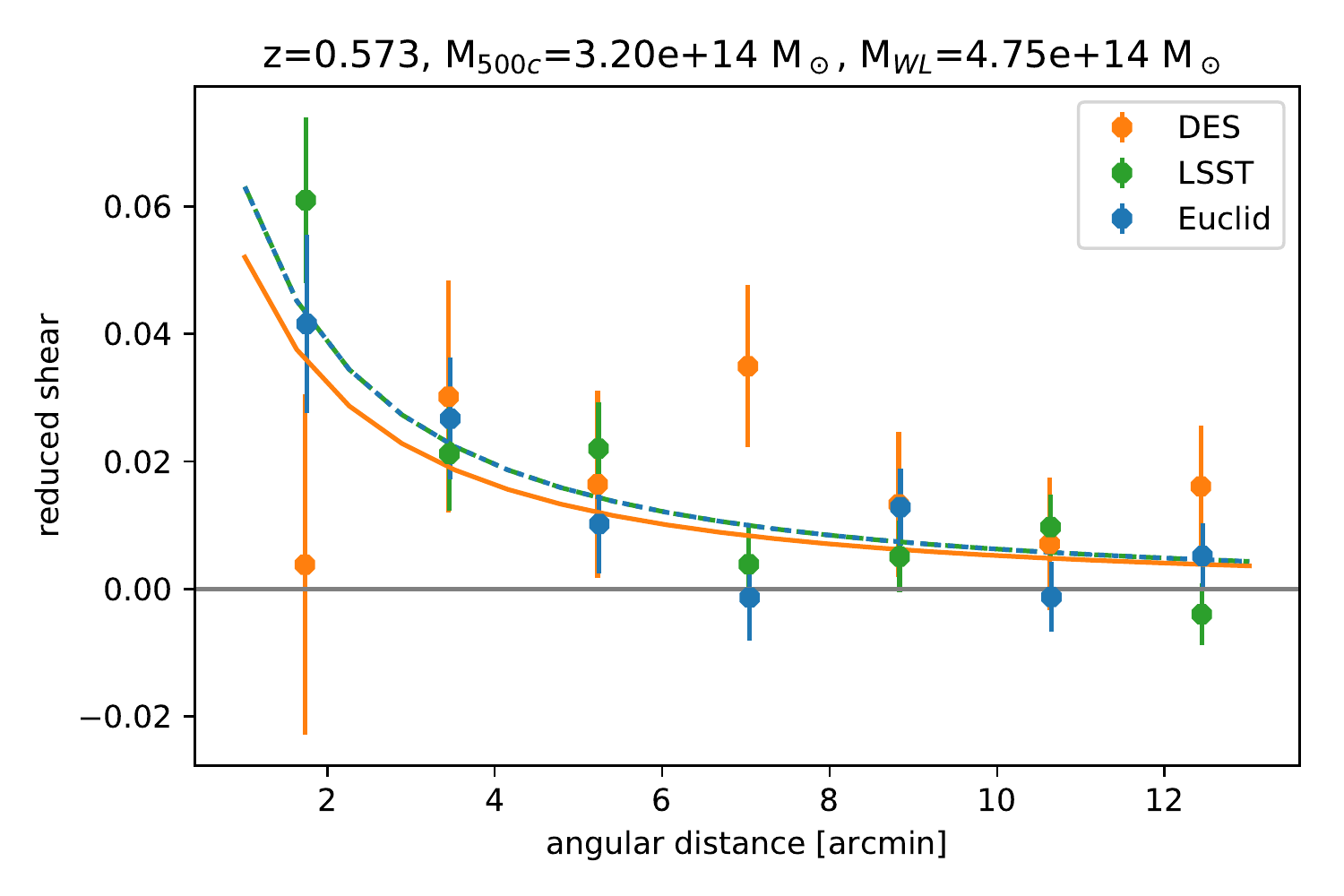}
	\includegraphics[width=0.49\textwidth]{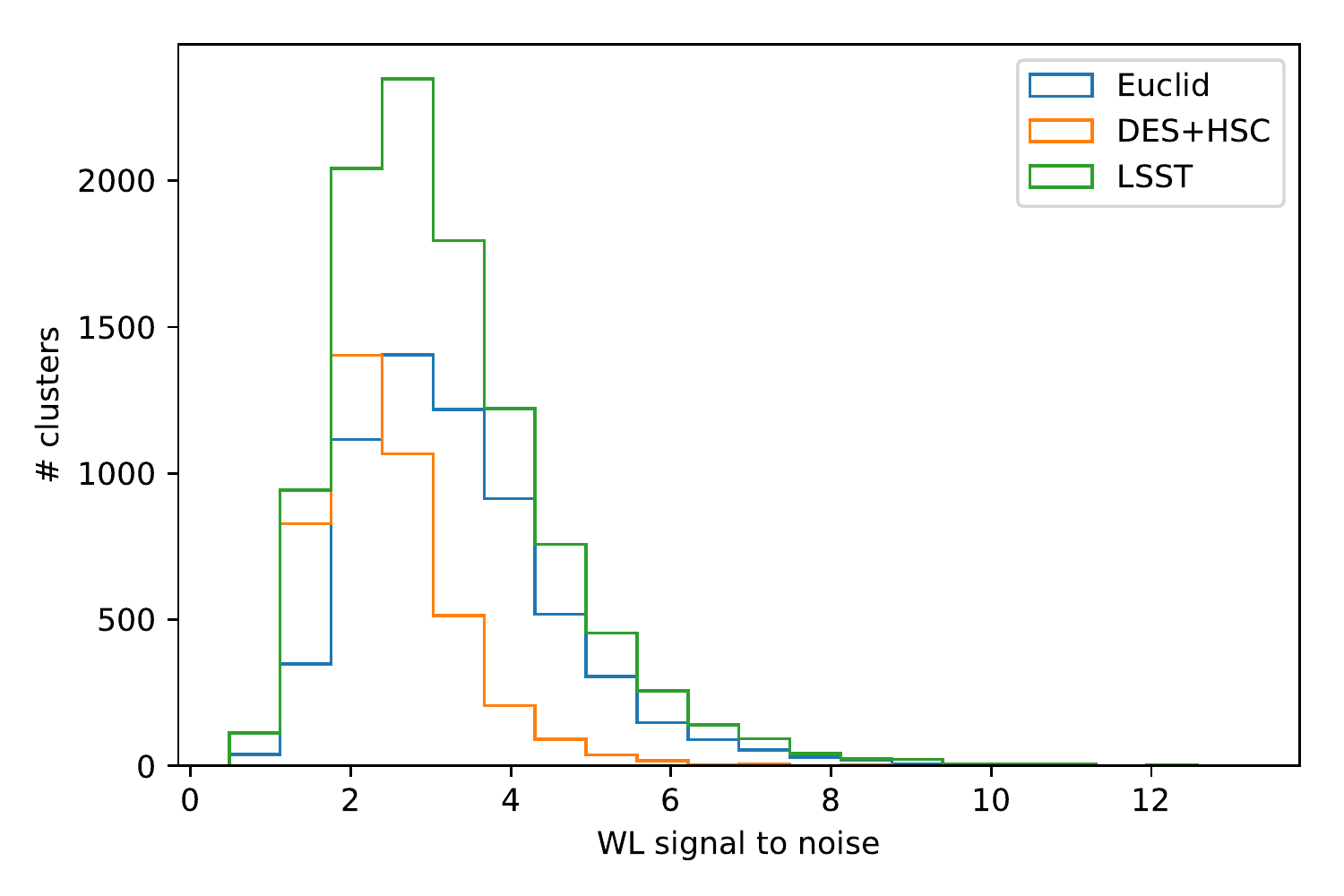}
	\vskip-0.1in
    \caption{\textit{Left:} Example of a shear profile in DES (orange), Euclid (blue) and LSST (green) data quality for 
    a cluster. 
    We show both the measured shear profile (dots with error bars) and the prediction (line). For all 
data quality cases, the measurement uncertainty is larger than the actual signal.
    \textit{Right:} Distribution of WL signal to noise for DES+HSC (orange), Euclid (blue) and LSST (green), computed for 
each single cluster from the measured shear profile and the covariance matrix. While Euclid and LSST provide 
both more objects and higher signal to noise, objects with a clear WL mass measurement (e.g., S/N$>5$) are rare 
for all datasets.}
    \label{fig:wl}
\end{figure*}

The actual redshift distribution behind a galaxy cluster is assumed to be the survey redshift distribution with the cut 
$N(z_\text{s}<z_\text{cl}+0.1) = 0$, where $z_\text{cl}$ is the cluster redshift. This cut is helpful in reducing the 
contamination of the background source galaxies by cluster galaxies (that are not distorted by the cluster potential).   
This cut also leads to a reduction 
of the source density $n_\epsilon(z_\text{s}>z_\text{cl}+0.1)$ used to infer the observational noise on the 
cluster shear signal.

Given a redshift distribution, the mean reduced shear signal can be estimated, following \citet{seitzschneider97}, 
as 
\begin{equation}\label{eq:gt}
g_\text{t} (\theta_i) = \frac{\gamma (\theta_i) }{1-\kappa (\theta_i)} \left(1+\kappa (\theta_i) \frac{\langle \beta^2 
\rangle}{\langle \beta \rangle ^2} \right),
\end{equation}
where $\gamma (\theta_i)$ and  $\kappa (\theta_i)$ are the shear and the convergence of an NFW mass 
profile, $\theta_i$ the angular bins corresponding to radii between $0.25$ and $5.0$~Mpc at the cluster redshift  
in our fiducial cosmology. 
This has the effect that low redshift clusters will have larger angular bins than high redshift clusters in 
to probe the similar physical scales. 
Also note that the inner radius, which we probe ($0.25$~Mpc), is smaller than in some previous studies 
\citep[$0.75$~Mpc in][]{applegateetal14, stern19, dietrich19}. 
While this will require a more precise treatment of systematic effects such a cluster member contamination, 
miscentering and the impact of intra-cluster light on the shape and redshift measurements, 
theoretical predictions for the resulting WL mass bias and WL mass scatter associated with these smaller inner radii have 
already been presented \citep{lee18}. Furthermore, \citet{gruen18} investigated the impact of intra-cluster light on the 
photometric redshift measurement of background galaxies. 
We therefore assume that ongoing and future studies will demonstrate the possibility of exploiting shear information at 
smaller cluster radii, thereby increasing the amount of extracted mass information.
 
Following \citet{bartelmann96}, the shear and the convergence can be computed analytically for any halo, given 
the mass,  the concentration, and the source galaxy redshift distribution $N(z_\text{s},z_\text{cl})$. 
Throughout this 
work, the concentration of any cluster will be derived from its halo mass, following the relation presented by 
\citet{duffy08}. The scatter in concentration at fixed halo mass is a contributor to the bias $b_\text{WL}$ and scatter $
\sigma_\text{WL}$ in the WL mass to halo mass relation (equation~\ref{eq:mwl}). 
The lensing efficiency $\beta=d_\mathrm{A}(z_\text{cl},z_\text{s}) / d_\mathrm{A}
(z_\text{s})$ is the ratio between the angular diameter distance $d_\mathrm{A}(z_\text{cl},z_\text{s})$ from 
the cluster to the source, and the angular diameter distance $d_\mathrm{A}(z_\text{s})$ from the observer to 
the source. In equation~(\ref{eq:gt}) the symbol $\langle \cdot  \rangle$ denotes averaging over the source redshift 
distribution 
$N(z_\text{s},z_\text{cl})$. 

The covariance of the measurement 
uncertainty on the reduced shear is
\begin{equation}
\mtrx{C}_{i,j} = \text{Cov} [g_\text{t} (\theta_i) , g_\text{t} (\theta_j) ] = \frac{\sigma^2_\epsilon}{\Omega_i 
n_\epsilon(z_\text{cl})}\delta_{i,j} + (\mtrx{C}_\text{uLSS})_{i,j}
\end{equation}
where $\delta_{i,j}=1$, if $i=j$, and $\delta_{i,j}=0$ else. The first term accounts for the shape noise in each radial bin,
estimated by scaling the intrinsic shape noise of the source galaxies $
\sigma_\epsilon=0.27$ by the number of source galaxies in each radial bin, taking into 
account the reduction of source galaxy density $n_\epsilon(z_\text{cl})=n_\epsilon(z_\text{s}>z_\text{cl}+0.1)$ and the 
angular area of the $i$-th radial bin $\Omega_i$.  We also add a contribution 
coming from uncorrelated large scale structure $(\mtrx{C}_\text{uLSS})_{i,j}$ \citep{hoekstra03}. We draw the 
measured reduced shear profile $\hat g_\text{t}$ from the Gaussian multivariate distribution with mean 
$g_\text{t}$ and covariance $\mtrx{C}$. 

For each cluster with WL information, we thus save the source redshift distribution $N(z_\text{s},z_\text{cl})$, the 
measured reduced shear profile $\hat g_\text{t}$, and the covariance $C$. We show an example for a 
measured reduced shear profile, both in DES, in Euclid and in LSST data quality in the left panel of 
Fig.~\ref{fig:wl}.

The WL signal around individual galaxy clusters derived from wide and deep photometric surveys is typically low 
signal to noise. In the right panel of Fig.~\ref{fig:wl}, we explore the distribution of WL signal to noise for the 
subsamples with DES+HSC WL data, Euclid WL data and LSST WL data. To this end we define the signal to noise as $
\text{S}/\text{N}=\sqrt {0.5\,\hat g_\text{t}^T\mtrx{C}^{-1}g_\text{t}}$. While the Euclid and LSST data provide a 
higher signal to noise on average, it rarely exceeds $\text{S}/\text{N}>5$. Thus, we confirm that WL mass 
calibration provides a low signal to noise, direct mass measurement for a large subset of our cluster catalog.

\begin{table}
	\centering
	\caption{Input parameters for our analysis. The exact definition of the parameters listed below is given in 
Section~\ref{sec:sampling}, Section~\ref{sec:scalingrelation} and Section~\ref{sec:wl_signal} for the 
cosmological parameters, the scaling relation parameters and the WL calibration parameters, respectively. 
\newline \textit{Comments: a)}~This value is determined to match $\sigma_8=0.768$ by \citet{dehaan16}. 
~\textit{b)}~We utilize here the value corresponding to the minimal model of a Cosmological Constant causing 
the accelerated expansion. \textit{c)}~This is the minimal value allowed by flavor neutrino oscillations, as 
reviewed by \citet{tanabashi18}. }
	\label{tab:input_parmas}
	\begin{tabular}{lcc} 
		\hline
		\multicolumn{3}{l}{Cosmological Parameters}\\
		\hline
		$H_\text{0}$ & 73.02 & \citet{riess16}\\
		$\omega_\text{b}$ & 0.02202 & \citet{cooke14}\\
		$\Omega_\text{M}$ & 0.306 & \citet{dehaan16}\\
		$A_\text{S}$ & 1.5792e-9 & a) \\
		$n_\text{S}$ & 0.9655 & \citet{planck16_cosmo} \\
		$w$ & -1.00 & b) \\
		$\sum m_\nu$ & 0.06 eV & c) \\
		$\Omega_\text{K}$ & 0. & \\ 
 		\hline
		\multicolumn{3}{l}{Luminosity--Mass--Redshift Relation}\\
		\hline
 		$\ln A_\text{L}$ & 1.52 & \citet{bulbul19} \\
 		$B_\text{L}$ & 1.95 & \\
 		$\gamma_\text{L}$ & -0.20 & \\
 		$\sigma_{L}$ & 0.237 & \\
 		\hline
		\multicolumn{3}{l}{Temperature--Mass--Redshift Relation}\\
		\hline
 		$\ln A_\text{T}$ & 1.83 & \citet{bulbul19} \\
 		$B_\text{T}$ & 0.849 & \\
 		$\gamma_\text{T}$ & -0.28 & \\
 		$\sigma_{T}$ & 0.177 & \\
 		\hline
		\multicolumn{3}{l}{WL Mass Bias and Scatter}\\
		\hline
 		$b_\text{WL}$ & 0.94 & \citet{dietrich19} \& \\
 		$\sigma_\text{WL}$ & 0.24 & \citet{lee18}\\
 		\hline
	\end{tabular}
\end{table}

\subsection{Fiducial cosmology and scaling relations}
\label{sec:fiducial}
Several steps in the above outlined creation of the mock data are cosmology sensitive. Therefore, the choice of 
input cosmology will impact the catalog properties. As an input cosmology, we choose the best fitting 
$\Omega_\text{M}$ and $\sigma_8$ results 
from the most recent SPT galaxy cluster cosmology analysis \citep{dehaan16}. 
We also assumed that dark energy can be described by a cosmological constant, i.e. that the dark energy equation of state 
parameter $w=-1$. Furthermore, we adopt the minimal neutrino mass allowed by flavor neutrino oscillation measurements, $
\sum m_\nu=0.06$ eV \citep{tanabashi18}. 
The parameter values are listed 
in Table~\ref{tab:input_parmas}. 

It is worth noting here that these input values for $\Omega_\text{M}$ and $\sigma_8$ are somewhat different (at less than 2$
\sigma$ significance) 
from the best fit values derived from the Planck CMB anisotropy measurements  \citep{planck16_cosmo}. This choice is 
intentional, as the masses of SPT clusters derived from a mass function fit with Planck CMB priors have been 
shown to be systematically high by studies of their WL signal \citep{dietrich19, stern19}, their 
dynamical mass \citep{capasso19} and their baryon content \citep{chiu18}. Furthermore, the input X-ray 
scaling relations by \citet{bulbul19}, adapted to determine the X-ray properties of our catalog entries, assume 
an SZE signature--mass--redshift scaling relation consistent with the best fit results from the SPT galaxy 
cluster cosmology analysis.  In summary, the input values for our analysis are chosen from the latest results of the 
SPT galaxy cluster sample, guaranteeing consistency between the assumed cosmology and the input X-ray scaling 
relations that we use to construct the mock eROSITA sample. 
Given that SPT covers a mass range of $M_{500\text{c}}\goa 3 \times 10^{14} M_{\sun}
$, and a redshift range of $z\in (0.20, 1.7)$, adopting SPT results within the eROSITA context implies only a 
modest extrapolation in mass and redshift.

On the other hand, the minimal neutrino mass is slightly inconsistent with recent results from joint fits to number counts of 
SPT selected clusters and Planck CMB measurements \citep{dehaan16, bocquet18}, 
which detect the neutrino mass at $2$-$3$ sigma. 
This detection is likely sourced by the slight inconsistency in the $(\Omega_\text{M}, \sigma_8$) plane discussed above. 
For the sake of this work, we adapt the minimal neutrino mass to predict improvement on the upper limits obtained, 
if cluster number counts and CMB measurements were in perfect agreement.

\section{Cosmology analysis method}
\label{sec:method}

In this section we describe the method we have developed for the cosmological analysis of an eROSITA cluster 
sample in the presence of WL mass calibration information.  This method builds upon a method developed 
and used for the analysis of the SPT SZE selected cluster sample \citep{bocquet15,dietrich19,stern19,bocquet18}.  We 
start with a description of the minimal scaling relation to describe the mapping of the selection observable to 
halo mass as a function of redshift (Section~\ref{sec:scalingrelation}), present the likelihoods in 
Section~\ref{sec:likelihoods} and discuss the likelihood sampling tool and our adopted priors in 
Sections~\ref{sec:sampling} and \ref{sec:priors}.

\subsection{Cluster selection scaling relation}\label{sec:scalingrelation}

The cosmological analysis of a galaxy cluster sample requires a model for the relation between the halo mass and 
the observable. In this work, we take an approach which is conceptually similar to the modeling of the scaling 
relation used for the SPT galaxy cluster sample first presented and applied to derive cosmological constraints 
by \citet{vanderlinde10} (for further applications, see for instance \citet{benson13, bocquet15, dehaan16,bocquet18}). 
We empirically calibrate a scaling relation between the selection observable, i.e. the eROSITA count rate $\eta$, 
and the halo mass and redshift. As motivated in Appendix~\ref{app:scaling_relations}, we adopt the following 
scaling of the count rate with mass and redshift:
\begin{equation}\label{eq:eta_mz}
\begin{split}
\frac{\eta}{\eta_0} =&e^{\ln{A_\text{X}}}\left(\frac{M_{500\text{c}}}{M_0}\right)^{B(z)} \left(\frac{E(z)}{E_0}\right)^{2} 
\left(\frac{d_\text{L}(z)}{d_{\text{L},0}}\right)^{-2}\\
& \left(\frac{1+z}{1+z_0}\right)^{\gamma_\text{X}} 
e^{\Delta_\text{X}},
\end{split}
\end{equation}
where the amplitude is $A_\text{X}$, the redshift dependent mass slope is given by
\begin{equation}
B(z) = B_\text{X} + B_\text{X}^\prime \ln \left(\frac{1+z}{1+z_0}\right), 
\label{eq:masstrend}
\end{equation}
the redshift trend describing departures from self-similar evolution is $\gamma_\text{X}$, 
and the deviation of a particular cluster from the mean scaling relation is described as 
$\Delta_\text{X}\sim\mathcal{N}(0, \sigma_\text{X}^2)$, with scatter $\sigma_\text{X}$ 
(i.e., log-normal scatter in observable at fixed halo mass). As pivot points we choose $M_0=2\times10^{14}\text{ M}_{\sun}$, 
$z_0=0.5$, $E_0=1.314$, $d_\mathrm{L,0}=2710$ Mpc, and $\eta_0=0.06$ cts s$^{-1}$.

Empirical calibration of the scaling relation has some major advantages compared to trying to measure accurate 
physical cluster quantities such as the flux. In doing the latter, the one might suffer biases \citep[e.g. 
the effect of substructures in the context of eROSITA found by][]{hofmann17} or additional sources of scatter from lack of
knowledge about the cluster physical state. 
Furthermore, any such biases might themselves have trends with mass or redshift.  An alternative approach, which has been
adopted with success within SPT, is to use mass calibration to empirically determine
the values of the scaling relation parameters.  In this approach, an unbiased solution is found assuming the correct likelihood 
is
adopted (see Section~\ref{sec:likelihoods}) and that the form of the observable mass scaling relation that is adopted has
sufficient flexibility to describe the cluster population.  One can examine this using goodness of fit tests 
\citep[see][]{bocquet15,dehaan16}.  
There is now considerable evidence in the literature that empirical calibration leads to a more robust cosmological 
experiment. 

In summary, our model for the rate mass scaling assumes that the rate is a power law in mass and redshift with 
log-normal intrinsic scatter that is independent of mass and redshift. 
Our model allows the mass slope to vary with redshift, which is required given the redshift dependence of the eROSITA
counts to physical flux conversion (see discussion in Appendix~\ref{app:scaling_relations}).  
Natural extensions of this model to, e.g., follow mass or redshift dependent
scatter are possible, but for the analysis presented here we adopt a scaling relation with the following five free 
parameters: $(\ln A_\text{X}, B_\text{X}, \gamma_\text{X}, \sigma_\text{X}, B_\text{X}^\prime)$.

\subsection{Likelihood functions}
\label{sec:likelihoods}

The likelihood functions we employ to analyze our mock eROSITA and WL data are hierarchical, 
Bayesian models, introduced in this form 
by \citet{bocquet15}.   The functions account self-consistently for (1) the Eddington and Malmquist bias, (2) the cosmological 
dependencies of both the direct mass measurements and of the cluster number counts, and (3) systematic uncertainties in
the halo mass of objects observed with a particular rate and redshift.  Given that we 
utilize a realistic mock catalog, these likelihoods constitute a prototype of the eROSITA cosmological analysis 
pipeline. Using this scheme, we design three likelihoods: (1) mass calibration with perfect masses, (2) mass 
calibration with WL observables and (3) number counts. In the following, to ensure a concise notation, we will 
refer to the halo mass $M_\text{500c}$ as $M$, and specify when we mean a mass defined w.r.t. any other 
overdensity. 

\subsubsection{Mass calibration with perfect masses}

The likelihood that a cluster of measured rate $\hat \eta$ and redshift $z$ has a given mass $M$ is given by 
\begin{equation} \label{eq:p_m_eta_z}
P(M|\hat\eta, z) \propto \int \text{d}\eta \,P(\hat\eta|\eta,z)\,P(\eta| M, z)  \frac{\text{d}N}{\text{d}M}(M,z) ,
\end{equation}
where 
\begin{enumerate}
\item $P(\hat\eta|\eta,z)$ is the probability density function (hereafter pdf) encoding the measurement error on the rate,
\item $P(\eta| M, z)$ is the pdf describing the scaling relation between rate 
and halo mass at a given redshift. We model it as a log-normal distribution with central value given by 
equation~(\ref{eq:eta_mz}) with scatter $\sigma_\text{X}$,
\item $\frac{\text{d}N}{\text{d}M}(M,z)$ is the derivative of the number of clusters w.r.t. to the mass at that redshift, 
which is the product of the halo mass function $\frac{\text{d}n}{\text{d}M}(M,z)$ by \citet{tinker08}, the co-
moving volume element $\frac{\text{d}V}{\text{d}z}(z)$ and the survey solid angle $\Omega_\text{DE}$.
\end{enumerate}
These quantities, with the exception of the rate measurement uncertainty kernel, 
depend on scaling relation parameters, 
mass function parameters and cosmological parameters.   Also note, that equation~(\ref{eq:p_m_eta_z}) needs to be 
properly normalized to be a pdf in halo mass $M$.

The total log-likelihood for mass calibration with perfect masses is then given by the sum of the natural logarithms 
of the likelihoods of the single clusters
\begin{equation}\label{eq:pfct_mss_lkl}
\ln\mathcal{L_\text{pfct}} = \sum_{j} \ln P(M^{(j)}|\hat\eta^{(j)}, z^{(j)}), 
\end{equation}
where $j$ runs over all clusters whose halo mass is known. Note that the perfect mass is only accessible in the 
case of a mock catalogue. This likelihood is thus not applicable to real data. Nevertheless, it is a function 
of the scaling relation and the cosmological parameters and can be used to extract the true 
underlying scaling relation from a mock dataset.

\subsubsection{WL mass calibration}

The likelihood that a cluster with measured rate $\hat \eta$ and redshift $z$ has an observed tangential shear 
profile $\hat g_\text{t}(\theta_i)$ can be computed as
\begin{equation}\label{eq:P_gt_eta_z}
P(\hat g_\text{t}|\hat\eta, z) = \int \text{d} M_\text{WL} \, P(\hat g_\text{t}(\theta_i) |M_\text{WL}, z_\text{cl}) 
P(M_\text{WL}|\hat\eta, z),
\end{equation}

where 

\begin{enumerate}

\item the probability of a cluster with measured rate $\hat \eta$ and redshift $z$ to have a WL mass 
$M_\text{WL}$ is 
\begin{equation}\label{eq:P_mwl_eta_z}
\begin{split}
P(M_\text{WL}|\hat\eta, z) \propto & \int \text{d} M \int \text{d} \eta  \, P(\hat\eta|\eta,z) P(M_\text{WL}, \eta| M, z) \\
& \frac{\text{d}N}{\text{d}M}(M,z),
\end{split}
\end{equation}

with $P(M_\text{WL}, \eta| M, z)$ being the joint pdf describing the scaling relations for the rate and the WL mass, 
given in equations~(\ref{eq:eta_mz} and \ref{eq:mwl}), respectively,

\item the probability of a cluster of WL mass $M_\text{WL}$ having an observed reduced shear profile $\hat 
g_{\text{t},i} = \hat g_\text{t}(\theta_i)$ is given by a Gaussian likelihood
\begin{equation}
\ln  P(\hat g_\text{t} | M_\text{WL}, z) = -\frac{1}{2}\ln \big(2 \upi \det \mtrx{C} \big) - \frac{1}{2} \Delta \hat g_\text{t}
^T \mtrx{C}^{-1} \Delta \hat g_\text{t},
\end{equation}
with $\Delta \hat g_\text{t} = \hat g_\text{t} - g_\text{t}$, where $g_\text{t}$ is the tangential shear profile computed 
following equation~(\ref{eq:gt}) for a cluster of mass $M_\text{WL}$ and the redshift distribution 
$N(z_\text{s},z_\text{cl}=z)$.

\end{enumerate}

The total log-likelihood for mass calibration with WL then reads
\begin{equation}
\ln\mathcal{L_\text{WL mssclbr}} = \sum_{j} \ln P(\hat g_\text{t}^{(j)}|\hat\eta^{(j)}, z^{(j)}),
\end{equation}
where $j$ runs over all clusters with WL information. 

\subsubsection{Number counts}\label{sec:number_counts}
We also model the observed number of clusters $\hat N$ in bins of measured rate $\hat \eta$ and redshift $z$. 
We predict this number by computing the expected number of clusters in each bin, given the scaling relation, 
halo mass function 
and cosmological parameters
\begin{equation}\label{eq:N_eta_z}
\begin{split}
N(\hat \eta, z) = & P(\text{det}| \hat\eta,z)\\
&\int \text{d} M \int \text{d} \eta \, P(\hat\eta|\eta,z) P(\eta| M, z) \frac{\text{d}N}{\text{d}M}(M,z),
\end{split}
\end{equation}
where $P(\text{det}| \hat\eta,z)$ is a binary function parameterizing if the bin falls within the selection criteria or 
not. Assuming a pure rate selection might be a simplification compared to the actual cluster selection function 
of the forthcoming eROSITA survey (for a study of this selection function, c.f. \citet{clerc18}). In summary, the 
expected number of clusters in observable space can be computed using the cosmology dependent 
halo mass function, volume-- redshift relation and observable--mass relation.

The number counts likelihood for the entire sample is the sum of the Poisson log-likelihoods in the individual 
bins
\begin{equation}
\ln\mathcal{L_\text{nmbr cts}} = \sum_\text{bins} \hat N \ln N - N.
\end{equation}
As above, this likelihood is a function of the scaling relation, halo mass function and the cosmological parameters.

\subsubsection{Validation}\label{sec:validation}

To validated these likelihoods, we create a mock which is ten times larger than the eROSITA mock (by considering 
the unphysical survey footprint $\text{Area}_\text{test} = 10\, \text{Area}_\text{DE}$). This leads to a reduction of 
the statistical uncertainties that enables us to better constrain systematic biases. We analyze this mock with 
the number counts and the Euclid WL mass calibration likelihood. We find that all parameters are 
consistent with the input values within less then two sigma. 
Scaling this up to the 
normal sized mock, we conclude that our code is unbiased at or below  $\sim{2\over3}$ sigma. We present
for inspection a plot showing the results of the validation run as 
Fig.~\ref{fig:validation} at the end of the paper.  The plot shows the marginal contours of the 
posterior distributions for the parameters with the input values 
marked.

Given that our mock catalog is a random realization of the stochastic processes modeled by the above described 
likelihoods, and that these likelihoods retrieve the input values even for a ten times larger mock, we take the 
liberty to shift best fit parameter values of the posterior samples presented in the following sections. These 
shifts are of the order of one sigma. Putting all posteriors to the same central value allows us to highlight the 
improvement of constraining power visible in the shrinking of the contours.

\subsection{Comments on sampling and model choice}
\label{sec:sampling}

Various combinations of the above described likelihood functions are sampled using \texttt{pymultinest} 
\citep{pymultinest}, a python wrapper of the nested sampling code \texttt{multinest} \citep{multinest}. Nested 
sampling was originally developed to compute the evidence, or marginal likelihood, but has the added 
advantage of providing a converged posterior sample in the process \citep{nestedsampling}.  

The parameters we sample depend on the specific application. In all cases considered, we sample the parameters 
of the X-ray selection scaling relation: $(\ln A_\text{X}, B_\text{X}, \gamma_\text{X}, \sigma_\text{X}, 
B_\text{X}^\prime)$. When the WL mass calibration likelihood is sampled in Section~\ref{sec:cosmo_constr}, 
also the parameters governing the WL mass scaling relation are sampled: $(b_\text{WL}, \sigma_\text{WL})$. 

We explore two different flat cosmological models:  (1) $\nu$-$\Lambda$CDM, and (2) $\nu$-$w$CDM.  
For both, we consider the following parameters: $H_\text{0}$, the current 
expansion rate of the Universe in units of km s$^{-1}$ Mpc$^{-1}$; $\omega_b$, the current day co-moving 
density of baryons w.r.t. the critical density of the Universe; $\Omega_M$, the current day density of matter 
w.r.t. the critical density; $A_\text{S}$, the amplitude of primordial curvature fluctuations; $n_\text{S}$, the 
spectral index of primordial curvature fluctuations;  and the sum of neutrino masses $\sum m_\nu$ in eV. 

The cosmological model where only these parameters are allowed to vary is called $\nu$-$\Lambda$CDM, because 
we allow for massive neutrinos of yet unknown mass, and assume that the 
agent of the late time accelerated expansion is a cosmological constant $\Lambda$. 

As a more complex model $\nu$-$w$CDM, we also consider the case that the late time acceleration is not caused by the 
cosmological constant, but by an as yet unknown form of energy, usually referred to as dark energy. The 
properties of dark energy are described here by a single equation of state parameter $w$. 

For better comparison, with other Large Scale Structure experiments,
in both models,
we also compute $\sigma_8$, the 
root mean square of linear matter fluctuations in a spherical region of 8 h$^{-1}$Mpc radius, as a derived quantity in each 
step of
the chain and present the posterior distribution in this quantity rather than in the 
primordial power spectrum fluctuation amplitude $A_\text{S}$. 

\subsection{Choice of priors}
\label{sec:priors}

\begin{table}
	\centering
	\caption{Priors used in our analysis. $\mathcal{U}(a, b)$ is a uniform flat prior in the interval $(a,b)$, $\ln 
\mathcal{U}(a, b)$ a uniform flat prior in log space, $\mathcal{N}(\mu, \sigma^2)$ refers to a Gaussian 
distribution with mean $\mu$ and variance $\sigma^2$, $\mathcal{N}_{>a}(\mu, \sigma^2)$ to a Gaussian 
distribution truncated for values smaller than $a$.\newline \textit{Comment: a)} Numerical stability when 
computing the equations~(\ref{eq:p_m_eta_z}, \ref{eq:P_gt_eta_z}, \ref{eq:P_mwl_eta_z} and 
\ref{eq:N_eta_z}), requires the scatter to be larger than the sampling size of the numerical integrals. }
	\label{tab:priors}
	\begin{tabular}{lll} 
		\hline
		\multicolumn{3}{l}{Cosmology for Number counts w/o CMB}\\
		\hline
		$H_\text{0}$ &  $\mathcal{U}(40, 120)$ & cf. Section~\ref{sec:cosmo_priors}\\
		$\omega_\text{b}$ & $\mathcal{U}(0.020, 0.024)$ & \\
		$\Omega_\text{M}$ & $\mathcal{U}(0.1, 0.5)$ & \\
		$A_\text{S}$ & $\ln \mathcal{U}(0.6\text{e}-9, 2.5\text{e}-9)$ & \\
		$n_\text{S}$ & $\mathcal{U}(0.94, 1.0)$ & \\
		$\sum m_\nu [eV]$ & $\mathcal{U}(0., 1.)$ & \\
		$w$ & $\mathcal{U}(-1.6, -0.6)$ & \\
 		\hline
		\multicolumn{3}{l}{Cosmology for Number counts w/ CMB}\\
		\hline
		& cf. Section~\ref{sec:cosmo_priors} & \\
		\hline
		\multicolumn{3}{l}{X-ray Selection Scaling Relation}\\
		\hline
 		$\ln A_\text{X}$ & $\mathcal{N}(-0.33, 0.23^2)$ & cf. Appendix~\ref{app:scaling_relations} \\
 		$B_\text{X}$ & $\mathcal{N}(2.00, 0.17^2)$ & \\
 		$\gamma_\text{X}$ & $\mathcal{N}(0.45, 0.42^2)$ & \\
 		$\sigma_{X}$ & $\mathcal{N}_{>0.1}(0.28, 0.11^2)$ & a) \\
 		$B_\text{X}^\prime$ & $\mathcal{N}(0.36, 0.78^2)$ & \\
  		\hline
		\multicolumn{3}{l}{DES/HSC WL}\\
		\hline
 		$b_\text{WL}$ & $\mathcal{N}(0.94, 0.051^2)$ & cf. Section~\ref{sec:WL_syst}\\
 		$\sigma_\text{WL}$ & $\mathcal{N}_{>0.1}(0.24, 0.02^2)$ & a) \\
 		\hline
		\multicolumn{3}{l}{Euclid WL}\\
		\hline
 		$b_\text{WL}$ & $\mathcal{N}(0.94, 0.013^2)$ & cf. Section~\ref{sec:WL_syst}\\
 		$\sigma_\text{WL}$ & $\mathcal{N}_{>0.1}(0.24, 0.008^2)$ & a) \\
 		\hline
 		\multicolumn{3}{l}{LSST WL}\\
		\hline
 		$b_\text{WL}$ & $\mathcal{N}(0.94, 0.015^2)$ & cf. Section~\ref{sec:WL_syst}\\
 		$\sigma_\text{WL}$ & $\mathcal{N}_{>0.1}(0.24, 0.008^2)$ & a) \\
 		\hline

	\end{tabular}
\end{table}

In general, any Bayesian analysis, and more specifically \texttt{pymultinest}, requires the specification of priors for 
all parameters one intends to sample. In the following, we present our choice of priors. If the parameter is 
not mentioned below, it has a uniform prior in a range that is larger than the typical posterior uncertainties of 
that parameter. The prior choices are summarized in Table~\ref{tab:priors}.

\subsubsection{Current priors on scaling relation}\label{sec:prior_SR}
As mentioned above-- and discussed in detail in Appendix~\ref{app:scaling_relations}-- the eROSITA count rate 
scaling relation is described by five parameters: $(\ln A_\text{X}, B_\text{X}, \gamma_\text{X}, \sigma_\text{X}, 
B_\text{X}^\prime)$. We put Gaussian priors on these parameters. The mean values are obtained in 
Section~\ref{sec:fid_SR_params} by determining the maximum likelihood points of the mass calibration 
likelihood when using perfect masses. The corresponding uncertainties in the priors are taken to match the 
uncertainties on the respective parameters presented in Table 5 of \citet{bulbul19} for the core included 
0.5-2.0~keV luminosity-mass-redshift relation when fit with the scaling relation of Form II.  These parameter 
uncertainties were extracted using a sample of 59 SPT selected galaxy clusters observed with XMM-{\it 
Newton} together with the SPT SZE-based halo masses calculated using the calibration from \citet[][see 
Table 3 results column 2]{dehaan16}.

When we extract cosmological constraints {\bf only} with these priors (i.e., without any WL information) 
we consider that a ``baseline'' result representing a currently achievable knowledge of the parameters of the 
eROSITA rate-mass relation.

\subsubsection{Priors on WL calibration}\label{sec:WL_syst}
The priors on the parameters of the WL mass -- halo mass relation reflect the understanding of both the 
observational and theoretical systematics of the WL mass calibration. In this work, we consider, the following 
sources of systematic uncertainty:

\begin{enumerate}
\item the accuracy of the shape measurement in the optical survey parameterized as the uncertainty on the 
multiplicative shear bias $\delta m$,

\item the systematic mis-estimation of the lensing efficiency $\langle \beta \rangle$ due to the bias in the 
photometric redshift estimation $b_{\hat z}$, 

\item the uncertainty in the estimation of the contamination by cluster members $f_\text{cl}$ which results from the 
statistical uncertainty of the photometric redshifts $\sigma_{\hat z}$ and the background galaxy selection,

\item the statistical uncertainty with which the theoretical bias and scatter of the WL mass $\delta b_\text{WL, sim}
$, and $\delta \sigma_\text{WL, sim}$, respectively, can be constrained with large structure formation 
simulations. 
\end{enumerate}

The first three effects do not directly induce a bias in the mass estimation, but affect the NFW fitting procedure. To 
estimate their impact on the WL mass estimate, we consider a shear profile for WL mass $3\times10^{14}$ 
M$_{\sun}$ and $z=0.4$, add the systematic shifts, and fit for the mass again. The difference in input and 
output masses is then taken as the WL mass systematic uncertainty induced by these effects. This technique 
provides an overall estimate of the systematic uncertainty level, while ignoring potential dependences on 
cluster redshift and mass.

For DES, we assume $\delta m=0.013$ \citep{zuntz18}. The bias on the photometric redshift estimation of the 
source galaxies is $b_{\hat z}=0.02$ \citep{cooke14} which, considering the source redshift distribution of 
DES (cf. Section~\ref{sec:wl_signal}), leads to an uncertainty on the lensing efficiency $\delta \langle \beta 
\rangle = 0.02$. For the uncertainty on the contamination, we project $\delta f_\text{cl} = 0.01$ based on 
\citet{dietrich19}. Taken all together, these uncertainties propagate to a WL mass uncertainty of $\delta 
b_\text{WL, obs, DES}=0.045$. 

The current uncertainty on the theoretical WL mass bias is $\delta b_\text{WL, 
sim, to day}=0.05$ in \citet{dietrich19}, when considering the effects of halo triaxiality, morphological variety, 
uncertainties in the mass-concentration relation and mis-centering. Due to larger available simulations \citep{lee18}, a 
better measurement of the mis-centering distribution and an improvement of the understanding of the mass--
concentration relation, for DES we project a reduction of this uncertainty by a factor 2, yielding $\delta 
b_\text{WL, sim, DES}=0.025$. The same scaling is applied to the uncertainty on the scatter, yielding $\delta 
\sigma_\text{WL, DES} = 0.02$. 

Given the level of observational uncertainty, this projection can also be read 
as a necessity to improve the understanding of the theoretical biases. The estimates above provide a 
total uncertainty of the bias of the WL mass
\begin{equation}
\begin{split}
\delta b_\text{WL, DES} &= \sqrt{ b_\text{WL, sim, DES}^2 + b_\text{WL, obs, DES}^2}\\
&=0.051,
\end{split}
\end{equation}
and an uncertainty on the scatter of the WL mass $\delta \sigma_\text{WL, DES} = 0.02$. This amounts to a $5.1\%$ 
mass uncertainty from systematic effects, which is a conservative assumption, given that 
\citet{mcclintock19} already achieved such a level of systematics control for DES cluster mass calibration. For sake of 
simplicity, we assume that the final level of systematics in HSC is of the same as in DES. This assumption will be inadequate 
for the actual analysis of the data. We postpone the discussion about the difference between the analysis methods to the 
respective future works.

The specifications for Euclid are given in \citet{laureijs11}. The requirement for the shape measurement is $
\delta m = 0.001$. For the bias on the photometric redshift estimation, the requirement is $b_{\hat z} = 0.002$, which 
translates into  $\delta \langle \beta \rangle = 0.0014$. For the projection of the uncertainty on the 
contamination, we assume that in the case of DES it has equal contribution from (1) the number of clusters used 
for to characterize it and (2) the photometric redshift uncertainty. Thus, for Euclid we estimate
\begin{equation}
\begin{split}
\delta f_\text{cl, Eu}^2 &= \frac{\delta f_\text{cl, DES}^2}{2} \frac{N_\text{DES}}{N_\text{Eu}} + \frac{\delta f_\text{cl, 
DES}^2}{2} \left(\frac{ \sigma_{\hat z\text{, Eu}}}{\sigma_{\hat z\text{, DES}}}\right)^2 \\
&= 0.0065^2,
\end{split}
\end{equation}
where $N_\text{DES}\approx 3.8$k, and $N_\text{Eu}\approx 6.4$k, are the number of clusters with DES and 
Euclid shear information in our catalog (cf. Section~\ref{sec:wl_signal}), $\sigma_{\hat z\text{, Eu}}=0.06$ is 
the photometric redshift uncertainty for Euclid \citep{laureijs11}, and $\sigma_{\hat z\text{, DES}}=0.1$ is the 
photometric redshift uncertainty for DES \citep{sanchez14}. 
Taking all the above mentioned values together, we find $\delta b_\text{WL, obs, Eu}=0.0085$ for Euclid. To match 
this improvement in data quality, we project an improvement in the understanding of the theoretical biases by 
a factor of 5, providing $\delta b_\text{WL, sim, Eu}=0.01$, and $\delta \sigma_\text{WL, Eu} = 0.008$. 
Thus, the total uncertainty on the WL mass bias for Euclid is 
\begin{equation}
\delta b_\text{WL, Eu} = 0.013.
\end{equation}

The specifications for LSST systematics are summarized in \citet{lsst_desc18}. The requirement for the shape 
measurement is $
\delta m = 0.003$, while the requirement for the bias on the photometric redshift estimation $b_{\hat z} = 0.001$, leading to  $
\delta \langle \beta \rangle = 0.0007$. Using $N_\text{LSST}\approx 11$k, and $\sigma_{\hat z\text{, LSST}}=0.02$, we find 
an uncertainty on the cluster member contamination of $\delta f_\text{cl, LSST}=0.0044$. Summing all the above mentioned 
values together, we get $\delta b_\text{WL, obs, LSST}=0.011$. We project the same understanding in theoretical 
systematics for LSST as for Euclid. Thus, the total uncertainty on the WL mass bias for LSST is
\begin{equation}
\delta b_\text{WL, LSST} = 0.015.
\end{equation}

These values are adopted throughout this work as priors for the WL mass scaling relation parameters, as 
summarized in Table~\ref{tab:priors}.  We note that the effort required to theoretically constrain 
the WL bias and scatter parameters with this accuracy is considerable.

\begin{figure*}
	\includegraphics[width=\textwidth]{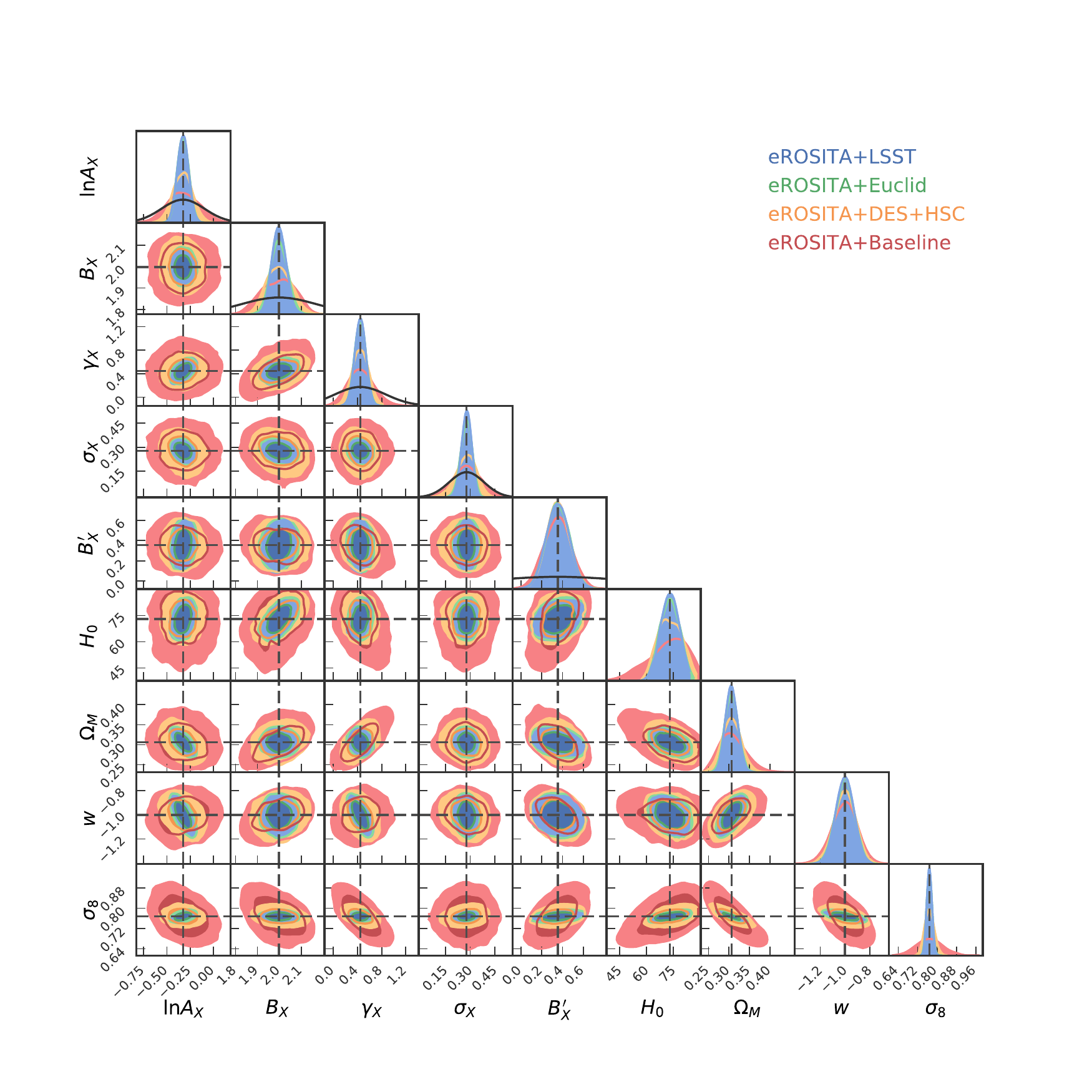}
	\vskip-0.1in
    \caption{Predicted constraints on the scaling relation and cosmological parameters in $w$CDM. In red the 
constraints from the number counts alone (eROSITA+Baseline), in orange the constraints from number counts 
and DES+HSC WL calibration (eROSITA+DES+HSC), in green number counts and Euclid WL calibration 
(eROSITA+Euclid), and in blue number counts and LSTT WL calibration 
(eROSITA+LSST). The median values, all statistically consistent with the input values, are shifted to the input 
values to better highlight the increase in constraining power.}
    \label{fig:cosmo_constr}
\end{figure*}

\subsubsection{Cosmological priors}\label{sec:cosmo_priors}
When sampling the number counts likelihood, we assume flat priors on all cosmological parameters except for 
$A_\text{S}$, for which we use a flat prior in log-space, as is good practice for strictly positive amplitudes. 
Similarly, we use priors on $\Omega_\text{M}$, $H_\text{0}$ and $w$ that are larger than the typical 
uncertainties on these parameters. For $\sum m_\nu$ we only explore the regime up to 1 eV, as current 
cosmological measurements, such as \citet{planck16_cosmo} give upper limits on the summed neutrino mass 
around and below that value.

For $\omega_b$ and $n_\text{S}$ we use tight flat priors around the measured values of these parameters by the 
CMB experiments \citep{planck16_cosmo} and Big Bang Nucleosynthesis constraints derived from deuterium 
abundances \citep{cooke14}. We confirm that cluster number counts are not sensitive to these parameters 
within these tight ranges \citep{bocquet18}. It is thus not necessary to use informative priors on these parameters, as 
previous 
studies have done \citep[see for instance][]{bocquet15, dehaan16}. 

In Section~\ref{sec:syn_CMB} we will consider the synergies between eROSITA number counts and WL mass 
calibration, and CMB temperature and polarization anisotropy measurements, which to date provide us with a  
significant amount of information about the cosmological parameters. In the models of interest, where either 
$w$ or $\sum m_\nu$ are free parameters, the CMB constraints from the Planck mission 
\citep{planck16_cosmo} display large degeneracies between the parameters we choose to sample.
\footnote{These degeneracies are partially due to our choice of sampling parameters. The CMB does not 
directly constrain $H_\text{0}$, which is a present day quantity. Consequently, also $\Omega_\text{M}$ is 
weakly constrained. 
The same holds for $w$, which has predominantly a late time impact on the expansion rate. In contrast, co-
moving densities like $\omega_\text{b}$, or primordial quantities like $A_\text{S}$ and $n_\text{S}$ are 
narrowed down with high precision.} For this reason, we cannot approximate the CMB posterior as a 
Gaussian distribution. To capture the non-Gaussian feature, we calibrate a nearest-neighbor kernel density 
estimator (KDE) on the publicly available\footnote{\url{https://pla.esac.esa.int/pla/\#cosmology}, where we 
utilized the \texttt{TTTEE\_lowTEB} samples.} posterior sample. We utilize Gaussian kernels and, for each 
model, we tune the bandwidth through cross calibration to provide maximum likelihood of the KDE on a test 
subsample. As discussed in Section~\ref{sec:fiducial}, our choice of input cosmology is slightly inconsistent 
with the CMB constraints. As we are only interested in the reduction of the uncertainties when combining 
CMB and eROSITA, we shift the CMB posteriors so that they are consistent with our input values at less than one sigma.
The resulting estimator reproduces the parameter uncertainties and the degeneracies 
accurately.

\section{Results}\label{sec:results}

In the following subsections we first calculate how accurately the observable--mass scaling relation parameters 
must be constrained to enable the best possible cosmological constraints from the sample 
(Section~\ref{sec:optimal}).  Thereafter we explore the impact of the WL mass calibration on the cosmological 
constraints that can be extracted from an analysis of the eROSITA galaxy cluster counts 
(Section~\ref{sec:eROSITA+WL}).  In Section~\ref{sec:eROSITA+CMB} we explore synergies of the 
eROSITA dataset with the CMB and in Section~\ref{sec:eROSITA+BAO} we examine the impact of combining 
the eROSITA dataset with BAO measurements from DESI.  In Section~\ref{sec:combined} we examine 
the constraints derived when combining with both these external data sets, and the final subsection focuses on 
the impact of an eROSITA sample where the minimum mass is allowed to fall from our baseline value of 
$M_{500\text{c}}\goa2\times 10^{14} \text{M}_{\sun}$ to $M_{500\text{c}}\goa5\times 10^{13} \text{M}_{\sun}$, 
corresponding to a sample that is $\sim3.5$ times larger.

\subsection{Optimal mass calibration}\label{sec:min_mssclbr}
\label{sec:optimal}

The number counts likelihood depends both on the scaling relation parameters, and-- through the 
mass function, the cosmological 
volume and their changes with redshift-- also on the cosmological parameters.  Furthermore, there are significant 
degeneracies between the mass scale of the cluster sample (i.e., the parameters of the observable mass relation) and the 
cosmological parameters, as demonstrated already in the earliest studies \citep{haiman01}.  A full self-calibration of the 
number counts (i.e., including no direct mass measurement information) that allows full cosmological and scaling relation 
freedom, results in only very weak cosmological constraints \citep[e.g.,][]{majumdar03, majumdar04}.
Thus, before forecasting the cosmological constraints from the eROSITA sample, we estimate how accurate 
the mass calibration needs 
to be so that the information contained in the number counts is primarily resulting in the reduction of uncertainties on the 
cosmological parameters rather than the observable mass scaling relation parameters. 

To estimate this required level of mass calibration, which we refer to as "optimal mass calibration", we quantify  
how much the number counts constrain the scaling relation parameters when the cosmological parameters are fixed to 
fiducial values.  In such a case, all the information contained in the number counts likelihood informs our 
posterior on the scaling relation parameters. If this level of information, or more, were provided by direct mass calibration, 
then the number counts information would predominantly constrain the cosmology. In this sense, the optimal mass 
calibration then provides a threshold or goal for the amount and precision of external mass calibration we should strive for
in our direct mass calibration through, e.g., weak lensing.  

We find that in fact the number counts alone do not contain enough information to meaningfully constrain all five scaling
relation parameters even in the presence of full cosmological information.  Our scaling relation parametrization includes two 
additional parameters beyond those explored in \citet{majumdar03}, the scatter $\sigma_X$ and the redshift evolution of the 
mass trend $B_\text{X}^\prime$.  Thus, as a next test, we examine 
the constraints from number counts with fixed cosmology while assuming priors only on $B_\text{X}^\prime$. 
Interestingly, in this case we find that the constraints lead to an upper limit on the scatter of the scaling relation 
$\sigma_X<0.44$ (at $95 \%$), which is weaker than our current knowledge of that parameter, which we infer from the 
scatter in the X-ray luminosity--mass relation from \citet[][see discussions in Section~\ref{sec:priors} and 
Appendix~\ref{app:scaling_relations}]{bulbul19}.
We therefore adopt this external prior
on the scatter parameter and allow full freedom for all other parameters (including $B_\text{X}^\prime$).
Results in this case are more interesting, providing constraints that we adopt as our estimate of optimal mass calibration.
The uncertainties are $\delta \ln A_\text{X} = 0.042$, $\delta B_\text{X} = 0.024$, $\delta \gamma_\text{X} 
= 0.053$, and $\delta B_\text{X}^\prime = 0.116$. 
We take this to mean that an optimal cosmological exploitation of the eROSITA cluster number counts will require that we 
know the parameters of the observable mass relation to at least these levels of precision. 
We will discuss in the following how this can be accomplished.

\subsection{Forecasts: eROSITA+WL}\label{sec:cosmo_constr}
\label{sec:eROSITA+WL}

\begin{table*}
	\centering
	\caption{Forecast parameter constraints for eROSITA number counts with current, best available calibration 
(eROSITA+Baseline), with DES+HSC WL calibration (eROSITA+DES+HSC), with Euclid WL calibration 
(eROSITA+Euclid), and with LSST WL calibration 
(eROSITA+LSST) are presented in two different models, $\nu$-$w$CDM and $\nu$-$\Lambda$CDM within three different 
scenarios.  From top to bottom they are eROSITA+WL alone, in combination with Planck CMB constraints 
(Pl15) and in combination with DESI BAO and Alcock-Pacynzki test constraints.  Also shown are the scaling 
relation parameter uncertainties for an optimal mass calibration. In addition to the five cosmological parameters who 
constraints are presented, each model includes the parameters $n_\mathrm{S}$ and $\omega_\mathrm{b}$ marginalized 
over weak priors (see Table~\ref{tab:priors}). 
The units of the column ``$\sum m_\nu$'' and ``$H_\text{0}$'' are eV and km s$^{-1}$ Mpc$^{-1}$, respectively.
\textit{Comments: a)} This parameter is not constrained within the prior ranges. When 
reporting upper limits ``<'', we refer to the 95th percentile, while lower limits ``>'' refer to the 5th percentile. 
When a parameter is kept fixed in that model, we use ``--''.}
	\label{tab:baseline_constraints}
	\begin{tabular}{llcccccccccc} 
		\hline
&		& $\Omega_\text{M}$ & $\sigma_8$ & $w$ & $\sum m_\nu$ & $H_\text{0}$ & $\ln A_\text{X}$ & 
$B_\text{X}$  & $\gamma_\text{X}$ & $\sigma_\text{X}$ & $B_\text{X}^\prime$\\
		\hline
		\hline
		 \multicolumn{3}{l}{optimal mass calibration\hfil} & & & & & 0.042 & 0.024 & 0.053 & & 0.116 \\
 		\hline
		\hline
		\multispan{12}{\hfil eROSITA + WL calibration\hfil}\\
		\hline
$\nu$-$w$CDM &		priors &  &  &  &  &  & 0.23 & 0.17 & 0.42 & 0.11 & 0.78 \\
&        eROSITA+Baseline & 0.032 & 0.052 & 0.101 & $^\text{a)}$ & 10.72 & 0.165 & 0.073 & 0.209 & 0.083 & 0.128 \\
&        eROSITA+DES+HSC & 0.023 & 0.017 & 0.085 & $^\text{a)}$ & 6.449 & 0.099 & 0.053 & 0.121 & 0.062 & 0.111\\
&        eROSITA+Euclid & 0.016 & 0.012 & 0.074 & $^\text{a)}$ & 5.210 & 0.059 & 0.037 & 0.090 & 0.034 & 0.107 \\
&        eROSITA+LSST & 0.014 & 0.010 & 0.071 & $^\text{a)}$ & 4.918 & 0.058 & 0.031 & 0.089 & 0.030 & 0.107 \\
 		\hline
$\nu$-$\Lambda$CDM & 		priors &  &  & -- &  &  & 0.23 & 0.17 & 0.42 & 0.11 & 0.78 \\
&        eROSITA+Baseline & 0.026 & 0.033 & -- & $^\text{a)}$ & 10.18 & 0.157 & 0.069 & 0.192 & 0.078 & 0.110 \\
&        eROSITA+DES+HSC & 0.016 & 0.014 & -- & $^\text{a)}$ & 5.664 & 0.091 & 0.049 & 0.103 & 0.059 & 0.104\\
&        eROSITA+Euclid & 0.011 & 0.007 & -- & $^\text{a)}$ & 4.691 & 0.040 & 0.035 & 0.065 & 0.033 & 0.104 \\
&        eROSITA+LSST & 0.009 & 0.007 & -- & $^\text{a)}$ & 4.691 & 0.039 & 0.032 & 0.058 & 0.029 & 0.104 \\
 		\hline 
 		\hline
		\multispan{12}{\hfil eROSITA + WL calibration + Pl15 (TTTEE\_lowTEB)\hfil}\\
		\hline
$\nu$-$w$CDM &		priors (incl. CMB) & <0.393 & 0.063 & 0.242 & <0.667 & >62.25 & 0.23 & 0.17 & 0.42 & 
0.11 & 0.78 \\
&        eROSITA+Baseline & 0.019 & 0.032 & 0.087 & <0.590 & 2.857 & 0.165 & 0.026 & 0.132 & 0.083 & 0.121 \\
&        eROSITA+DES+HSC & 0.018 & 0.019 & 0.085 & <0.554 & 2.206 & 0.099 & 0.024 & 0.118 & 0.062 & 0.107\\
&        eROSITA+Euclid & 0.014 & 0.010 & 0.074 & <0.392 & 1.789 & 0.059 & 0.020 & 0.090 & 0.034 & 0.107 \\
&        eROSITA+LSST & 0.013 & 0.009 & 0.069 & <0.383 & 1.662 & 0.058 & 0.018 & 0.080 & 0.030 & 0.103 \\
 		\hline
$\nu$-$\Lambda$CDM & 		priors (incl. CMB) & 0.024 & 0.035 & -- & <0.514 & 1.723 & 0.23 & 0.17 & 0.42 & 
0.11 & 0.78 \\
&        eROSITA+Baseline & 0.016 & 0.018 & -- & <0.425 & 1.192 & 0.122 & 0.025 & 0.101 & 0.077 & 0.110 \\
&        eROSITA+DES+HSC & 0.013 & 0.015 & -- & <0.401 & 1.067 & 0.086 & 0.023 & 0.098 & 0.060 & 0.104\\
&        eROSITA+Euclid & 0.011 & 0.007 & -- & <0.291 & 0.978 & 0.039 & 0.020 & 0.065 & 0.033 & 0.103 \\
&        eROSITA+LSST & 0.009 & 0.007 & -- & <0.285 & 0.767 & 0.038 & 0.020 & 0.054 & 0.030 & 0.103 \\
 		\hline
 		\hline
		\multispan{12}{\hfil eROSITA + WL calibration + DESI (BAO)\hfil}\\
		\hline
$\nu$-$w$CDM &		priors (incl. BAO) & 0.007 & $^\text{a)}$ & 0.086 & $^\text{a)}$ & $^\text{a)}$ & 0.23 & 
0.17 & 0.42 & 0.11 & 0.78 \\
&        eROSITA+Baseline & 0.007 & 0.030 & 0.063 & $^\text{a)}$ & 1.987 & 0.164 & 0.043 & 0.139 & 0.083 & 0.128 \\
&        eROSITA+DES+HSC & 0.006 & 0.010 & 0.051 & $^\text{a)}$ & 1.597 & 0.086 & 0.037 & 0.110 & 0.056 & 0.101\\
&        eROSITA+Euclid & 0.006 & 0.005 & 0.047 & $^\text{a)}$ & 1.463 & 0.040 & 0.030 & 0.086 & 0.032 & 0.096 \\
&        eROSITA+LSST & 0.006 & 0.005 & 0.043 & $^\text{a)}$ & 1.403 & 0.040 & 0.026 & 0.076 & 0.029 & 0.095 \\
 		\hline
$\nu$-$\Lambda$CDM & 		priors (incl. BAO) & 0.006 & $^\text{a)}$ & -- & $^\text{a)}$ & $^\text{a)}$ & 0.23 & 
0.17 & 0.42 & 0.11 & 0.78 \\
&        eROSITA+Baseline & 0.006 & 0.015 & -- & $^\text{a)}$ & 0.943 & 0.094 & 0.041 & 0.109 & 0.078 & 0.110 \\
&        eROSITA+DES+HSC  & 0.006 & 0.010 & -- & $^\text{a)}$ & 0.925 & 0.074 & 0.040 & 0.077 & 0.055 & 0.104\\
&        eROSITA+Euclid & 0.006 & 0.005 & -- & $^\text{a)}$ & 0.910 & 0.040 & 0.029 & 0.054 & 0.032 & 0.089 \\
&        eROSITA+LSST & 0.006 & 0.005 & -- & $^\text{a)}$ & 0.910 & 0.035 & 0.025 & 0.053 & 0.027 & 0.089 \\
 		\hline
 		\hline
 		\multispan{12}{\hfil eROSITA + WL calibration + DESI + Pl15\hfil}\\
		\hline

$\nu$-$w$CDM &		priors (incl. CMB+BAO)  & 0.007 & 0.027 & 0.049 & <0.284 & 1.118 & 0.23 & 0.17 & 
0.42 & 0.11 & 0.78 \\
&        eROSITA+Baseline & 0.006 & 0.026 & 0.049 & <0.281 & 1.103 & 0.161 & 0.023 & 0.079 & 0.083 & 0.128 \\
&        eROSITA+DES+HSC & 0.006 & 0.011 & 0.048 & <0.245 & 1.050 & 0.085 & 0.023 & 0.071 & 0.061 & 0.104\\
&        eROSITA+Euclid & 0.005 & 0.006 & 0.047 & <0.241 & 1.023 & 0.039 & 0.017 & 0.064 & 0.032 & 0.095 \\
&        eROSITA+LSST & 0.005 & 0.006 & 0.039 & <0.223 & 0.870 & 0.038 & 0.017 & 0.064 & 0.029 & 0.089 \\
 		\hline
$\nu$-$\Lambda$CDM & 		priors (incl. CMB+BAO)  & 0.004 & 0.020 & -- & <0.256 & 0.255 & 0.23 & 0.17 & 
0.42 & 0.11 & 0.78 \\
&        eROSITA+Baseline & 0.004 & 0.016 & -- & <0.254 & 0.253 & 0.093 & 0.024 & 0.067 & 0.074 & 0.110 \\
&        eROSITA+DES+HSC & 0.004 & 0.009 & -- & <0.218 & 0.251 & 0.072 & 0.021 & 0.062 & 0.051 & 0.095\\
&        eROSITA+Euclid & 0.003 & 0.004 & -- & <0.211 & 0.148 & 0.035 & 0.020 & 0.050 & 0.033 & 0.071 \\       
&        eROSITA+LSST & 0.002 & 0.003 & -- & <0.185 & 0.145 & 0.033 & 0.017 & 0.050 & 0.033 & 0.069 \\
	\hline
 		\hline

 	\end{tabular}
\end{table*}

\subsubsection{$\nu$-$w$CDM constraints}

As a first cosmological model we investigate $\nu$-$w$CDM, a flat cold Dark Matter cosmology with dark energy 
with constant but free equation of state parameter $w$ and massive neutrinos. In this Section, we present the 
constraints on the cosmological parameters for three different cases: number counts alone combined with baseline 
priors on the X-ray observable mass scaling relation that we derive from the latest analysis within SPT \citep{bulbul19}
(eROSITA+Baseline), number counts with DES+HSC WL mass calibration (eROSITA+DES+HSC), number counts 
with Euclid WL mass calibration (eROSITA+Euclid), and number counts 
with LSST WL mass calibration (eROSITA+LSST) . The respective marginal contour plot is shown in 
Fig.~\ref{fig:cosmo_constr}, and the corresponding uncertainties are listed in Table~\ref{tab:baseline_constraints}. 

Considering the current knowledge of the X-ray scaling relation, we find that eROSITA number counts constrain 
$\Omega_\text{M}$ to $\pm0.032$, $\sigma_8$ to $\pm 0.052$, $w$ to $\pm 0.101$, and $H_\text{0}$ to $\pm 
10.72$ km s$^{-1}$ Mpc$^{-1}$, while marginalizing over the summed neutrino mass $\sum m_\nu<1$ eV without 
constraining it. We 
also find no constraints on $\omega_\text{b}$ and $n_\text{S}$ within the prior ranges that we assumed.

The addition of mass information consistently reduces the uncertainties on the cosmological parameters: the 
knowledge on $\Omega_\text{M}$ is improved by factors of $1.4$,  $2.0$ and $2.3$ when adding DES+HSC, 
Euclid, and LSST WL information, respectively; for $\sigma_8$ the improvements are $3.1$, $4.3$ and $5.2$, whereas 
for the dark energy equation of state parameter they are $1.2$, $1.4$ and $1.4$, respectively. In summary, weak 
lensing calibration provides the strongest improvement of the determination of $\sigma_8$, followed by $
\Omega_M$. The improvements on the dark energy equation of state parameter $w$ are clearly weaker.

\subsubsection{$\nu$-$\Lambda$CDM constraints}

We also investigate a model in which the equation of state parameter $w$ is kept constant: $\nu$ $
\Lambda$CDM. The corresponding uncertainties are shown in Table~\ref{tab:baseline_constraints}. In this model, 
we find that the constraints on $\Omega_\text{M}$ and $\sigma_8$ are $0.019$ and $0.032$, respectively, 
which is tighter than in the $\nu$-$w$CDM model. However, the constraint on $H_\text{0}$ is comparable in 
the two models.

We also find that the addition of WL mass information improves the constraints on $\Omega_\text{M}$ by factors 
of $1.6$, $2.4$ and $2.9$ for DES+HSC, Euclid and LSST, respectively. The determination of $\sigma_8$ improves by 
factors 
$2.4$, $4.7$ and $4.7$. It is especially worth highlighting how eROSITA with Euclid or LSST WL information will be 
able to determine $\sigma_8$ at a sub-percent level. Nevertheless, also in this simpler model we find that 
eROSITA number counts do not constrain the summed neutrino mass in the sub-eV regime.

\begin{figure}
	\includegraphics[width=\columnwidth]{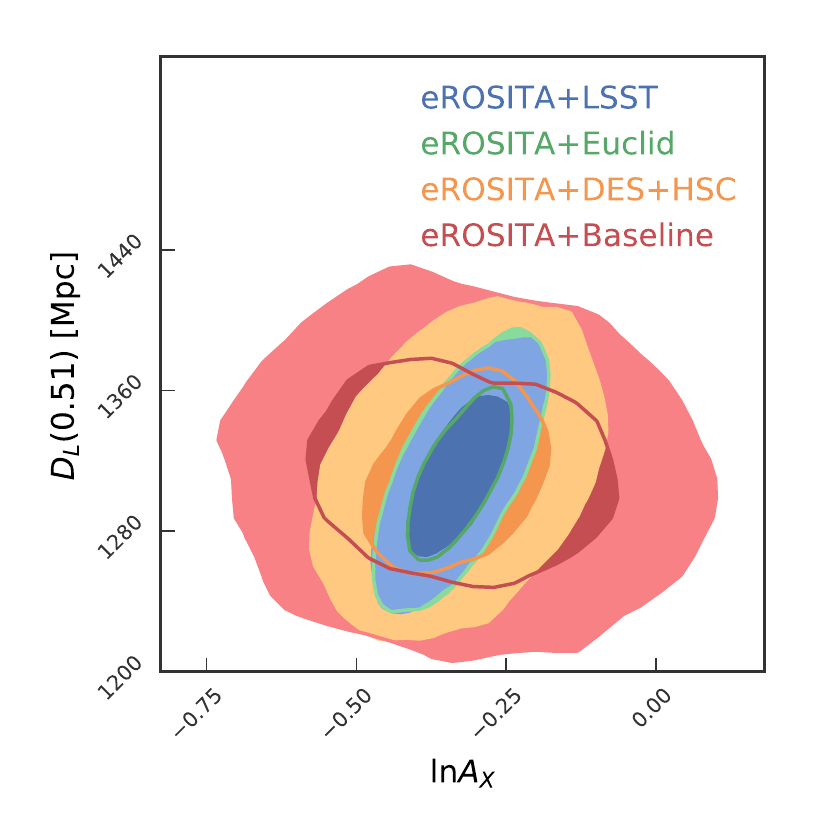}
	\vskip-0.10in
    \caption{Two dimensional marginalized posterior sample of the amplitude of the scaling relation $A_\text{X}$ 
and the luminosity distance to the median redshift of our sample $D_\text{L}(0.51)$ in Mpc, as derived from 
the cosmological parameters in the posterior sample in the $w$CDM model. In red, orange, green and blue we 
present the constraints from the number counts alone (eROSITA+Baseline), from number counts and DES+HSC 
WL calibration (eROSITA+DES+HSC), Euclid WL calibration (eROSITA+Euclid), and LSST WL calibration (eROSITA+LSST),  
respecitvely. When no direct mass information is 
present, as in the case of number counts only, the two quantities are not degenerate with each other. As 
mass information is added, the underlying parameter degeneracy between the amplitude of the X-ray 
observable mass relation and the cosmological distance information emerges.}
    \label{fig:dist_deg}
\end{figure}

\subsubsection{Limiting parameter degeneracy}
\label{sec:degeneracy}

We have studied the causes of the weaker improvement in $w$ when calibrating with Euclid or LSST WL, and we 
have discovered an interesting degeneracy due to the $w$ sensitivity of the distance.  Remember that our WL 
calibration dataset consists of observations of the shear profiles and the redshift distributions of the 
background galaxies.  To turn these into masses, one needs the cosmology sensitive angular diameter 
distances (see discussion below equation~\ref{eq:gt}).  Moreover, our selection observable is the eROSITA 
count rate (similar to X-ray flux) that is related to the underlying X-ray luminosity through the luminosity 
distance (see equation~\ref{eq:eta_mz}).  This leads to a degeneracy between $w$, governing the redshift 
evolution of distances, and the amplitude and redshift trend of the selection observable--mass relation.

The degeneracy between $w$ and ($\ln A_\text{X}$, $\gamma_\text{X}$) can be easily understood by considering 
the parametric form of the rate mass scaling relation in equation~(\ref{eq:eta_mz}). Ignore for a moment the 
distance dependence of the mass.  Then for a given redshift $z$ and rate $\eta$, a shift in $w$ leads to a 
shift in the luminosity distance $D_\text{L}(z)$, and, to a minor degree, to a shift in the co-moving expansion 
rate $E(z)$. Such a shift can be compensated by a shift in $\ln A_\text{X}$ and $\gamma_\text{X}$, resulting 
in the same mass, and consequently the same number of clusters, making it indiscernible.  The distance 
dependence of the shear to mass mapping and the power law dependence of the rate on mass leads to a 
somewhat different dependence, and so the parameter degeneracy is not catastrophic.

This effect is demonstrated in Fig.~\ref{fig:dist_deg}, where the joint posterior of the luminosity distance to the median 
cluster redshift $D_\text{L}(0.51)$ and of the amplitude of the scaling relation $\ln A_\text{X}$ is shown. In the 
case of no direct mass information, when we fit the number counts with priors on the scaling relation 
parameters, the median distance and the amplitude are uncorrelated. As one adds more mass information, e.g., the
+DES-HSC WL, and +Euclid WL or +LSST WL cases, the underlying correlation between the median distance and the 
amplitude 
becomes apparent. This degeneracy provides a limitation to improving the $w$ constraint from the number counts by means 
of mass calibration. Given that it affects the halo masses directly, and not only the WL signal, we expect these 
degeneracies to be present also in other mass calibration methods, although to a different extent, given the 
different scaling of the selection observables with mass. 

As a side note, these degeneracies highlight the importance of fitting for mass calibration and number counts 
simultaneously and self consistently. A mass calibration done at fixed cosmology would miss these 
correlations and lead to underestimated uncertainties on the scaling relation parameters.  More worrisome, modeling 
mass calibration by simply adopting priors on the observable mass scaling relation parameters would 
miss the underlying physical degeneracies altogether \citep[e.g.,][]{sartoris16,pillepich18}.

The degeneracies between the distance redshift relation and the scaling relation parameters in the mass 
calibration explain why the impact of WL mass calibration in weaker in the $\nu$-$w$CDM model, compared 
to the $\nu$ $\Lambda$CDM model: in the latter $w$ is kept fixed, and the redshift evolution of distances and 
critical densities is controlled predominantly by a single variable: $\Omega_\text{M}$. With one degenerate 
degree of freedom less, WL mass calibration can put tighter constraints on $\ln A_\text{X}$ and $
\gamma_\text{X}$ in the $\nu$-$\Lambda$CDM than in the $\nu$-$w$CDM model. 

\subsection{Synergies with Planck CMB}\label{sec:syn_CMB}
\label{sec:eROSITA+CMB}

It is customary in observational cosmology to combine the statistical power of different experiments to further 
constrain the cosmological parameters. An important part of these improvements is due to the fact that each 
experiment has distinctive parameter degeneracies that can be broken in combination with constraints from 
another experiment. This is especially true for CMB temperature and polarization anisotropy measurements, 
which constrain the cosmological parameters in the early Universe, but display important degeneracies on 
late time parameters such as $\Omega_\text{M}$, $\sigma_8$ and $w$ \citep[for a recent study applicable to 
current CMB measurements, see][]{Howlett12}. We will discuss in the following the synergies between the 
Planck cosmological constraints from temperature and polarization anisotropy \citep{planck16_cosmo} and 
those from the eROSITA cluster counts analysis. 

\begin{figure}
	\vskip-0.2in
	\includegraphics[width=\columnwidth]{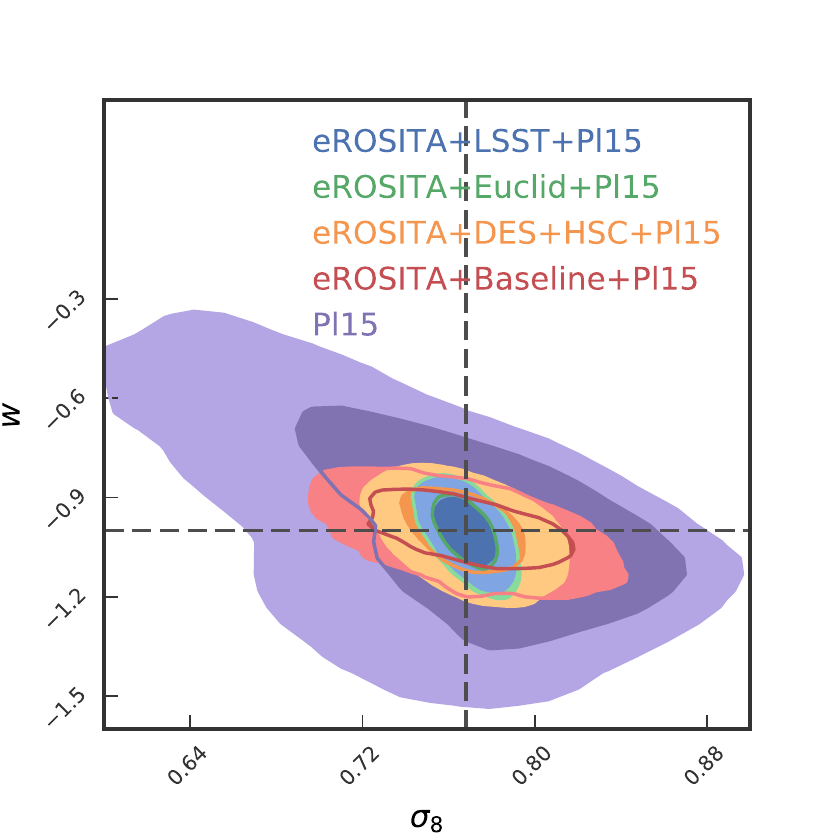}
	\vskip-0.1in
    \caption{Marginalized posterior sample of $\sigma_8$ and $w$ in the $w$CDM model. In purple the constraints 
from Planck CMB alone (Pl15), in red the constraints from the number counts and Planck 
(eROSITA+Baseline+Pl15), in orange the constraints from the addition of DES+HSC WL calibration 
(eROSITA+DES+HSC+Pl15), in green for the addition Euclid WL calibration (eROSITA+Euclid+Pl15), in blue for the addition 
LSST WL calibration (eROSITA+LSST+Pl15). 
Cluster information breaks the inherent CMB degeneracies and allows to constrain the late time parameters 
to high precision.}
    \label{fig:cmbdegen}
\end{figure}

\subsubsection{$\nu$-$w$CDM constraints}

In the $\nu$-$w$CDM model, the CMB suffers from the so called \textit{geometrical degeneracy} 
\citep{efstathiou99}, 
that arises because the CMB anisotropy primarily constrains the ratio of the sound horizon at recombination and the angular 
diameter distance to that epoch.  As a consequence, for example, the 
current day expansion rate $H_\text{0}$ is degenerate with the equation of state parameter 
$w$. This uncertainty in the expansion history of the Universe leads to large uncertainties on late time 
properties such as $\Omega_\text{M}$ and $\sigma_8$. Addition of a late time probe that constrains these 
quantities allows one to break the degeneracies and put tighter constraints on $w$. This can be nicely seen 
for the case of eROSITA in Fig.~\ref{fig:cmbdegen}, where the red CMB degeneracy between $\sigma_8$ 
and $w$ is broken by the addition of cluster information. The corresponding uncertainties are shown in 
Table~\ref{tab:baseline_constraints}.

While in this model the CMB alone is not able to determine $\Omega_\text{M}$, the addition of eROSITA number 
counts allows a constraint of $\pm0.019$.
Inclusion of WL mass information further reduces the uncertainty to $0.018$, $0.014$ and $0.013$ for DES+HSC, Euclid and 
LSST, 
respectively. The uncertainty in $\sigma_8$ is reduced from $0.065$ when considering only the CMB, to 
$0.032$ with number counts, $0.019$ with number counts and DES+HSC WL, and $0.010$ with number counts 
and Euclid, and $0.009$ with LSST WL. Noticeably, the determination of the equation of state parameter $w$ is improved 
from 
$0.242$ from CMB data alone, to $0.087$ when adding number counts. Even more remarkable is the fact 
that WL calibrated eROSITA constraints on $w$ are only margimally improved by the addition of CMB information.

\begin{figure}
	\vskip-0.2in
	\includegraphics[width=\columnwidth]{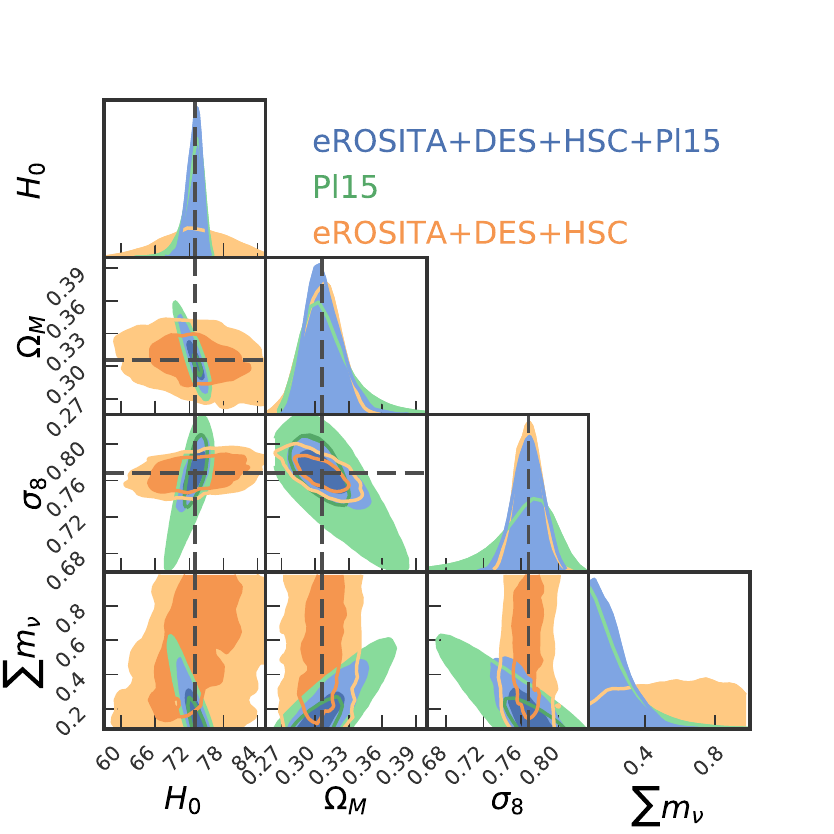}
	\vskip-0.1in
    \caption{Marginalized posterior sample of $H_\text{0}$, $\Omega_\text{M}$, $\sigma_8$ and $\sum m_\nu$ in the $\nu$-$
\Lambda$CDM model. In red the constraints from Planck CMB alone (Pl15) and the constraints from 
eROSITA number counts and DES WL calibration without CMB priors in blue (eROSITA+DES), and with CMB 
priors in purple (eROSITA+DES+Pl15). By measuring $\sigma_8$ and $\Omega_\text{M}$ independently of 
the sum of neutrino masses, WL calibrated cluster number counts break the degeneracy among these 
parameters in the CMB posteriors.}
    \label{fig:cmbdegen_mnu}
\end{figure}

\subsubsection{Constraints on sum of the neutrino masses}\label{sec:sum_mnu}

We showed earlier that cluster number counts, even when they are WL calibrated, provide little information about 
the sum of the neutrino masses in the regime $<1$ eV. On the other hand, the CMB posteriors on $
\sigma_8$ and $\Omega_\text{M}$ are strongly degenerate with the neutrino mass, as can be seen in 
Fig.~\ref{fig:cmbdegen_mnu}. Contrary to the CMB, the number counts of galaxy clusters are only weakly 
affected by the sum of the neutrino mass.  Recent studies have shown that the halo mass function is a 
function of the power spectrum of baryons and dark matter only \citep{costanzi13,castorina14}. Effectively,
this means that number counts can be used to constrain the density $\Omega_\text{coll}$ and fluctuation amplitude  $
\sigma_{8, \text{coll}}$ of baryons and dark matter  independently of the neutrino mass. 
If one considers matter as cold dark matter, baryons and neutrinos, as is customarily done, then $\Omega_\text{M}
=\Omega_\text{coll}+\Omega_\nu$ and $\sigma_8^2 = \sigma_{8, \text{coll}}^2 +\sigma_{8, \nu}^2$, where $
\Omega_\nu$ is the density parameter of neutrinos and $\sigma_{8, \nu}^2$ is the amplitude of their clustering on 
8$h^{-1}$~Mpc scales.  The counts derived constraints on $\Omega_\text{coll}$ and $ \sigma_{8, \text{coll}}$ then lead to 
only very weak
degeneracies between the sum of the neutrino masses and $\Omega_\mathrm{M}$ and $\sigma_8$, respectively, because 
neutrinos constitute a tiny fraction of the total matter density and the total matter fluctuations on 8$h^{-1}$~Mpc scales.
In Fig.~\ref{fig:cmbdegen_mnu} we can see how these very different parameter degeneracies in the CMB and cluster counts 
manifest themselves.
Combining these weaker degeneracies arising from eROSITA+DES WL with the more 
pronounced degeneracies in the CMB posteriors allows us to break the latter and to better constrain the sum 
of the neutrino masses. 

Consistently, we find that in the $\nu$-$\Lambda$CDM model, the addition of CMB priors only marginally improves 
the constraints eROSITA will put on $\sigma_8$ and $\Omega_\text{M}$. However, while the CMB alone puts 
an upper limit of $\sum m_\nu<0.514$ eV (at 95\%) we determine that the combination of Planck CMB and 
eROSITA number counts will constrain the neutrino masses to $<0.425$ eV, which will improve to 
$<0.401$ eV, $<0.291$ eV and $<0.285$ eV with the addition of WL information from DES+HSC, Euclid and LSST, 
respectively.

\begin{figure*}
	\includegraphics[width=\columnwidth]{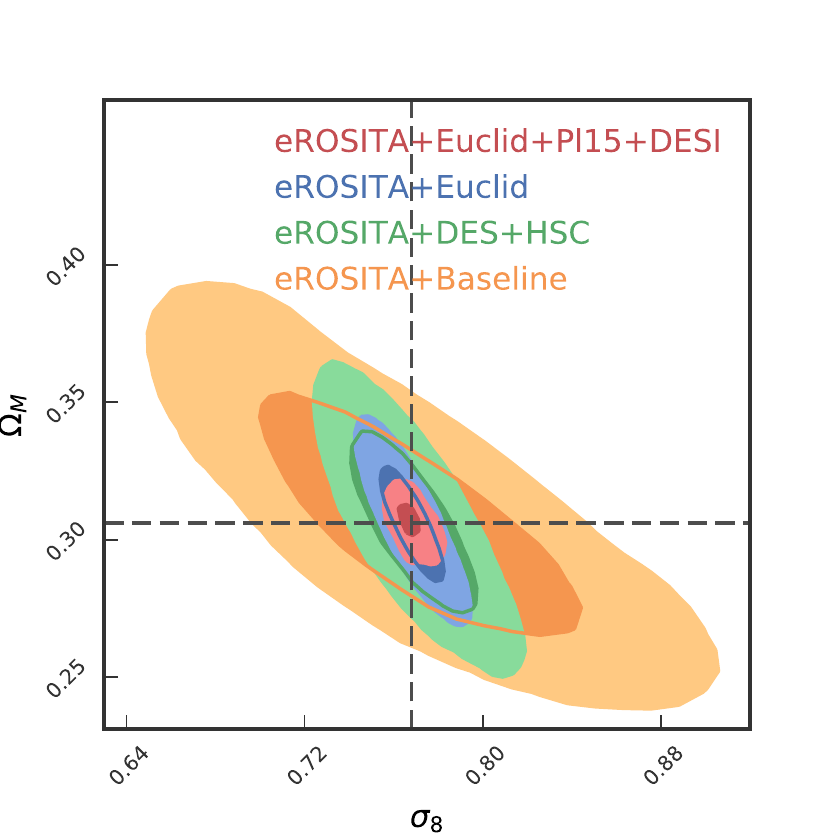}
	\includegraphics[width=\columnwidth]{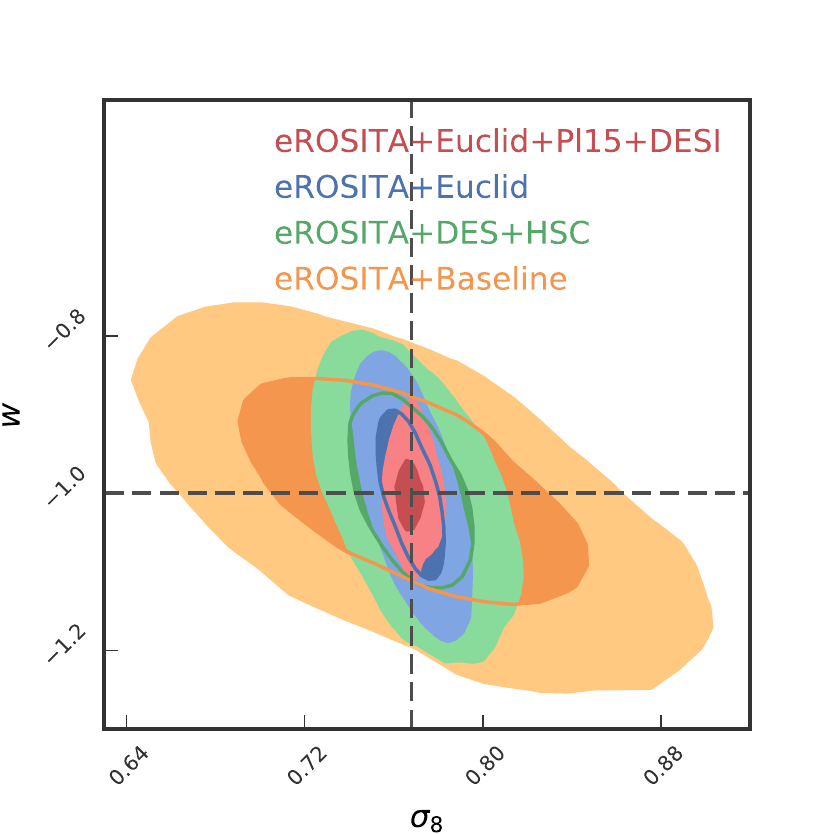}
	\vskip-0.15in
    \caption{2 dimensional marginal contours of the posteriors in ($\Omega_\text{M}$,$\sigma_8$) (left panel) and 
($w$,$\sigma_8$) (right panel), showing the incremental improvement of constraining power when first 
adding WL information and second combining with external cosmological data sets (``Pl15'' stands for the 
CMB fluctuation measurements by the Planck satellite, while ``DESI'' refers only to the BAO constraints). 
These posteriors are derived while simultaneously marginalizing over the summed neutrino mass.}
    \label{fig:plbao}
\end{figure*}

\subsection{Synergies with DESI BAO measurements}
\label{sec:eROSITA+BAO}

From the discussion in Section~\ref{sec:degeneracy}, it is apparent that the flux based X-ray selection and the 
distance dependent WL mass information lead to an inherent degeneracy between distances to the clusters 
and scaling relation parameters that ultimately limits the constraint on $w$. 
It would be desirable to utilize CMB independent constraints on the distance 
redshift relation, to allow for more stringent consistency checks between cluster derived constraints and CMB 
constraints. Some previous cosmological studies of X-ray clusters have used the distance information 
gleaned from the assumption of constant intracluster medium (ICM) mass fraction with redshift 
\citep{Mantz15,Schellenberger17}.  While these results are encouraging, a challenge with this method is that 
it only provides accurate distance information if in fact the ICM mass fraction is constant at all redshifts. It has 
been established for decades now that the ICM mass fraction varies with cluster mass \citep[e.g.,][]{mohr99}, 
but direct studies of how the ICM mass fraction varies over the redshift range of the eROSITA survey (i.e., 
extending beyond $z=1$) have only recently been undertaken \citep{lin12,chiu16,chiu18,bulbul19}.  The 
evolution is consistent with constant ICM mass fraction, but the uncertainties are still large.  Further study is 
clearly needed.  Another interesting eROSITA internal prospect for better constraining the distance redshift relation
 is to utilize the clustering of clusters to 
determine the BAO scale \citep[for a recent application, see][and references therein.]{marulli18}

As an alternative, we consider constraints from other low redshift experiments, more precisely the measurement of 
the Baryonic Acoustic Oscillations (hereafter BAO) in future spectroscopy galaxy surveys. In this work, we 
consider the forecast for the constraints provided by the Dark Energy Spectroscopy 
Instrument\footnote{\url{https://www.desi.lbl.gov}} \citep[DESI;][]{levi13} as the relative error on the transversal 
BAO measurement $d_\mathrm{A}/r_\text{S}$ and the radial BAO measurement $H(z)\, r_\text{S}$ as 
functions of redshift, where $d_\mathrm{A}$ is the angular diameter distance, $H(z)$ the expansion rate, and 
$r_\text{S}$ is the sound horizon.
The values adopted in this work are reported in Table V of \citet{Font-Ribera14}. Furthermore, we follow the 
authors indications and assume that in each redshift bin, the measurement error on the two quantities are 
correlated with correlation coefficient $\rho=0.4$. Using this information we perform an importance sampling 
of the posterior samples presented above and summarize the resulting uncertainties in 
Table~\ref{tab:baseline_constraints}.

When considering the uncertainties on the different parameters obtained by sampling these observables, we find 
that the BAO measurement dominates the uncertainty on $\Omega_\text{M}$. The addition of number counts, 
or number counts and WL information does not lead to major improvements on this parameter either in $
\nu$-$w$CDM or in $\nu$-$\Lambda$CDM. However, the uncertainty on the dark energy equation of state parameter $w$ is 
reduced from $0.086$ in the BAO only case, to $0.065$ when adding just number counts, $0.054$ and 
$0.047$ when adding DES+HSC and Euclid WL information, respectively. Remarkably, eROSITA counts with BAO 
priors on the expansion history outperforms eROSITA counts with CMB priors when it comes to constraining the 
parameters $\Omega_\text{M}$, $\sigma_8$ and $w$, while simultaneously marginalizing over the summed 
neutrino mass.  The latter is unconstrained by eROSITA+BAO, even when considering WL mass 
information. Furthermore, eROSITA+BAO allows us to measure the Hubble constant $H_\text{0}$ to varying 
degrees of precision, depending on the quality of the WL data. While these constraints never go below the 
present precision from other methods \citep[see for instance][]{riess16}, they will provide a valuable 
systematics cross-check \citep[for an example of systematics in SNe Ia that impact local $H_0$ measurements, see, e.g.,][]
{rigault13,rigault15,rigault18}.

\subsection{Combining all datasets}
\label{sec:combined}

It is current practice in cosmology to first test consistency of constraints from different data sets as a check on systematics 
and to then combine the constraints as possible to provide the most precise
cosmological parameter constraints possible.  In 
the case of a forecast work like this, agreement is guaranteed by the choice of input cosmology for the mock 
creation, while statistical independence can be assumed for eROSITA with WL data, DESI and the CMB 
measurement from Planck.

We provide the results of this combination at the bottom of Table~\ref{tab:baseline_constraints}.  In $\nu$-$w$CDM, already 
the combination of Planck CMB measurements and DESI BAOs allows us to 
determine $\Omega_\text{M}$ and $w$ to $0.007$ and $0.049$, respectively, while simultaneously putting an 
upper limit of $<0.284$ eV on the summed neutrino mass. Addition of eROSITA+Euclid WL only 
marginally improves these constraints to $0.005$ and $0.047$ for $\Omega_\text{M}$ and $w$, respectively, 
and leads to the 95\% confidence upper limit $\sum m_\nu<0.241$ eV. 
In this configuration, however, the added value of eROSITA 
number counts and WL mass calibration lies in the ability to constrain $\sigma_8$: while CMB and BAO put a 
constraint of $0.027$, addition of eROSITA improves this to $0.026$, $0.011$ and $0.006$, when considering 
the baseline mass information, DES+HSC WL, Euclid WL or LSST WL, respectively. In summary, using BAO and 
CMB priors together increases the constraining power of eROSITA cluster cosmology considerably, as can be seen in 
the shrinking of the 2 dimensional marginal contours in ($\Omega_\text{M}$,$\sigma_8$) and ($w$,$\sigma_8$) 
space, shown in the left and the right panel of Fig.~\ref{fig:plbao}, respectively.

\subsection{Inclusion of low mass clusters and groups}
\label{sec:low_mass}

In this work, we have taken the conservative approach of excluding all systems with a halo mass $\loa 2 \times 10^{14} 
\text{M}_{\sun}$ by means of increasing the eROSITA cluster count rate threshold at low redshift (cf. 
Section~\ref{sec:x-ray_mock} and Appendix~\ref{app:selection}). There are several good reasons to do so, all 
of them related, in one way or another, to an increase in systematic uncertainty when going to lower mass 
systems that are not as well studied. However, to enable comparison to previous work, and as a 
motivation to further investigate and 
control the systematic uncertainties in low mass clusters and groups, we also examine the impact of WL 
mass calibration on the constraining power for a cluster sample where the count rate threshold is reduced at low redshift so
that only clusters with masses $M_\text{500c} \lessapprox 5\times 
10^{13} \text{M}_{\sun}$ are excluded.

\begin{table*}
	\centering
	\caption{Parameter uncertainties, for number counts (eROSITA+Baseline), number counts and DES+HSC WL 
calibration (eROSITA+DES+HSC), number counts and Euclid WL calibration (eROSITA+Euclid), and number counts and 
LSST WL calibration (eROSITA+LSST) in the 
$\nu$-$w$CDM model when including low mass clusters. 
The units of the column ``$\sum m_\nu$'' and ``$H_\text{0}$'' are eV and km s$^{-1}$ Mpc$^{-1}$, respectively.
\textit{Comments: a)} This parameter is not constrained 
within the prior ranges. When reporting upper limits ``<'', we refer to the 95th percentile, while lower limits ``>'' 
refer to the 5th percentile. When a parameter is kept fixed in that model, we use ``--''.}
	\label{tab:lowmass_constraints}

	\begin{tabular}{llcccccccccc} 
		\hline
&		& $\Omega_\text{M}$ & $\sigma_8$ & $w$ & $\sum m_\nu$ & $H_\text{0}$& $\ln A_\text{X}$ & 
$B_\text{X}$  & $\gamma_\text{X}$ & $\sigma_\text{X}$ & $B_\text{X}^\prime$\\
		\hline
		\hline
		 \multicolumn{3}{l}{optimal mass calibration\hfil} & & & & & 0.028 & 0.021 & 0.050 & & 0.116 \\
 		\hline
		\hline
		\multispan{12}{\hfil eROSITA + WL\hfil}\\
		\hline
$\nu$-$w$CDM &		priors &  &  &  &  &  & 0.23 & 0.17 & 0.42 & 0.11 & 0.78 \\
&        eROSITA+Baseline & 0.025 & 0.038 & 0.079 & $^\text{a)}$ & 8.081 & 0.113 & 0.071 & 0.202 & 0.078 & 0.086 \\
&        eROSITA+DES+HSC & 0.012 & 0.012 & 0.069 & $^\text{a)}$ & 4.572 & 0.081 & 0.028 & 0.097 & 0.052 & 0.072\\
&        eROSITA+Euclid & 0.009 & 0.007 & 0.056 & $^\text{a)}$ & 3.762 & 0.042 & 0.019 & 0.073 & 0.027 & 0.058 \\
&        eROSITA+LSST & 0.007 & 0.006 & 0.050 & $^\text{a)}$ & 2.707 & 0.042 & 0.016 & 0.068 & 0.023 & 0.051 \\
 		\hline
 		\hline
		\multispan{12}{\hfil eROSITA + WL + Pl15 (TTTEE\_lowTEB)\hfil}\\
		\hline
$\nu$-$w$CDM &		priors (incl. CMB) & <0.393 & 0.063 & 0.242 & <0.667 & >62.25 & 0.23 & 0.17 & 0.42 & 
0.11 & 0.78 \\
&        eROSITA+Baseline & 0.017 & 0.028 & 0.078 & <0.580 & 2.745 & 0.131 & 0.026 & 0.128 & 0.083 & 0.087 \\
&        eROSITA+DES+HSC & 0.010 & 0.012 & 0.069 & <0.542 & 1.587 & 0.092 & 0.017 & 0.102 & 0.052 & 0.065\\
&        eROSITA+Euclid & 0.007 & 0.006 & 0.060 & <0.381 & 1.401 & 0.046 & 0.013 & 0.076 & 0.021 & 0.054 \\
&        eROSITA+LSST & 0.006 & 0.005 & 0.051 & <0.365 & 1.317 & 0.045 & 0.012 & 0.065 & 0.021 & 0.050 \\
 		\hline
 		\hline
		\multispan{12}{\hfil eROSITA + WL + DESI (BAO)\hfil}\\
		\hline
$\nu$-$w$CDM &		priors (incl. BAO) & 0.007 & $^\text{a)}$ & 0.086 & $^\text{a)}$ & $^\text{a)}$ & 0.23 & 
0.17 & 0.42 & 0.11 & 0.78 \\
&        eROSITA+Baseline & 0.006 & 0.016 & 0.051 & $^\text{a)}$ & 1.703 & 0.136 & 0.036 & 0.090 & 0.068 & 0.070 \\
&        eROSITA+DES+HSC & 0.006 & 0.009 & 0.048 & $^\text{a)}$ & 1.425 & 0.080 & 0.025 & 0.084 & 0.050 & 0.059 \\
&        eROSITA+Euclid & 0.005 & 0.005 & 0.038 & $^\text{a)}$ & 1.379 & 0.036 & 0.016 & 0.063 & 0.021 & 0.050 \\
&        eROSITA+LSST & 0.004 & 0.005 & 0.038 & $^\text{a)}$ & 1.303 & 0.036 & 0.014 & 0.061 & 0.021 & 0.049 \\
 		\hline
 		\hline
 		\multispan{12}{\hfil eROSITA + WL + DESI + Pl15\hfil}\\
		\hline

$\nu$-$w$CDM &		priors (incl. CMB+BAO) & 0.007 & 0.027 & 0.049 & <0.284 & 1.118 & 0.23 & 0.17 & 
0.42 & 0.11 & 0.78 \\
&        eROSITA+Baseline & 0.005 & 0.015 & 0.046 & <0.279 & 1.114 & 0.134 & 0.022 & 0.079 & 0.067 & 0.067 \\
&        eROSITA+DES+HSC & 0.005 & 0.010 & 0.044 & <0.242 & 1.040 & 0.078 & 0.014 & 0.067 & 0.049 & 0.056\\
&        eROSITA+Euclid & 0.005 & 0.005 & 0.037 & <0.237 & 1.015 & 0.039 & 0.012 & 0.058 & 0.021 & 0.049 \\
&        eROSITA+LSST & 0.004 & 0.005 & 0.034 & <0.224 & 0.790 & 0.039 & 0.010 & 0.053 & 0.021 & 0.047 \\
 		\hline
 		\hline

 	\end{tabular}
\end{table*}

\subsubsection{Systematics of low mass clusters and groups}

There are several important systematic concerns.  For instance, \citet{bocquet16_mfct} find in a study using 
hydrodynamical structure formation simulations that for masses below $10^{14} \text{M}_{\sun}$, baryonic 
feedback effects reduce the halo mass function by up to 10\% compared to halo mass functions extracted from
dark matter only simulations. 
The magnitude of this effect depends on the feedback model, and therefore needs be treated as a systematic 
uncertainty in the cosmological modeling. The magnitude of this uncertainty awaits further study.

Baryonic feedback effects also impact the mass profiles of clusters. \citet{lee18} show how active galactic nuclei 
feedback induces a deficit of mass in the cluster center when compared to gravity only simulations. The 
partial evacuation of baryons is strong enough to modify also the matter profile. \citet{lee18} demonstrate how 
this effect impacts the WL bias $b_\text{WL}$ and the WL scatter $\sigma_\text{WL}$, making them mass 
dependent. Such effects will need to be taken into account, especially when considering lower mass systems.

Similarly, the thermodynamic structure of low mass systems, generally called groups, is more complex than for 
massive galaxy clusters, showing a larger impact of non gravitational physics \citep{eckmiller11, 
Bharadwaj14, barnes17}. \citet{lovisari15} showed that the mass slope of the luminosity mass relation is 
significantly steeper for groups than for clusters. \citet{Schellenberger17} demonstrate how such a break in 
the power law might bias the cosmological results derived from an X-ray selected cluster sample. We have thus chosen
the conservative approach of excluding these systems from our primary eROSITA forecasts, thereby
reducing the sensitivity of the forecast cosmological 
parameter constraints to these important complications at low masses.

\subsubsection{Improvement of the constraints}

Nevertheless, the controlled environment of mock data analysis allows us to investigate how much constraining power 
could ideally be gained  by lowering the mass limit if all the above described systematics where well 
understood and controlled. To this end, we select a low mass sample by imposing an observable selection with redshift that 
enforces $M_{500\text{c}}\goa5\times 10^{13} \text{M}_{\sun}$, assuming that the scaling relation  and the mass 
function used for the fiducial sample are still valid also at this lower mass scale. This increases the sample size to 
43k clusters, with a median redshift $\bar z=0.31$ and a median halo mass of $\bar M_{500\text{c}} = 
1.4\times 10^{14} \text{M}_{\sun}$. The resulting constraints on the parameters of the $\nu$-$w$CDM model 
are shown in Table~\ref{tab:lowmass_constraints}. The constraints both on the cosmological parameters, as well as on 
the scaling relation parameters show a strong improvement compared to those from the higher mass sample. 
For eROSITA number counts we determine that the uncertainty on $\Omega_\text{M}$, $\sigma_8$ and $w$ 
will be reduced by factors of $1.3$, $1.4$ and $1.3$, respectively. When calibrating masses with DES+HSC, we 
find improvements of factor $1.9$, $1.4$ and $1.2$, when considering Euclid the inclusion of low 
mass systems will reduce the uncertainties by $1.8$, $1.7$ and $1.3$, 
while using LSST leads to reductions by  $2.0$, $1.7$ and $1.4$.
In absolute terms, eROSITA including 
low mass systems, calibrated with Euclid will provide constraints on $\Omega_\text{M}$, $\sigma_8$ 
and $w$ of $0.009$, $0.007$ and $0.056$, respectively. We emphasize that these tight constraints can 
only be obtained if the aforementioned systematic effects are adequately controlled.
   
\section{Discussion}\label{sec:discussion}

The above presented results on the constraining power of the eROSITA cluster sample demonstrate its value as a 
cosmological probe. They also underline the crucial impact of WL mass calibration on the constraining power 
of cluster number counts. However, they also give some clear indications of how this impact manifests itself in 
detail. 

In the following subsections we discuss first how the constraints on the scaling relation parameters are affected by 
the addition of better WL data, by the choice of the model and by the choice of cosmological priors, resulting 
in an assessment of the conditions under which we can attain an optimal mass calibration. We then 
determine the sensitivity of our observable to the different input parameters. Finally, we compare our 
prediction to  the constraints from current and future experiments.

\subsection{Impact of WL on scaling relation parameters}
\label{sec:WLimpact}

In the previous section we discussed in detail the impact of WL mass calibration on the eROSITA cosmological 
parameter constraints. Naturally, adding WL information will also improve the constraints on the scaling 
relation parameters. The resulting uncertainties are reported in Table~\ref{tab:baseline_constraints}. In the 
following, we will focus on two interesting aspects of these results: first, we assess under which 
circumstances eROSITA will be optimally calibrated; second, we comment on the constraints on the scatter in 
observable at fixed mass.

\subsubsection{Which mass calibration is optimal?}

In Section~\ref{sec:optimal}, we introduced the concept of the $\textit{optimal}$ mass calibration. Comparing the 
bounds on the parameter uncertainties derived there to the forecasts for DES+HSC, we find that, independent of 
the presence of external cosmological priors and in both models we consider, DES WL will not provide an 
optimal calibration of the eROSITA observable mass relation. Only the calibration of the mass slope 
$B_\text{X}$ when considering CMB and BAO data is an exception to this. This is not to say that, as shown 
above, the inclusion of DES+HSC WL information does not improve the cosmological constraints. It is to say that 
some part of the information contained in the number counts is used to constrain the scaling relation 
parameters instead of the cosmological parameters.

The optimal nature of the Euclid or LSST mass calibration is more subtle. When the dark energy equation of state 
parameter is kept fixed in the $\nu$-$\Lambda$CDM model, Euclid provides an optimal mass 
calibration on the amplitude of the scaling relation, 
both with and without external cosmological priors from CMB or BAO observations. However, in the 
$\nu$-$w$CDM model without external priors, Euclid or LSST WL does not constrain the scaling relation 
parameters optimally. The amplitude is calibrated optimally after the inclusion of BAO data. On the other 
hand, including CMB priors makes an optimal calibration of the mass trend possible. In the presence of dark 
energy with free but constant equation of state, the redshift slope is never calibrated optimally.
Nevertheless, as demonstrated in the previous section, even in the limit of sub-optimal mass calibration, the 
eROSITA dataset provides cosmological information complementary to these other cosmological 
experiments.
Furthermore, the calibration of the redshift trend could be improved by complementary direct mass calibration 
methods. At high redshift, the most promising options would be pointed observations of high-z clusters 
\citep{schrabback18a, schrabback18b} and CMB-WL calibration \citep{baxter15, planck16cluster_cosmo, baxter18}. 

\subsubsection{Scatter in the count rate to mass relation}

\begin{figure*}
	\includegraphics[width=\textwidth]{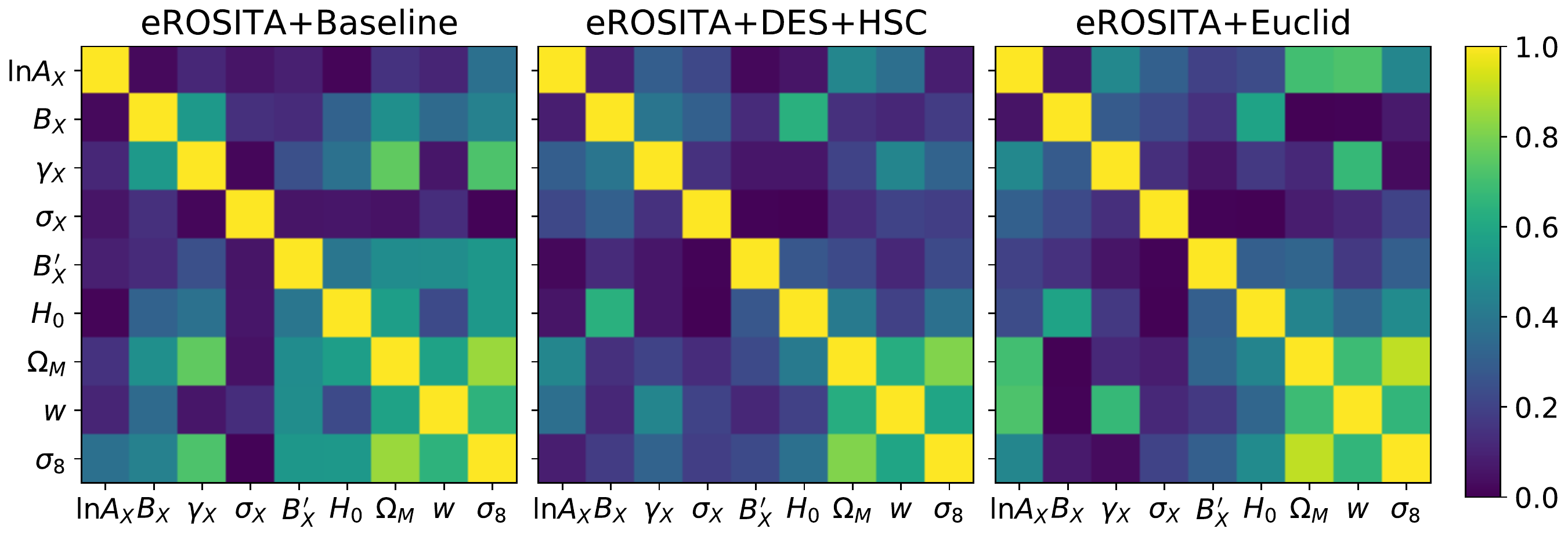}
	\vskip-0.10in
    \caption{Absolute values of the correlation matrices of the posterior samples in the $w$CDM model, for number 
count (eROSITA+Baseline), number counts with DES+HSC WL information (eROSITA+DES+HSC) and number counts 
with Euclid WL information (eROSITA+Euclid). Noticeably, we find that the initial correlations between 
the pairs $(\Omega_\text{M}, \sigma_8)$ and $(B_\text{X}, \gamma_\text{X})$ (in eROSITA+Baseline) is gradually 
broken by the addition of better mass information (eROSITA+DES+HSC and eROSITA+Euclid). However, the better 
the mass information, the clearer the inherent correlations between $w$, $\Omega_\text{M}$, $A_\text{X}$ 
and $\gamma_\text{X}$. They indicate the degeneracies among these parameters stemming from the cosmology 
dependence of the rate mass mapping, as discussed in Section~\ref{sec:params_sens}. }
    \label{fig:covs}
\end{figure*}

One may imagine that the inclusion of low scatter mass proxies in the number counts and mass calibration 
analysis may tighten the constraints on the scatter and thereby reduce the uncertainties on the cosmological 
parameters. Our present work does not seem to support this hypothesis. First, we show that even an 
arguably weak constraint on the scatter can be considered an optimal calibration (cf. 
Section~\ref{sec:min_mssclbr}). In other words, even at fixed cosmology, the number counts are unable to 
constrain the scatter. Consequently, our ability to constrain the cosmology using the number counts is not 
expected to depend strongly on the knowledge of the scatter. This can also be seen by the fact that there is 
little correlation between the scatter and any other parameter of interest in the $\nu$-$w$CDM posterior 
sample, as shown in Fig.~\ref{fig:cosmo_constr} and Fig.~\ref{fig:covs}. We conclude that constraining the 
scatter to high precision, although of astrophysical interest, is not required to perform an optimal cosmological 
analysis.

Furthermore, our results indicate that DES+HSC, Euclid and LSST WL mass calibration will be able to determine the 
scatter to $0.062$, $0.034$, and $0.030$, respectively (see Table~\ref{tab:baseline_constraints}). This may seem surprising, 
because WL mass calibration has large observational uncertainties and a large intrinsic scatter when 
compared to typical low scatter mass proxies such as the ICM mass or temperature. However, the final 
constraining power stems in our analysis from the large number of cluster with WL information and the 
relatively small prior uncertainty on the intrinsic WL scatter $\sigma_\text{WL}$. In summary, given that the 
knowledge of the scatter does not impact the constraints on the cosmological parameters, and that WL mass 
calibration is able to constrain the scatter directly, it is not clear that a dedicated scatter calibration through 
the inclusion of low scatter mass proxies like the ICM mass will significantly impact eROSITA cluster 
cosmology constraints.  Further study would be required to confirm this.

\subsection{Parameter sensitivities}\label{sec:params_sens}

\begin{figure*}
	\includegraphics[width=\textwidth]{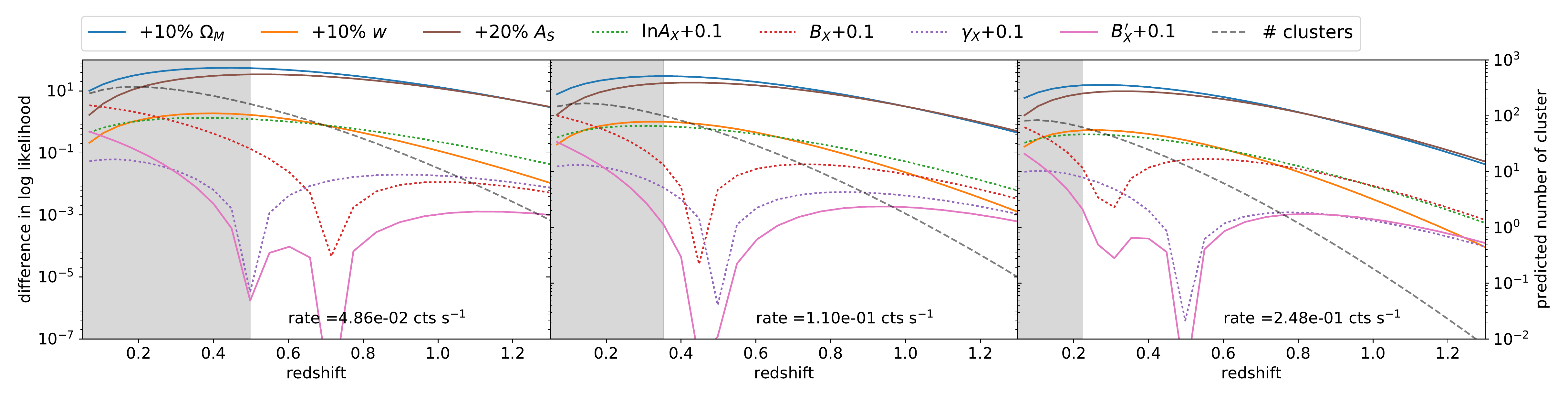}
	\vskip-0.1in
    \caption{Sensitivity in terms of change in log likelihood of the number counts likelihood  to various parameters as a 
function of redshift.  From left to right, each panel represents a higher count rate bin. 
 The total number of clusters for the 
  fiducial parameter values is shown as a dashed line.  The parameters are varied from the fiducial 
values as noted at the top of the figure. 
The grey area shows the redshift range where we exclude low mass clusters by raising the selection 
threshold.  Notably, we find that the number counts likelihood is most sensitive to the parameters $\Omega_\text{M}$ and
$A_\text{S}$ with comparable sensitivity to $w$ and $A_\text{X}$. }
    \label{fig:cosmo_sens_nobs}
\end{figure*}

\begin{figure*}
    \vskip-0.25in
	\includegraphics[width=\textwidth]{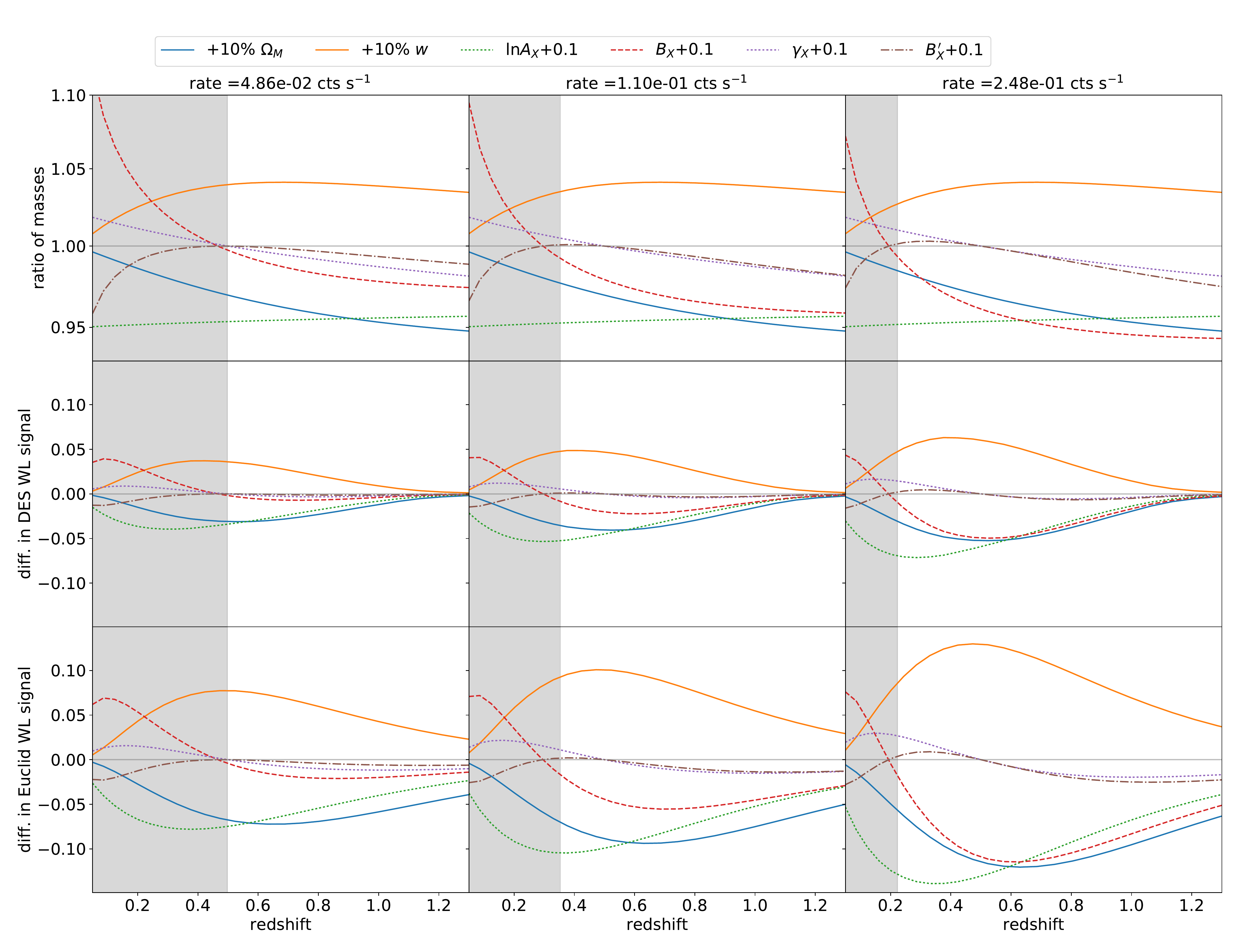}
	\vskip-0.1in
    \caption{Sensitivity of different mass observables to the parameters considered in this work. On the x-axis, we 
plot the redshift and each column represents a different count rate. The parameters are varied around the 
input values. The grey area shows the observable range which is excluded by the approximate mass cut. 
From the top, the first row shows the fractional change in mass. The second and third rows show the 
difference in tangential shear for a single cluster, weighted by the observational WL uncertainty for a single cluster at that 
redshift, for DES and 
Euclid, respectively. We also see that both for the halo masses and for the shear signal, $\Omega_\text{M}$ 
and $w$ lead to changes comparable to the change in amplitude $\ln A_\text{X}$ and $\gamma_\text{X}$. 
We conclude that these parameters must be degenerate with each other.}
    \label{fig:cosmo_sens}
\end{figure*}

To investigate in more detail how our observables-- i.e. the number counts of clusters as a function of rate and 
redshift together with the WL mass calibration information-- depend on the model parameters, we perform the 
following experiment: we vary the model parameters one by one and examine how the number counts, the 
masses and the WL signals change. The results of this test are shown in Figs.~\ref{fig:cosmo_sens_nobs} 
and~\ref{fig:cosmo_sens}. At three different fixed rates (increasing from left to right in the three columns), 
we investigate the sensitivity as a function of redshift 
with respect to the input parameter of the likelihood of the number counts (Fig.~\ref{fig:cosmo_sens_nobs}), as well as the 
masses, and the WL signals (Fig.~\ref{fig:cosmo_sens}). We grey out the part of rate--redshift space that 
is rejected due to our mass cut.

\subsubsection{Number counts}

Fig.~\ref{fig:cosmo_sens_nobs} shows the sensitivity of the number counts with respect to shifts in the input 
parameters. We decide here to plot the difference in log likelihood between the fiducial number counts 
$N_\text{fid}$ and the number counts $\tilde N$ if one parameter is varied. The difference in log likelihood in 
each bin reads 
\begin{equation}
\delta \ln L = N_\text{fid} \ln \left(\frac{N_\text{fid}}{\tilde N} \right) - N_\text{fid} + \tilde N, 
\end{equation}
which can be simply obtained by taking the Poisson log likelihoods in that bin. We find that the number counts are 
most sensitive to the parameters $\Omega_\text{M}$ and $A_\text{S}$. The sensitivity to the parameters 
$w$, $A_\text{X}$, $B_\text{X}$, and $\gamma_\text{X}$ is much lower. This is reflected also in our results 
for the parameter uncertainties (Tables~\ref{tab:baseline_constraints} and \ref{tab:lowmass_constraints}).  
The number counts do put tighter 
constraints on $\Omega_M$ and $\sigma_8$, than on $w$, consistent with results from the 
first forecast studies for large scale cluster surveys \citep{haiman01,holder01}.

For comparison we also plot the total number of objects $N_\text{fid}$ (dashed line), on a scale proportional to the 
difference in log likelihood. We can readily see that the difference in log likelihood is not simply proportional to 
the number of objects: the rarer, higher redshift, and consequently, at fixed rate, higher mass objects 
contribute more log likelihood per cluster than the lower redshift, lower mass systems. This trend is especially 
true for the constraints on $\Omega_M$ and $\sigma_8$ ($A_\mathrm{S}$), as noted in 
previous studies of cluster number counts \citep{haiman01, majumdar04}.  The sensitivity to 10\% shifts in $w$ and 
$A_\mathrm{X}$ are comparable.  The more similar the shapes of the sensitivity curves for two parameters, the stronger the 
parameter degeneracy one could expect between those parameters.

\subsubsection{Masses and WL observables}

The first row of Fig.~\ref{fig:cosmo_sens} shows how much the masses are impacted by changes in input 
parameters. To this end, we plot the ratio between the input mass and the mass determined at the shifted 
parameters. In the range of interest for our study, the white area, we find that all parameters (except for 
$A_\text{S}$, of course, which we do not include in this figure) have a 
comparably large impact on the masses. Most remarkably, both shifts in $
\Omega_\text{M}$ and $w$ change the masses associated to a given rate and redshift. This is because the 
rate mass relation has a strong distance dependence and also some critical density dependence.  Both $\Omega_\text{M}$ 
and $w$ alter the redshift dependence of distances and critical densities. 

More precisely, the shift to more positive $w$ leads to a shift to higher masses, which mirrors the effect of changing the
amplitude of the scaling relation $\ln A_\text{X}$ and the redshift slope $\gamma_\text{X}$. Similarly, the 
redshift dependent mass shift induced by $\Omega_\text{M}$ could be compensated by a shift in the redshift 
slope $\gamma_\text{X}$ and $\ln A_\text{X}$. We therefore conclude that within the context of the masses corresponding to 
a fixed eROSITA count rate, the parameters 
$w$ and $\Omega_\text{M}$  are degenerate with a combinations of $\ln A_\text{X}$ and $
\gamma_\text{X}$. This degeneracy impacts the predicted halo masses. The mass slope parameter 
$B_\text{X}$, however, seems to impact the masses in a distinctively different way, leading to no obvious 
parameter degeneracy.  The same can be said for its redshift trend $B_\text{X}^\prime$.

In our main experiment, we do not consider perfect halo masses, but WL signal. 
Therefore, we explore also the 
sensitivity of the WL signal for a single cluster to the input parameters. For the sake of simplicity, we do not consider the 
entire 
profile, but just assume one large radial bin spanning the fixed metric range corresponding to 0.25 -- 5.0 Mpc 
in our fiducial cosmology.
Given the constant metric size of the area considered, the WL measurement uncertainty for a single cluster  due to 
shape noise can be computed by considering the background source density as a function of cluster redshift 
$n_\epsilon(z_\text{cl})=n_\epsilon(z_\text{s}>z_\text{cl}+0.1)$.
In addition, the mapping from halo mass to tangential shear is non-linear and cosmology dependent. 
Consequently, the shear signal associated with a given rate and redshift is expected to have strong 
dependencies on cosmological parameters and, through the mass, also on the scaling relation parameters.

We visualize these effects in the second and third rows of Fig.~\ref{fig:cosmo_sens} by plotting the difference 
between the WL signal for a single cluster in the fiducial model and the shifted model, divided by the expected 
magnitude of the 
shape noise for a single cluster. Indeed, one can readily see how the sensitivity per cluster of DES WL (second row) is 
generally lower, but 
also decreases more quickly with redshift than the sensitivity of Euclid WL (third row). This 
is due to the larger Euclid source galaxy sample and its extension to higher redshift as compared to DES.  
The trends we discuss above for the difference in halo mass do apply also to the sensitivity of the WL 
signal as a function of redshift. 

We find the same degeneracies in the covariance matrices of our posterior samples in the $\nu$-$w$CDM model 
for the three cases of eROSITA+baseline, eROSITA+DES+HSC, and eROSITA+Euclid, shown in 
Fig.~\ref{fig:covs}. In the case of number counts alone, we find a strong correlation between the pairs $
(\Omega_\text{M}, \sigma_8)$ and $(B_\text{X}, \gamma_\text{X})$. The latter degeneracy is strongly reduced by the 
addition of WL mass information, and is not present in the case of Euclid WL calibration. This is due to 
the fact that WL is quite sensitive to $B_\text{X}$. This is in line with improvements of both the 
$(\Omega_\text{M}, \sigma_8)$ and $B_\text{X}$ constraints when adding WL mass information. However, 
when $w$ is free to vary, the degeneracies between $w$, $\ln A_\text{X}$ and $\gamma_\text{X}$ lead to 
stronger correlations between these parameters for better mass information. They are most pronounced in 
the case of number counts with Euclid WL mass calibration. 

\subsection{Comparison to previous work}

Finally, we compare our results to the constraints of recent and future experiments, with the intention of exploring 
how competitive eROSITA will be.

\subsubsection{Current probes}

The most up to date number counts analysis of an X-ray selected sample with WL mass calibration has been 
presented by \citet[called Weighing the Giants, hereafter WtG]{Mantz15}. It consists of 224 clusters, 51 of 
which have a WL mass measurement, and 91 of which have ICM mass measurements. The analysis method 
is similar to the one described in this paper, with the exception that we did not consider cosmological 
constraints from the measurement of the ICM mass fraction. In the $w$CDM model (i.e. fixing the neutrino mass), when 
considering only X-
ray and WL data, the uncertainties on $\Omega_\text{M}$, $\sigma_8$ and $w$ are 0.036, 0.031 and 0.15, 
respectively. The direct comparison to our work is made difficult by the addition of the distance sensitive gas 
fraction measurements, which by themselves constrain $\delta \Omega_\text{M}=0.04$ and $\delta w = 0.26$ 
\citep{Mantz14}. This measure clearly dominates the error budget on $\Omega_\text{M}$ and provides 
valuable distance information. Nevertheless, eROSITA cluster cosmology is evenly matched with WtG when 
considering just the number counts. It will outperform the constraining power of WtG when calibrated with 
DES+HSC WL information. In the case of LSST WL calibration, we project that the uncertainties on $
\Omega_\text{M}$, $\sigma_8$ and $w$ are smaller by factors $2.6$, $3.1$ and $2.1$, respectively. These 
projections ignore distance information from the eROSITA clusters and AGN, which would further improve the constraints.

Another recent cluster cosmology study has been presented by \citet{dehaan16}. Therein, the cosmological 
constraints from 377 Sunyaev-Zeldovich selected clusters detected by the South Pole Telescope (hereafter 
SPT) above redshift $>0.25$ are determined. From the number counts alone, the dark energy equation of state parameter 
is constrained to a precision of $\delta w = 0.31$, which is a factor $3.1$ worse than our prediction for the 
number counts from eROSITA alone. Furthermore, \citet{dehaan16} find $\delta \Omega_\text{M} = 0.042$ and $\delta 
\sigma_8 = 0.039$, while keeping the summed neutrino mass fixed at its minimal value. 
By comparison, in the baseline configuration eROSITA will improve the constrain on $\Omega_\text{M}$ and $\sigma_8$ by 
a factor $1.5$ and $1.2$, however while marginalizing over the summed neutrino mass. 
Also note that the priors used for the \citet{dehaan16} analysis encode the 
mass uncertainty over which \citet{bulbul19} marginalized when deriving the uncertainties on the X-ray 
scaling relation parameters we employ as our eROSITA+Baseline.

When the SPT number counts are combined with the CMB constraints from Planck, \citet{dehaan16} report 
constraints on $\sigma_8$ and $w$ of $0.045$ and $0.17$ respectively. We find that eROSITA number 
counts alone, in combination with Planck, will do better by a factor $2.8$ on $\sigma_8$ and a factor $2.0$ on 
the equation of state parameter $w$, while additionally marginalizing over the summed neutrino mass.
These numbers improve even more, if we consider the WL mass 
calibration by DES+HSC, Euclid and LSST. 

Comparing our forecasts on the improvement of the upper limit on the summed neutrino mass to previous results from the 
combination of Planck CMB measurements with either SPT cluster number counts or WtG is complicated by several factors.
First, we considere the full mission results for Planck \citep{planck16_cosmo}, 
while SPT \citep[][]{dehaan16} used the half mission data \citep{planck13cosmo} in addition to BAO data, 
and WtG \citep[][]{Mantz15} additionally added ground based CMB measurements and supernova data.
SPT reports the measurement $\sum m_\nu = 0.14 \pm 0.08$ eV, which is impacted to some degree 
by the statistically insignificant shift between their constraint and the CMB constraints in the $(\Omega_\text{M}, \sigma_8)$ 
plane.
Comparison to this result is complicated by our choice to use the minimal neutrino mass as input value. 
On the other hand, WtG reports $\sum m_\nu \le 0.22$ at 95\% confidence, which is comparable with our result from 
eROSITA number counts, DES+HSC WL, Planck CMB and DESI BAO. 

The latest cosmological constraints from measurements of the Large Scale Structure (LSS) of the Universe were 
presented by the \citet{desY1_3x2pt} for the first year of observations (Y1), where the joint constraints from 
the cosmic shear and photometric galaxy angular auto- and crosscorrelation functions are derived. In the $
\nu$-$w$CDM model, the uncertainties on $\Omega_\text{M}$, $\sigma_8$ and $w$ are 0.036, 0.028 and 
0.21, respectively. This is better than the constraints from eROSITA number counts alone, except for the dark 
energy equation of state parameter, which will be constrained better by eROSITA. However, utilizing DES+HSC to 
calibrate the cluster masses, we forecast that eROSITA will outperform the DES-Y1 analysis. In 
combination with Planck CMB data, DES-Y1 puts a 95\% upper limit of 0.62 eV on the sum of the neutrino 
masses, whereas we forecast an upper limit of 0.424 (0.401) when combining eROSITA number counts (and 
DES+HSC WL calibration) with Planck data. Considering that our DES WL analysis assumes year 5 data, it will be 
interesting to see whether the DES Y5 LSS measurements or eROSITA with DES WL calibration will provide 
the tighter cosmological constraints. 

As can be seen from Table~\ref{tab:baseline_constraints}, eROSITA will clearly outperform Planck CMB measurements 
on several cosmological parameters.  
In the $\nu$-$\Lambda$CDM model, eROSITA with WL mass information will outperform 
Planck on the parameters $\Omega_\text{M}$ and $\sigma_8$, and in the 
$\nu$-$w$CDM eROSITA with WL case will also outperform Planck on the equation of state 
parameter $w$.  However, for constraints on the sum of the neutrino 
mass, Planck alone offers much more than eROSITA alone.
 Given, however, that eROSITA and Planck extract their constraints at low redshift and high redshift, 
respectively, the true benefit of these two experiments lies in assessing the mutual consistency and thereby 
probing whether our evolutionary model of the Universe is correct. If this is the case, their joint constraints will 
tightly constrain the cosmological model, and provide improved constraints on the sum of neutrino masses.

\subsubsection{Previous forecasts for eROSITA}

This work elaborates further on the forecast of the eROSITA cosmological constraints first presented in 
\citet{merloni12}, and subsequently discussed in more detail in P18. The direct 
comparison to the latter is complicated by several diverging assumptions, including that we only consider the 
German half of the sky.   Perhaps the most significant difference is their approach of using 
Fisher matrix estimation and modeling mass calibration as simply being independent priors on the 
various scaling relation parameters, whereas we have developed a working prototype for
the eROSITA cosmology pipeline and used it to analyze a mock sample with shear profiles in a self-consistent manner.

Other differences include their use of different input scaling relations from older work at lower redshift 
and different fiducial cosmological parameters.  P18 includes constraints from the angular clustering 
of eROSITA clusters, although these constraints
are subdominant in comparison to counts except for parameters associated with non-Gaussianity in the initial density
fluctuations \citep[see][]{pillepich12}.  In our analysis, we marginalize 
over the sum of the neutrino mass as well as relatively weak priors on $\omega_\mathrm{b}$ and $n_\mathrm{S}$.  

Following what P18 call the \text{pessimistic} case with an approximate 
limiting mass of $5\times 10^{13} M_{\sun} h^{-1}$, they predict 89 k clusters, which is in good agreement 
with our forecast of 43 k clusters when including clusters down to masses of $5\times 10^{13} M_{\sun}$. 
Under the assumption of a 0.1 \% amplitude prior, 14 \% mass slope prior and 42 \% redshift slope prior, they
forecast a constraint of $0.017$, $0.014$ and $0.059$ on $\sigma_8$, $\Omega_\text{M}$ and $w$, 
respectively. P18 also consider an \textit{optimistic} case, in which clusters down to masses of $1\times 
10^{13} M_{\sun} h^{-1}$ are used under the assumption of 4 times better priors on the scaling relation 
parameters. For this case, the constraints on  $\sigma_8$, $\Omega_\text{M}$ and $w$ are $0.011$, $0.008$ 
and $0.037$, respectively. 

A quantitative comparison to our work is complicated by the fact that we find a constraint on the amplitude of the 
scaling relation (through direct modeling of the WL calibration from Euclid or LSST) that is worse than their 
\textit{pessimistic} case, but our constraint on the mass and redshift trends is better than their \text{optimistic} 
case.  Consistently, we predict tighter constraints of $\sigma_8$ and $\Omega_\text{M}$, which are sensitive 
to the mass and redshift trends of the scaling relation, while we predict lower precision on $w$, which we 
demonstrate to be degenerate with the amplitude of the scaling relation through the amplitude distance 
degeneracy.  Important here is the realization that the observed shear profiles map into cluster mass 
constraints in a distance dependent fashion \citep[this is true for all direct mass constraints;][]{majumdar03}.  
It is not straightforward to capture this crucial subtlety by simply 
adopting priors on observable mass scaling 
relation parameters.

\subsubsection{Euclid cosmological forecasts}

The Euclid survey will not only provide shear catalogs to calibrate the masses of clusters, but will also allow the 
direct detection of galaxy clusters via their red galaxies \citep{sartoris16}, and the measurement of the auto- 
and cross-correlation of red galaxies and cosmic shear \citep{Giannantonio14}. For the optically selected 
Euclid cluster sample, \citet{sartoris16} forecast $2\times 10^6$ galaxy clusters with limiting mass of $7 \times 
10^{13} M_{\sun}$ up to redshift $z=2$, yielding constraints on $\Omega_\text{M}$, $\sigma_8$, and $w$ of 
$0.0019$ ($0.0011$), $0.0032$ ($0.0014$), and $0.037$ ($0.034$), respectively, when assuming no 
knowledge on the scaling relation parameter (perfect knowledge of the scaling relation parameters). Under 
these assumptions, the number counts and the angular clustering of Euclid selected clusters would outperform 
eROSITA cluster cosmology.  Nevertheless, cross comparisons between the X-ray based eROSITA selection 
and the optically based Euclid cluster selection will provide chances to validate the resulting cluster 
samples.

\citet{Giannantonio14} forecast that the auto- and cross-correlations between red galaxies and cosmic shear in the 
Euclid survey will provide constraints on $\Omega_\text{M}$, $\sigma_8$, and $w$ of $0.005$, $0.033$ and 
$0.050$, respectively. Such a precision on $\sigma_8$ would be achieved by the baseline eROSITA+Euclid 
analysis, too. However, to achieve similar precisions in $\Omega_\text{M}$ and $w$, it would be necessary to 
consider eROSITA detected clusters down to masses of $ 5\times 10^{13} \text{M}_{\sun}$.

\section{Conclusions}\label{sec:conclusions}
In this work, we study the impact of WL mass calibration on the cosmological constraints from an eROSITA cluster 
cosmology analysis. To this end, we create a mock eROSITA catalog of galaxy clusters. We assign luminosities and ICM 
temperatures to each cluster using the latest measurements of the X-ray scaling relations over the relevant redshift range 
\citep{bulbul19}.
Considering the eROSITA ARF, we then compute the eROSITA count rate for all 
clusters in this sample. We apply a selection on the eROSITA count rate, corresponding to a $\sim6\sigma$ detection 
limit given current background estimates, to define a sample for a cosmological forecast.  This detection limit ensures both 
high 
likelihood of existence and angular extent, and -- through raising the detection threshold at low redshift -- also excludes low 
mass objects at low redshift.  
We assume all cluster redshifts are measured photometrically using red sequence galaxies \citep[see discussion in, e.g.][]
{klein18,klein19}. 
We forecast that in the 14,892~deg$^2$ of the low Galactic extinction sky
accessible to the eROSITA-DE collaboration, when raising the detection threshold at low redshift to exclude 
clusters with $M_{500\text{c}} \lessapprox 2\times 10^{14} \text{M}_{\sun}$, 
we predict that eROSITA will detect 13k clusters.  This baseline cosmology sample has a median mass of 
$\bar M_{500\text{c}} = 2.5\times 10^{14} \text{M}_{\sun}$ and a median redshift of $\bar z = 0.51$. 
For the case where we adjust the low redshift detection threshold to exclude clusters with
$M_{500\text{c}} \lessapprox 5\times 10^{13} \text{M}_{\sun}$, we predict 43k clusters. This sample has a median 
mass $\bar M_{500\text{c}} = 1.4\times 10^{14} \text{M}_{\sun}$, and a median redshift $\bar z = 0.31$.  
Both samples extend to high redshift with $\sim 400$ clusters at $z>1$.

We then analyze these mock samples using a prototype of the eROSITA cluster cosmology 
code that is an extension of the code initially
developed for SPT cluster cosmology analyses \citep{bocquet15,dehaan16,bocquet18}.  This codes employs a 
Bayesian framework for simultaneously
evaluating the likelihoods of cosmological and scaling relation parameters given the distribution of 
clusters in observable and redshift together 
with any direct mass measurement information.  The scaling relation between 
the selection observable (eROSITA count rate) and the mass and redshift is parametrized as a power law with log-normal 
intrinsic scatter.  Final 
parameter constraints are marginalized over the uncertainties (systematic and statistical) in the parameters of the mass--
observable scaling relation.

We first estimate the optimal level of mass calibration necessary for the number counts of eROSITA clusters 
to mainly inform the constraints on the cosmological parameters. This requires a calibration of the amplitude 
of the mass observable relation at 4.2\%, the mass trend of the scaling relation at 2.4\%, and the redshift 
trend at 5.3\%. These numbers are derived using current knowledge of the scatter 
around the mass luminosity relation. Furthermore, we determine that the mass trend of the rate mass relation 
has to be allowed to vary with redshift to enable the recovery of unbiased cosmological results.

We then examine cosmological constraints in three different cluster mass calibration contexts:  (1) using "baseline" 
constraints 
existing today that are taken from the recent SPT analysis of the X-ray luminosity and temperature mass relations 
\citep{bulbul19}, 
(2) using WL information from the DES+HSC survey and (3) using WL information from the future Euclid and LSST survey.
For the subset of the two catalogs that overlap the DES, HSC, Euclid or LSST survey footprints, we produce tangential shear 
profiles with 
appropriate characteristics for these surveys.
We also estimate the level of systematic mass uncertainties in the WL masses that 
would result from the data quality of these two surveys and from theoretical uncertainties in the impact of mis-centering and 
mis-fitting
the shear profiles.  We adopt mass uncertainties of 5.1\%,
1.3\% and 1.5\% for DES+HSC, Euclid, and LSST, respectively. These levels of systematic mass uncertainty will require that 
our understanding of the theoretical mass bias from simulations be improved by factors of 2 and 5 for DES+HSC 
and Euclid/LSST, respectively, in comparison to current work \citep{dietrich19}.  We note that achieving these improvements 
will 
require a significant investment of effort.

Throughout this 
work, we allow the summed neutrino mass to vary. All results are thus marginalized over 
the summed neutrino mass. In the $\nu$-$w$CDM model, we forecast that eROSITA number counts will 
constrain the density of matter in the Universe $\Omega_\text{M}$ to 0.032, the amplitude of fluctuation $\sigma_8$ to 0.052, 
and the equation of state parameter of the dark energy $w$ to 0.101. 
Calibrating the masses of eROSITA clusters with DES+HSC (Euclid; LSST) WL will reduce these uncertainties to 0.023 
(0.016; 0.014), 0.017 (0.012; 0.010), and 0.085 (0.074; 0.071), respectively. 
We also find that eROSITA clusters alone will not provide 
appreciable constraints on the sum of the neutrino masses.

eROSITA number counts will be able to break several degeneracies in current CMB constraints, especially on late 
time parameters such as $\Omega_\text{M}$, $\sigma_8$ and $w$. In combination with Planck constraints 
from the measurement of the angular auto- and crosscorrelation functions of CMB temperature and 
polarization anisotropies, we determine that eROSITA will constrain these parameters to 0.019, 0.032 and 
0.087 when adopting "baseline" priors on the scaling relation parameters. These uncertainties shrink to 0.018 
(0.014; 0.013), 0.019 (0.010; 0.009) and 0.085 (0.074; 0.069) when calibrating the masses with DES+HSC (Euclid; LSST) 
WL 
information. 

When considering the $\nu$-$\Lambda$CDM model, the upper limit on the neutrino mass of 0.514~eV from CMB 
alone can be improved to a constraint of 0.425~eV when utilizing number counts with the "baseline" priors, 0.404~eV 
when also considering DES WL calibration, and to 0.291~eV when calibrating with Euclid WL,
and 0.285~eV when calibrating with LSST WL.

We find that the constraining power of eROSITA cluster cosmology, even when calibrated with high quality shear 
profiles, is limited by a degeneracy between the scaling relation parameters and the cosmological distance to 
the clusters. This degeneracy arises, because the luminosity distance is necessary to transform observed 
count rates into luminosities, whose absolute and redshift dependent scaling with mass needs to be fitted 
simultaneously with the cosmological parameters that alter the redshift distance relation. This leads to the 
assessment that even the Euclid or LSST WL mass calibration will, by itself, not reach what we have defined as 
optimal levels in the $\nu$-$w$CDM model. 

However, we demonstrate that, with the inclusion of BAO measurements that constrain the redshift distance 
relation, the Euclid or LSST WL dataset can be used to calibrate cluster masses at an optimal level. 
Considering DESI-like BAO measurements, we project that eROSITA with Euclid WL mass 
calibration will constrain $\sigma_8$ to 0.005 and $w$ to 0.047, while the uncertainty on $\Omega_\text{M}$ 
will be dominated by the BAO measurement. 

Furthermore, we investigate the impact of lowering the mass limit to $M_\text{500c} \goa 5\times 10^{13} 
\text{M}_{\sun}$. Given the larger number of low mass clusters or groups, the eROSITA counts with 
Euclid WL can optimistically be used to determine $\Omega_\text{M}$ to 0.009, $\sigma_8$ to 0.007, and $w$ to 
0.056, if these low mass systems are simple extrapolations of the high mass systems.  The expected additional
complexity of these low mass systems would have to be modeled, and this additional modeling would likely 
weaken the cosmological constraints.

In summary, WL mass calibration from DES+HSC, Euclid, and LSST will significantly improve cosmological constraints from 
eROSITA cluster number counts, enabling a precise and independent cross-check of constraints from other measurements. 
The constraining power on $w$ suffers from an inherent degeneracy between the distance redshift relation and 
the scaling relation between the X-ray observable, mass and redshift. This degeneracy can be lifted by 
inclusion of other cosmological measurements, such as BAO or CMB measurements. In turn eROSITA 
cluster cosmology can break degeneracies in these other observations, underscoring the synergies between 
different cosmological experiments.
 
\section*{Acknowledgements}

We thank Hermann Brunner for help in accessing the eROSITA ARF, Thomas Reiprich, Tim Schrabback,
Andrea Merloni, Peter Predehl and Cristiano Porciani for the useful comments, and Matteo Costanzi, Steffen Hagstotz, David 
Rapetti, Tommaso 
Giannantonio, and Daniel Gruen for helpful conversations.  
We acknowledge financial support from the MPG Faculty Fellowship program, the DFG Cluster of Excellence 
``Origin and Structure of the Universe'', the DFG Transregio program TR33 ``The Dark Universe'',  and the 
Ludwig-Maximilians-Universit\"at Munich.
AS is supported by the ERC-StG ``ClustersXCosmo'' grant agreement 71676, and by the FARE-MIUR grant 
``ClustersXEuclid'' R165SBKTMA.
Numerical computations in this work relied on the \texttt{python} packages \texttt{numpy} \citep{numpy} and 
\texttt{scipy} \citep{scipy}. The plots were produced using the package \texttt{matplotlib} \citep{matplotlib}. The 
marginal contour plots were created using \texttt{pyGTC} \citep{pygtc}.


\bibliographystyle{mnras}
\bibliography{manuscript} 


\appendix

\section{Comments on selection}\label{app:selection}


\begin{figure*}
	\includegraphics[width=\textwidth]{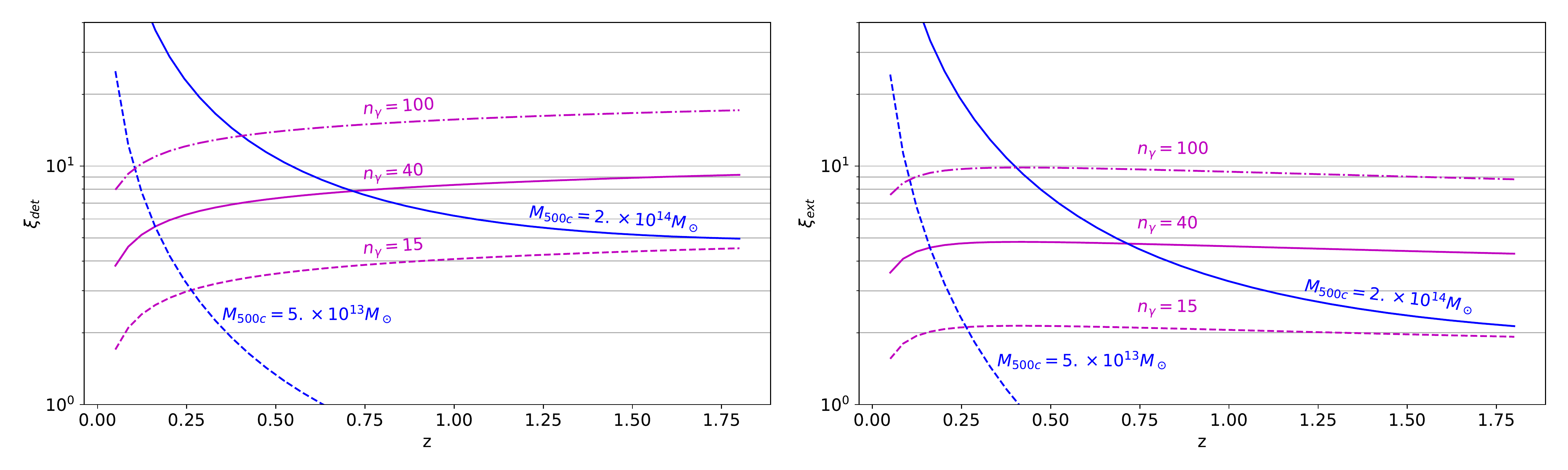}
	\vskip-0.15in
    \caption{Estimated significance of detection (left panel) and significance of extent (right panel) as functions of 
redshift for a cluster with 15, 40, and 100 photons, and clusters of halo mass $M_{500c}= 5\times 10^{13}$ 
M$_{\sun}$, and $ 2\times 10^{14}$ 
M$_{\sun}$ for median exposure time and background brightness. We find that 40 (15, 100) photon counts 
corresponds, at least, to a 8$\sigma$ (3$\sigma$, >10$\sigma$) detection and a $3.5 \sigma$ (2 $\sigma$, 9$\sigma$) 
significance of extent, 
rather independently of the cluster redshift.}
    \label{fig:selection}
\end{figure*}

In this work, we assume two selection criteria for our X-ray cluster sample: 

\begin{itemize}
\item a cut in measured number of photons $\hat n_\gamma > 40$ cts, which for the median eROSITA field with 
an exposure time of $1.6$ ks translates into a measured rate cut $\hat \eta > 2.5 \times 10^{-2}$ cts s$^{-1}$,

\item a cut in the observed mass 
\begin{equation}
\begin{split}
M_\text{obs} = & M_\text{0}\, e^{-\frac{\ln A_X}{B(z)}} \left(\frac{\hat \eta}{\eta_0}\right)^{\frac{1}{B(z)}} 
\left(\frac{E(z)}{E_0}\right)^{-\frac{2}{B(z)}} 
 \left(\frac{D_\text{L}(z)}{D_\text{L}(z_0)}\right)^{\frac{2}{B(z)}} \\
 & \left(\frac{1+z}{1+z_0}\right)^{-\frac{\gamma(z)}{B(z)}},
\end{split}
\end{equation}
which is derived from the rate-mass scaling relation equation~(\ref{eq:eta_mz}). It is evaluated for the fiducial 
cosmology and the fiducial scaling relation parameters derived in Appendix~\ref{app:scaling_relations}. The 
cut $M_\text{obs}>2\times 10^{14} \text{M}_{\sun}$, or $M_\text{obs}>5\times 10^{13} \text{M}_{\sun}$ is thus provided 
by a function of redshift, which is 
independent of cosmology and of the scaling relation parameters,
and leads to the above mentioned cuts $M_{500\text{c}}\goa2\times 10^{14} \text{M}_{\sun}$ or $M_{500\text{c}}
\goa5\times 10^{13} \text{M}_{\sun}$.
The low mass cut is thus implemented as a redshift dependent cut in observables \citep[as for instance also in] []
{vikhlinin09b, pillepich12}.

\end{itemize}

The cut at  $\hat n_\gamma > 40$ cts is justified by the following considerations on X-ray cluster detection. 
Detection of galaxy clusters hinges on the assumption that galaxy clusters are extended sources in the 
extragalactic X-ray sky, as discussed, for instance, by \citet{vikhlinin98} in the case of the ROSAT PSPC, 
\citet{pacaud06} in the case of XMM-{\it Newton}, and \citet{clerc18} in the context of eROSITA. For this 
reason, their extraction is usually divided into two steps: first all X-ray sources are identified, then, among the 
identified source, those who are extended are selected. As outlined in \citet{clerc18}, eROSITA will follow a 
similar procedure. 

Following \citet{pacaud06}, consider an X-ray image with a number of photons in each pixel $i$ given by $\hat n_i$, 
and define the following three likelihoods:

\begin{itemize}
\item the likelihood that the image is simply background with background brightness $\mu_\text{bkg}$, which reads
\begin{equation}
\ln\mathcal{L_\text{bkg}} = \sum_i \hat n_i \ln (\mu_\text{bkg}\, \text{A}_i) - \mu_\text{bkg}\, \text{A}_i, 
\end{equation}
where $\text{A}_i$ is the area of each pixel $i$;

\item  the likelihood of being a point source centered in $\vctr{x}_\text{ps}$ with total number of photons 
$n_\text{ps}$, given by
\begin{equation}
\begin{split}
\ln\mathcal{L_\text{ps}} = & \sum_i \hat n_i \ln (n_\text{ps} \text{PFS}(\vctr{x_\text{ps}})_i+\mu_\text{bkg}\, \text{A}_i) \\
& - n_\text{ps} \text{PFS}(\vctr{x_\text{ps}})_i-\mu_\text{bkg}\, \text{A}_i
\end{split}
\end{equation}
where $\text{PFS}(\vctr{x})_i$ is the value in the pixel $i$ of the average survey point spread function (PSF) 
centered in $\vctr{x}$; 

\item the likelihood of being a cluster with total number of photons $n_\gamma$, modeled as the convolution of a $
\beta$-model \citep{beta_model} with the PSF, which, for a cluster position $\vctr{x}$ and cluster core 
radius $\theta_c$, reads
\begin{equation}
\begin{split}
\ln\mathcal{L_\text{cl}} = & \sum_i \hat n_i \ln (n_\gamma S(\vctr{x}; \theta_c)_i+\mu_\text{bkg}\, \text{A}_i) \\
& - n_\gamma S(\vctr{x}; \theta_c)_i-\mu_\text{bkg}\, \text{A}_i, 
\end{split}
\end{equation}
where $S(\vctr{x}; \theta_c)_i$ stands for the value of the PSF convolved beta-profile with center $\vctr{x}$ and 
core radius $\theta_c$.
\end{itemize}

For the median eROSITA field, we expect $\mu_\text{bkg}=3.6$ cts arcmin$^{-2}$ \citep{clerc18}. To estimate the 
approximate significances of clusters with $n_\gamma = (15, 40, 100)$ at different redshifts, we use the rate-
mass relation derived in this work (c.f. Section~\ref{sec:scalingrelation}). Furthermore, we assume that the 
core radius is given by $\theta_c = 0.2 \theta_{500c}$, where $\theta_{500c}$ is the angular extent of the 
radius inclosing an over density 500 times the critical density of the Universe. We take the PSF to be a 
gaussian with half energy width of $24$ arcsec. Assuming $\beta=2/3$, we create an X-ray 10-by-10 arcmin 
image of the expected number of photons $\hat n$, by computing $\hat n_i=n_\gamma S(\vctr{x}; \theta_c)_i+
\mu_\text{bkg}\, \text{A}_i$. We intentionally do not draw a Poisson realization of the model, in order to 
capture the mean behavior of the extraction procedure.

On this image, the three likelihoods are then maximized by varying ($n_\text{ps}$, $\vctr{x}_\text{ps}$), and 
($n_\gamma$, $\theta_c$, $\vctr{x}$), respectively. We shall denote the maximum likelihood 
$\ln\hat{\mathcal{L}}_\alpha$ for $\alpha \in$ (bkg, ps, cl). 
To create an analogy to the SZE case\footnote{In the SZE case, the 
likelihoods above take the form of $\chi^2$ thanks to the Gaussian nature of the noise. Their maximization 
w.r.t. to the amplitude of the templates simplifies the problem to a maximization of the signal to noise as a 
function of scale and position. Maximizing the signal to noise is thus formally equivalent to maximizing the 
likelihood. As can readily be seen, this simplification does not apply to the Poissonian case of X-ray images. 
In both cases, however, we can define the maximum signal to noise, or significance, as 
$\xi = \sqrt{2 \ln \hat{\mathcal{L}}}$.}, 
which can be interpreted as "sigmas", both for detection 
$\xi_\text{det}=\sqrt{2(\ln \hat{\mathcal{L}}_\text{cl}-\ln\hat{\mathcal{L}}_\text{bkg})}$, 
and for extent 
$\xi_\text{ext} = \sqrt{2(\ln\hat{\mathcal{L}}_\text{cl}-\ln\hat{\mathcal{L}}_\text{ps})}$. 
Note that for well detected point sources, the best fit core radius $ \theta_c\approx 0$, 
such that 
$\ln \hat{\mathcal{L}}_\text{cl} \approx\ln\hat{\mathcal{L}}_\text{ps}$ 
and 
$\xi_\text{ext}\approx 0$, while $\xi_\text{det} \gg 0$. 
In contrast, for well detected extended sources, 
$\xi_\text{det} \gg 0$ and  $\xi_\text{ext}\gg 0$. 
For practical purposes, the region in which the likelihood is maximized is often pre-selected.

We repeat this exercise for different redshifts. The results are shown in Fig.~\ref{fig:selection}. We find that a 
source with 40 (15, 100) photons will be detected, at least, at 8$\sigma$ (3$\sigma$, 10$\sigma$) for redshifts above 0.5.  
 Furthermore, 
we determine that these clusters will have a significance of extent of 3.5$\sigma$ (2$\sigma$, 9$\sigma$).
The significance of extension at, e.g., 15 or 40 photons, is rather low.  Extent acts as a secondary 
selection on a sample which contains approximately 10\% clusters and 90\% point sources. 
Considering that 3.5$\sigma$ (2$\sigma$) corresponds to a p-value of $2.3\times10^{-4}$ ($2.3\times10^{-2}$), 
the extent cut would create a cluster sample with an approximate contamination of 0.2\%
(17\%) by point sources after the X-ray selection.
With the optical followup of a tool like MCMF, any X-ray cluster candidate without an associated 
overdensity of red galaxies can be easily removed from the sample \citep{klein18,klein19}.

Fig.~\ref{fig:selection} also shows the redshift evolution of the significances for a cluster of fixed mass. The 
characterization of this evolution, especially its dependence on distance both through the flux and the angular 
extent, might be worth further investigation, as its knowledge would allow us to use the significance as a the 
primary X-ray observable. Such a study is currently limited by the somewhat simplistic assumption that the 
core radius is a multiple of the virial radius $\theta_{500c}$. Furthermore, a study of the measurement 
uncertainty on the significance would be necessary.


\section{Comments on the count rate to mass relation}\label{app:scaling_relations}

To obtain unbiased cosmological results, the parametric form of the scaling relation needs to provide 
accurate mass predictions over the mass and redshift range of interest for the sample considered. This is 
necessary, as systematic biases in mass lead to a systematic misestimation of the abundance of clusters in a 
given observable redshift range. These misestimations of cluster abundance will then be compensated by 
shifts in the cosmological parameters, which would generically lead to systematic biases in the best 
fitting cosmological parameters. Therefore, it is of paramount importance to utilize a parametric 
form for the scaling relation that has adequate freedom to describe the sample being modeled.

\begin{figure}
	\vskip-0.2in
	\includegraphics[width=\columnwidth]{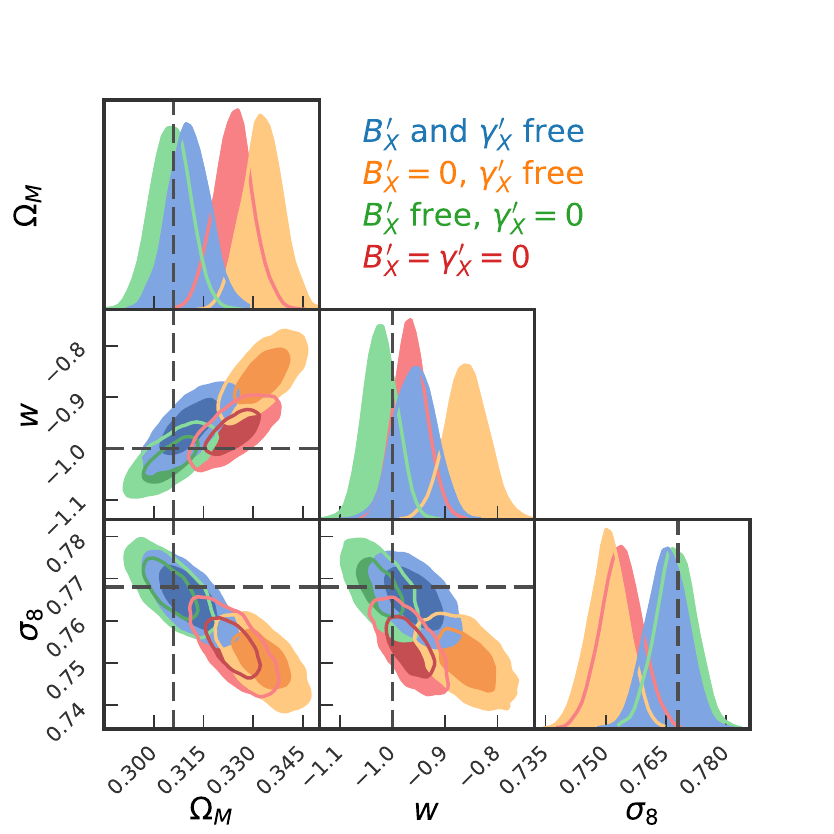}
	\vskip-0.10in
    \caption{Cosmological constraints derived from a ten times larger sample in four cases: free $B_\text{X}
^\prime$ and $\gamma_\text{X}^\prime$, free $B_\text{X}^\prime$ and $\gamma_\text{X}^\prime=0$,
$B_\text{X}^\prime=0$ and free $\gamma_\text{X}^\prime$,  and $B_\text{X}^\prime=\gamma_\text{X}
^\prime=0$. Noticeable shift in the inferred cosmological parameters occur once $B_\text{X}^\prime=0$, 
whereas the case of free $B_\text{X}^\prime$ and $\gamma_\text{X}^\prime=0$ is very similar to the maximal 
case.}
    \label{fig:cosmo_params_shift}
\end{figure}

\begin{figure*}
	\includegraphics[width=\textwidth]{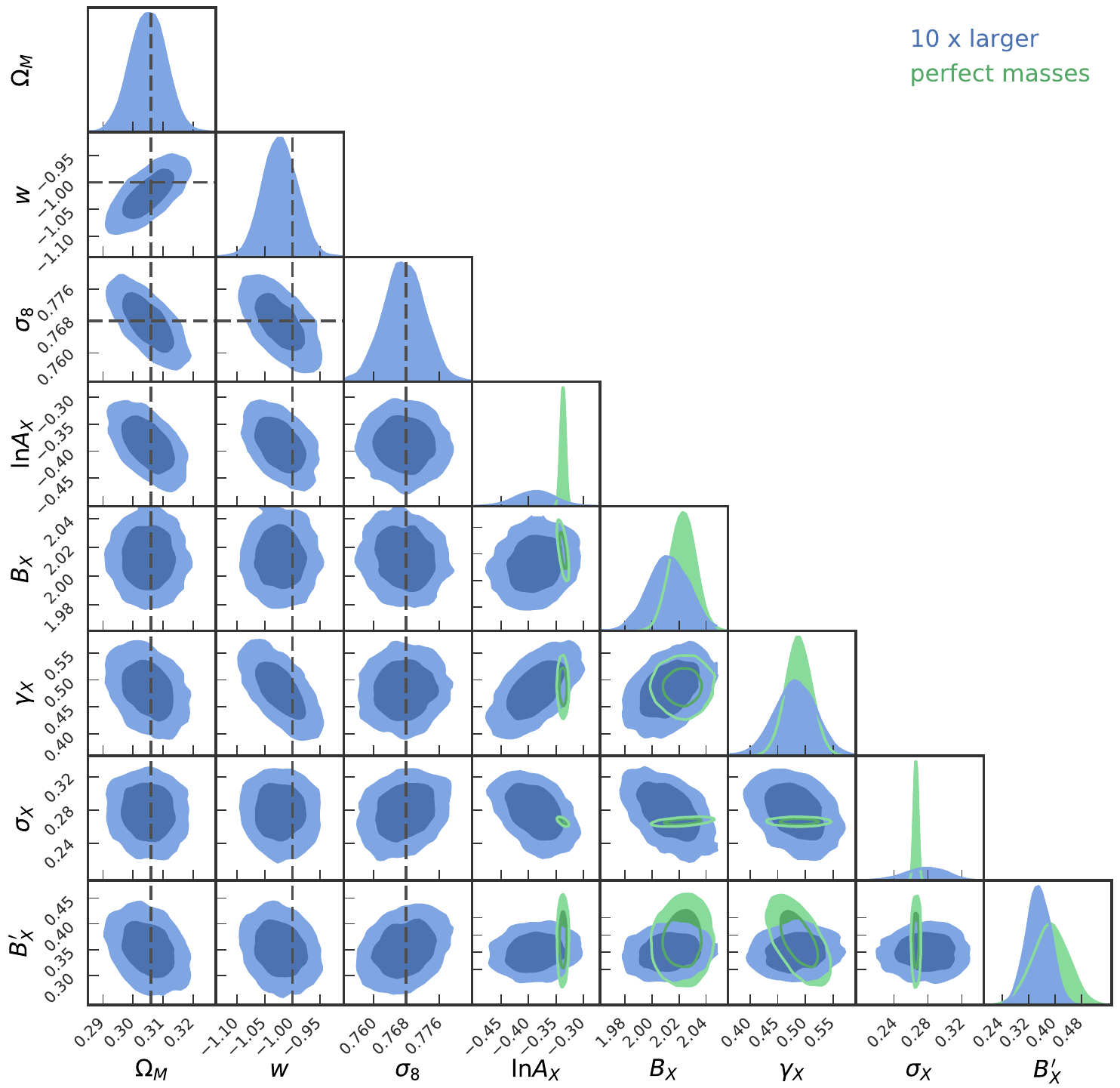}
	\vskip-0.15in
    \caption{Posterior constraints of the number counts and the Euclid WL mass calibration of the 10 times 
larger validation mock (blue), and the constraints on the scaling relation parameters from the perfect masses 
mass calibration (green). All values are consistent within two sigma, indicating that our likelihoods are unbiased 
at a level which is small compared to the Poisson noise in our mock eROSITA sample.}
    \label{fig:validation}
\end{figure*}

\subsection{Scaling relation form}\label{sec:fid_SR_params}

It is worth noting that
even though the observed X-ray luminosity to mass and temperature to mass relations show no evidence 
supporting a redshift dependent mass trend or a changing power law in the redshift trend \citep{bulbul19}, 
the mass and redshift dependences of the eROSITA count rate to flux conversion could introduce 
these additional trends into the count rate to mass scaling relation (equation~\ref{eq:eta_mz}).
Here we explore whether these additional freedoms are needed in the forecasts we perform.

The ratio between the eROSITA count rate of a cluster and its rest 
frame 0.5-2~keV flux, which we shall call $K = \eta / f_\text{X}$, parameterizes the response of 
the eROSITA cameras to a specific spectral form and the k-correction necessary to 
account for the transformation from rest frame 0.5-2~keV to the observed frame.  As such, one would expect it
to be both redshift and temperature dependent.
The rate can be written as
\begin{equation}
    \eta = K \, f_\text{X} = K\, \frac{L_X}{4 \upi d_\mathrm{L}^2(z)}.
\end{equation}
Thus, conceptually, the rate mass redshift scaling equals the luminosity mass redshift scaling, with the addition of 
the luminosity distance dependence and the mass and redshift trends of the flux to rate conversion.

$K$ has a noticeable but weak redshift dependence, and this redshift dependence is different for clusters of different 
temperature. The redshift dependence leads to the 
difference between the redshift trends of the rate and luminosity to mass relations (i.e., parameters $\gamma_\text{X}$ 
and $\gamma_\text{L}$), while the temperature dependence of this 
redshift trend combined with the temperature to mass redshift scaling leads to a non vanishing redshift variation in 
the mass trend of the rate scaling relation.

The baseline scaling relation for our analysis (equation~\ref{eq:eta_mz}) already has a redshift dependent mass trend 
parameter $B_\text{X}^\prime$ (equation~\ref{eq:masstrend}).  Here we introduce a generalization to 
equation~(\ref{eq:eta_mz}) by allowing a redshift trend $\gamma^\prime_\text{X}$ in the redshift slope 
\begin{equation}
\gamma(z) = \gamma_\text{X} + \gamma_\text{X}^\prime \ln \left(\frac{1+z}{1+z_0}\right), 
\end{equation}
which allows for the power law index in redshift to change with redshift.    

We examine the importance of this additional redshift dependence by fitting the scaling relation described by
the full parameter set $(\ln A_\text{X}, B_\text{X}, \gamma_\text{X}, \sigma_\text{X}, B_\text{X}^\prime,  
\gamma_\text{X}^\prime)$ using both true masses and Euclid WL mass constraints
 and allowing the parameter values to vary within large priors. 
Such an analysis is straightforward in the controlled regime of a mock catalog analysis where true masses and
therefore the underlying form of the scaling known. In analyzing the real 
eROSITA dataset, one must use direct mass constraints like those
from WL or dynamical masses to carry out a goodness of fit test for any proposed scaling relation form 
\citep[as done in the SPT analyses;][]{bocquet15,dehaan16}.

We analyze the number counts with perfect and with Euclid WL constraints for the 10 times larger validation mock (cf. 
Section~\ref{sec:validation}) using effectively four different mass observable relations. As a baseline we sample 
both $B_\text{X}^\prime$ and $\gamma_\text{X}^\prime$, the two parameters that allow for redshift variation of the mass
and redshift trends. 
We then also consider the cases where either or both of the two extra parameters are set to 
zero. The resulting constraints on the cosmological parameters of interest are shown in 
Fig.~\ref{fig:cosmo_params_shift}. We find that the run with $\gamma_\text{X}^\prime=0$ has constraints 
comparable to the baseline case where both $B_\text{X}^\prime$ and $\gamma_\text{X}^\prime$ are free. On the other 
hand, the run with $B_\text{X}^\prime=0$ produces biased cosmological parameters. This finding is consistent with the 
conclusions of \citet{pillepich12}, who-- although not using a rate mass scaling with empirical calibration, noted nonetheless 
that the effective rate mass scaling they derived had a mass slope which varied with redshift. We thus conclude that the 
parameter $B_\text{X}^\prime$ needs to be sampled, while the parameter $\gamma^\prime_\text{X}$ can be 
fixed to zero without biasing the cosmological inference at a level that is important, given the statistical uncertainties.  
In principle this parameter could also be left free to float without any bias implications, but adding more free parameters
than needed in the mass observable relation tends to reduce the precision of the constraints on all parameters.

\begin{table}
	\centering
	\caption{Results for the scaling relation parameters when using the perfect halo mass $M_\text{500c}$ to 
calibrate the mass observable relation. These values are used as fiducial values for the scaling relation 
parameters in the rest of the work.}
	\label{tab: pfctmssclbr}
	\begin{tabular}{ccccc} 
		\hline
		$\ln A_\text{X}$ & $B_\text{X}$ & $\gamma_\text{X}$ & $\sigma_{X}$ & $B^\prime_\text{X}$ \\
		\hline
 		-0.328 & 1.997 & 0.446 & 0.278 & 0.355 \\ 
 		\hline
 	\end{tabular}
\end{table}

\subsection{Fiducial parameter values}

To determine the fiducial values for the parameters of the scaling between rate and mass, we 
sample the mass calibration likelihood with perfect masses.  We adopt the form justified in the previous section, 
where $\gamma^\prime_\text{X}=0$ (see equation~\ref{eq:eta_mz}).
The best fit scaling relation parameters when perfect masses are used are reported in Table~\ref{tab: 
pfctmssclbr}. They are used as fiducial values in several occasions during this work. The value of $\ln 
A_\text{X}$ can be rescaled arbitrarily by changing the pivot points $\eta_\text{0}$, $M_\text{0}$ and 
$z_\text{0}$. $B_\text{X}$ and $\sigma_\text{X}$ take values very similar to the mass slope of the luminosity 
mass relation $B_\text{L}$ and the scatter around that relation $\sigma_\text{L}$, respectively.  For the values presented
in Table~\ref{tab: pfctmssclbr}, we have sampled the mass calibration likelihood for perfect masses discussed in 
Eq.~(\ref{eq:pfct_mss_lkl}). The value for $\gamma_\text{X}$ is larger then the redshift slope of the luminosity--mass 
scaling. Also, there is a clear preference
for redshift evolution of the mass trend.


\bsp	
\label{lastpage}
\end{document}